\documentclass[journal]{IEEEtran}
\usepackage{graphicx}
\usepackage{epstopdf}

\usepackage{subcaption}
\usepackage{color}
\usepackage[usenames,dvipsnames, table, xcdraw]{xcolor}
\usepackage{epsf,exscale,times}
\usepackage[english]{babel}
\usepackage{amsmath}
\usepackage{fancyhdr}
\usepackage{hyperref}
\usepackage{glossaries}
\usepackage{comment}
\usepackage{multirow}
\usepackage{amssymb}
\usepackage{algorithm}
\usepackage[utf8]{inputenc}
\usepackage[noend]{algpseudocode}

\usepackage{cite}
\usepackage{mathtools}
\usepackage{array}

\newcolumntype{P}[1]{>{\centering\arraybackslash}p{#1}}
\newcolumntype{M}[1]{>{\centering\arraybackslash}m{#1}}

\DeclareMathOperator*{\subjectto}{subject~to}
\DeclareMathOperator*{\minimize}{minimize}

\DeclareMathOperator*{\maximum}{max}

\DeclareMathOperator*{\argmin}{arg\,min}
\DeclareMathOperator*{\Diag}{Diag}

\newcommand{\ignore}[1]{}
\newcommand{\bfeta}{\mbox{\boldmath{$\eta$}}}
\newcommand{\bfgamma}{\mbox{\boldmath{$\gamma$}}}

\newtheorem{remark}{Remark}

\newacronym{Pd}{Pd}{detection probability}
\newacronym{BW}{BW}{bandwidth}
\newacronym{MI}{MI}{mutual interference}
\newacronym{FM}{FM}{frequency modulated}
\newacronym{APAS}{APAS}{Almost Perfect auto-correlation Sequences}
\newacronym{MPS}{MPS}{Minimum Peak Sidelobe} 
\newacronym{PMCW}{PMCW}{Phase Modulated Continuous Wave}
\newacronym{MF}{MF}{Merit Factor}
\newacronym{SNR}{SNR}{Signal to Noise Ratio}
\newacronym{INR}{INR}{Interference to Noise Ratio}
\newacronym{SIR}{SIR}{Signal-to-Interference Ratio}
\newacronym{SINR}{SINR}{Signal to Interference plus Noise Ratio}
\newacronym{AF}{AF}{Ambiguity Function}
\newacronym{MIMO}{MIMO}{Multiple Input Multiple Output}
\newacronym{SISO}{SISO}{Single Input Single Output}
\newacronym{CD}{CD}{Coordinate Descent}
\newacronym{BCD}{BCD}{Block Coordinate Descent}
\newacronym{GD}{GD}{Gradient Descent}
\newacronym{MM}{MM}{Majorization-Minimization}
\newacronym{FMCW}{FMCW}{Frequency Modulated Continuous Wave}
\newacronym{CDM}{CDM}{Code Division Multiplexing}
\newacronym{DFT}{DFT}{Discrete Fourier Transform}
\newacronym{FFT}{FFT}{Fast Fourier Transform}
\newacronym{MVDR}{MVDR}{Minimum Variance Distortionless Response}
\newacronym{MBI}{MBI}{Maximum Block Improvement}
\newacronym{RFPA}{RFPA}{Radio Frequency Power Amplifier}
\newacronym{BPSK}{BPSK}{Binary Phase Shift Keying}
\newacronym{QPSK}{QPSK}{Quadrature Phase Shift Keying}
\newacronym{ULA}{ULA}{Uniform Linear Array}
\newacronym{DOF}{DOF}{Degrees of Freedom}
\newacronym{PSK}{PSK}{Phase Shift Keying}
\newacronym{PSL}{PSL}{Peak Sidelobe Level}
\newacronym{ISL}{ISL}{Integrated Sidelobe Level}
\newacronym{PSLR}{PSLR}{Peak Sidelobe Level Ratio}
\newacronym{ISLR}{ISLR}{Integrated Sidelobe Level Ratio}
\newacronym{LFM}{LFM}{Linear Frequency Modulation}
\newacronym{HPM}{HPM}{Hybrid Phased MIMO}
\newacronym{MPSK}{MPSK}{M-ary Phase Shift Keying}
\newacronym{LPI}{LPI}{Low Probability of Intercept}
\newacronym{RoC}{RoC}{Radar-on-Chip}
\newacronym{RF}{RF}{Radio-Frequency}
\newacronym{PAR}{PAR}{Peak-to-Average Power Ratio}
\newacronym{LTE}{LTE}{Long Term Evolution}
\newacronym{DL}{DL}{Down Link}
\newacronym{UL}{UL}{Up Link}
\newacronym{iid}{i.i.d.}{independent and identically distributed}
\newacronym{BS}{BS}{Base Station}
\newacronym{BSUM}{BSUM}{Block Successive Upper-bound Minimization}
\newacronym{SDR}{SDR}{Software-Defined Radio}
\newacronym{OTA}{OTA}{Over-The-Air}
\newacronym{USRP}{USRP}{Universal Software Radio Peripheral}
\newacronym{FoV}{FoV}{field of view}
\newacronym{CPI}{CPI}{Coherent Pulse Interval}
\newacronym{LoS}{LoS}{Line of Sight}
\newacronym{Tx}{Tx}{Transmitter}
\newacronym{Rx}{Rx}{Receiver}
\newacronym{LNA}{LNA}{low-noise amplifier}
\newacronym{TI}{TI}{Texas Instrument}
\newacronym{PRT}{PRT}{pulse repetition time}
\newacronym{TBP}{TBP}{time bandwidth product}
\newacronym{XO}{XO}{Crystal Oscillators}
\newacronym{GUI}{GUI}{graphical user interface}
\newacronym{ADC}{ADC}{analog-to-digital converter}
\newacronym{TDM}{TDM}{time division multiplexing}
\newacronym{FDM}{FDM}{frequency division multiplexing}
\newacronym{GPS}{GPS}{global positioning system}
\newacronym{CFAR}{CFAR}{constant false alarm rate}
\newacronym{COTS}{COTS}{commercial-off-the-shelf}
\newacronym{OOP}{OOP}{object oriented programming}
\newacronym{MISL}{MISL}{monotonic minimizer for integrated sidelobe level}
\newacronym{MM-PSL}{MM-PSL}{Monotonic Minimizer for the $\ell_p$-norm of autocorrelation sidelobes}
\newacronym{CAN}{CAN}{Cyclic Algorithm New}
\newacronym{PECS}{PECS}{Polynomial phase Estimate of Coefficients for unimodular Sequences}
\newacronym{LS}{LS}{Least Squares}
\newacronym{BSD}{BSD}{Blind Spot Detection}
\newacronym{ACC}{ACC}{Adaptive Cruise Control}
\newacronym{SRR}{SRR}{Short Range Radar}
\newacronym{PRF}{PRF}{pulse repetition frequency}
\newacronym{JRC}{JRC}{Joint Radar Communication}
\begin{document}

\title{Designing Interference-Immune Doppler-Tolerant Waveforms for  Radar Applications}
%
\author{Robin Amar,~
        ~Mohammad Alaee-Kerahroodi,~
        Prabhu Babu,~
        and~Bhavani Shankar M. R.~
\thanks{Robin Amar, ~ SnT, University of Luxembourg, Luxembourg.\\
	email: robin.amar@uni.lu }
\thanks{Mohammad Alaee-Kerahroodi and Bhavani Shankar M. R. are with SnT, University of Luxembourg, Luxembourg.}
\thanks{Prabhu Babu is  with  IIT Delhi.}%
\thanks{This work was supported by FNR (Luxembourg) through the CORE project ``SPRINGER: Signal Processing for Next Generation Radar’’ under grant  C18/IS/12734677/SPRINGER.}
\thanks{Manuscript received December 16, 2021; revised mm dd, yyyy.}}
%
\markboth{IEEE Transactions on Aerospace and Electronic Systems}%
{Shell \MakeLowercase{\textit{et al.}}: Bare Demo of IEEEtran.cls for IEEE Journals}
\maketitle
\begin{abstract}
%
Dynamic target detection using  {FMCW} waveform   is challenging in the presence of interference
for different  radar applications. Degradation in {SNR} is irreparable and interference is difficult to mitigate in time and frequency domain.
In this paper, a waveform design problem is addressed using the Majorization-Minimization (MM) framework by considering PSL/ISL cost functions, resulting in a code sequence with Doppler-tolerance characteristics of an FMCW waveform and interference immune characteristics of a tailored PMCW waveform (unique phase code + minimal ISL/PSL).
The optimal design sequences possess polynomial phase behavior of degree $Q$ amongst its sub-sequences and obtain optimal ISL and PSL solutions with guaranteed convergence. 
By tuning the optimization parameters such as degree $Q$ of the polynomial phase behavior, sub-sequence length $M$ and the total number of sub-sequences $L$, the optimized sequences can be as Doppler tolerant as {FMCW} waveform in one end, and they can possess small cross-correlation values similar to random-phase sequences in PMCW waveform on the other end. 
If required in the event of acute interference, new codes can be generated in the runtime which have low cross-correlation with the interferers. 
The performance analysis 
indicates that the proposed method outperforms the state-of-the-art counterparts. 
\end{abstract}
\IEEEpeerreviewmaketitle
\section{Introduction}
RADAR has traditionally been associated with military and law enforcement applications due to its development, but it is now a common solution for civil situations.
Currently, high-resolution 
\gls{FMCW}
radar sensors 
operating at $60$GHz, $79$GHz, and $140$GHz with sometimes finer than 
$10$cm range resolution 
are becoming integral in a variety of applications ranging from automotive safety and autonomous driving \cite{6127923,7870764}, indoor positioning \cite{6908039, 7758620} to infant and elderly health monitoring \cite{9266404}.
As the penetration of these low-cost high-performance sensors grows, the likelihood of radar signal interference and the associated ghost target problems also grows \cite{9266283, 9466178}.

The transmit waveform that is commonly used in a large number of modern millimeter wave (mmWave) radar sensors is based on \gls{LFM}.
\gls{LFM} is well-known in radar literature due to its distinctive properties of large time-bandwidth product (high pulse compression ratio) and high Doppler-tolerance  \cite{richards2010principles,984612}. In automotive \gls{FMCW} radars, \gls{LFM} is typically used as the waveform modulation scheme,  since  it can be compressed with a very low-cost and efficient technique known-as \textit{de-chirping} operation\footnote{Also referred to as \emph{stretch processing}.} \cite{jankiraman2018fmcw}. 
The primary benefit of de-chirping is that the received signal can be sampled at much lower rates in comparison to its bandwidth \cite{4760820,7446254,7455584}. Not surprisingly, this advantage has been motivating many manufactures to build their radar system based on \gls{FMCW} technology. While \gls{FMCW} remains the most prevalent modulation scheme for mmWave radar sensors, alternate modulation schemes such as \gls{PMCW} \cite{4775877,7485114} have been proposed as a way to mitigate interference for different applications \cite{levanon1994comparison, 7485114, 8891682}.
Using \gls{PMCW} technology, 
Doppler-tolerant property can be obtained by using 
a class of codes characterized by a systematic generation
formula \cite{levanon1994comparison}, such as, Frank \cite{1057798}, P1, P2, P3, and P4 \cite{4102706}, Golomb \cite{256535}, Chu \cite{1054840}, PAT \cite{6172584},  etc (refer Appendix \ref{ap:Chirplike} for more details).
These codes typically exhibit small \gls{PSL} and \gls{ISL} values in their aperiodic autocorrelation function; however, they are unfortunately sometimes constrained to the specific lengths \cite{Levanon_Radar}. Also, due to a lack of uniqueness in these codes, they are equally vulnerable to  \gls{MI} similar to \gls{FMCW} technology.
We aim to overcome this limitation in this paper by proposing a framework for designing constant modulus polyphase Doppler-tolerant sequences that are also immune to interference for variety of radar applications.
\vspace{-4mm}
\subsection{Doppler-Tolerant Waveforms}
When a target is stationary, i. e., the Doppler distortions of the returned radar signals can be neglected, then correlation or matched-filter processing is relatively straight-forward \cite{1446820}. In those applications where high resolution requirements and high target speeds combine, the distortions in the waveform lead to severe
degradations, and a knowledge of the Doppler shift of the target should be available, or a bank of mismatched filter needs to be considered in the receive side to compensate loss in \gls{SNR}
\cite{4644058}. However, using so-called Doppler-invariant waveforms in the transmit side is a simpler alternative approach to dealing with high Doppler shifts. In this case, even in the presence of an arbitrarily large Doppler shift, the received signal remains matched to the filter, but a range-Doppler coupling may occur as an unintended consequence \cite{1446820}.

\begin{figure}
	\centering
	\includegraphics[width=.95\linewidth]{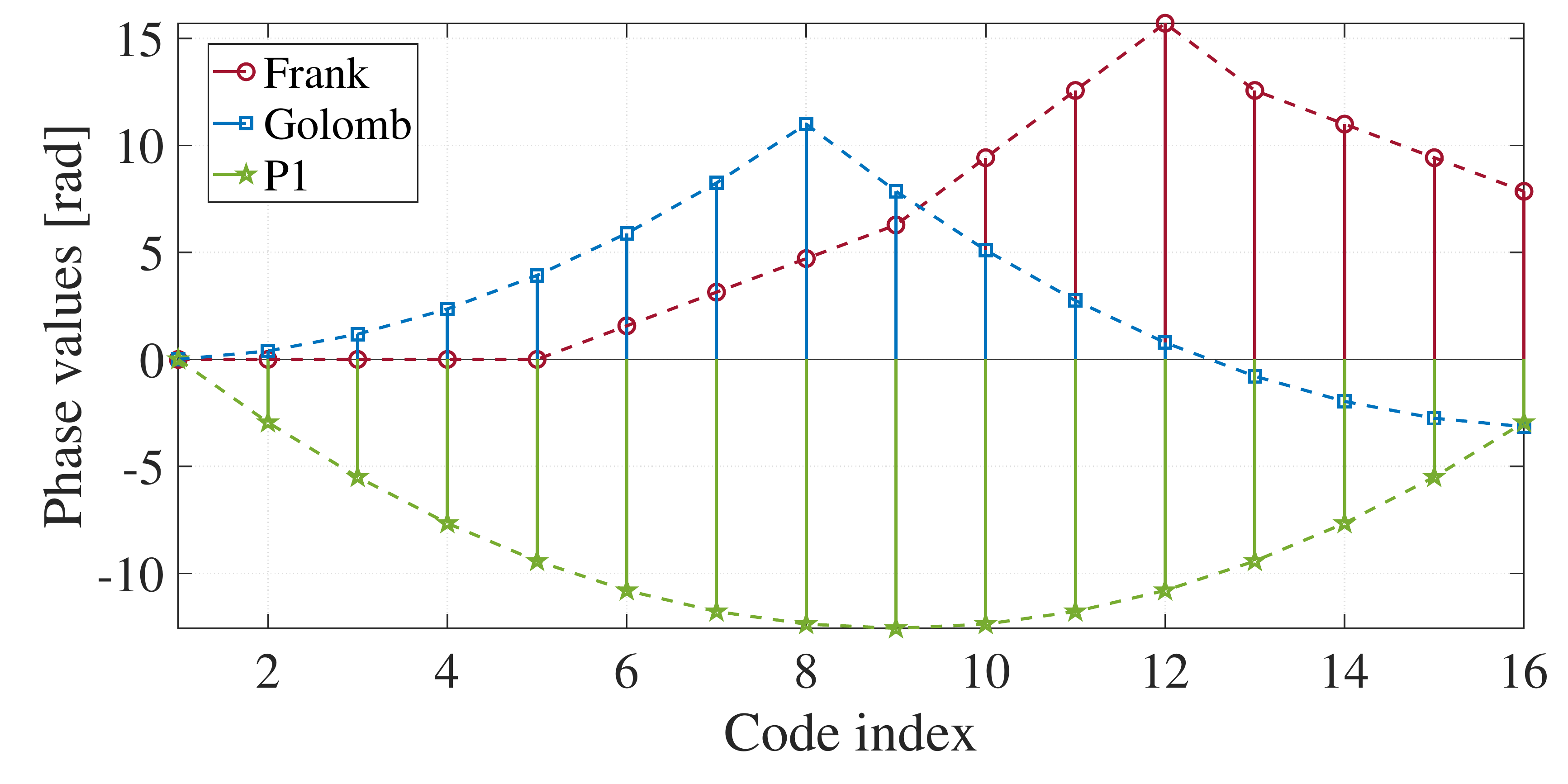}
	\caption{The unwrapped phase values of three polyphase codes of length $N = 16$: Frank, Golomb, and P1.}
	\label{fig:PhaseValues}
\end{figure}

It is known that \gls{FMCW} waveform has the Doppler-tolerant property by its nature. Thus, a possible solution to build \gls{PMCW} with Doppler-tolerant properties is to
mimic the behaviour of \gls{FMCW} waveforms by using the phase history of a pulse with linearly varying frequency
, and building \textit{chirplike polyphase sequences} \cite{Levanon_Radar}. 
Indeed, because frequency is the derivative of phase, in order to have linear frequency properties like \gls{FMCW}, the phase variation of \gls{PMCW} sequences should be quadratic throughout the length of the sequence, as shown in \figurename{~\ref{fig:PhaseValues}} for Frank, Golomb, and P1 sequences of length $N = 16$. 
Interestingly, chirplike polyphase sequences are known to typically have small
\gls{PSL} and \gls{ISL} values\footnote{\gls{PSL} shows the maximum autocorrelation sidelobe of a transmit waveform. If this value is not small, then either a false detection or a miss detection may happen. Similar properties hold for \gls{ISL} of transmitting waveforms where the energy of autocorrelation sidelobes should be small to mitigate the deleterious effects of distributed targets, which are the case for automotive applications.}, the metrics which are 
strictly connected to the sharpness of the code autocorrelation function \cite{4644058,4749273,7362231,7967829,8168352,9440807,9469031,SANKURU2022108348,JYOTHI2021103142}.

Recently, several studies have considered the design of polyphase sequences with good Doppler tolerance  properties \cite{FENG2017143,6907982,7469294,8335406,8682216,9018020,9114756,9173387}. 
In \cite{FENG2017143} Linear-FM Synthesized (LFM-Syn) waveform has been proposed 
to achieve low range sidelobes and high Doppler tolerance in the synthesized waveform. The study 
combines the random noise waveform and the conventional \gls{LFM} waveform. In \cite{8682216} a \gls{CD}-based approach is proposed for designing waveform with high Doppler tolerance properties. In both 
\cite{FENG2017143} and \cite{8682216}, 
a template-matching objective function is established and optimized to obtain waveforms with low range sidelobes and high Doppler tolerance properties.
Unlike the template matching approaches, in \cite{6952459} 
a computational approach is proposed for designing polyphase sequences with \textit{piecewise linear phase properties}. Despite the fact designing Frank-like sequences was considered in this paper, the overall quadratic phase trend of this sequence was not taken into account during the proposed optimization framework. 
As a result, the optimization method mentioned in it does not result in Doppler-tolerant waveforms.
Different from designing waveform with Doppler tolerance properties, several studies like 
\cite{6563125,7707372,9091055,9104358}
have considered the problem 
of shaping \gls{AF} by 
minimizing sidelobes on range-Doppler plane around the origin to improve the detection performance. However, the optimized waveform by these studies is not  resilient to Doppler shifts.

\subsection{Interference in FMCW Radars}
Interference is a well-known and unavoidable problem in many fields.
In emerging automotive radar application, the radar sensors use \gls{FMCW} as the standard sensing waveform.
Vast deployment of \gls{FMCW} radars in multiple passenger and commercial vehicles has led to a significant vehicle-to-vehicle radar interference. 
The interference in all these cases is largely due to simultaneous use of shared spectrum when operating in the detection range of the other sensor and the inherent lack of coordination between radars resulting from the lack of a centralized control and resource allocation mechanism
\cite{4106078,5760761,8828037}.

Further, in this context, \gls{FMCW} radars experience similar-slope and sweeping slope interference \cite{9266283}, \cite{9466178} from other  sensors operating in the \gls{FoV}.
Even though sweeping interference can be avoided or repaired (using original signal reconstruction), similar-slope interference is hard to manage \cite{9466178}. Mitigation techniques are able to address both the interference\footnote{Although a comprehensive solution to mutual interference still remains an open issue, there has been some prior work in the design of mitigation techniques to diminish the problem of interference \cite{MOSARIM2012,8835681,9454981}.} but enhances the average noise resulting in the decrement of \gls{SNR} \cite{9127843}.
As all the available degrees of freedom (time, operating frequency, bandwidth) to design the \gls{FMCW} radars has already been exploited, \gls{PMCW} based sensors are being looked upon as a promising solution to match the sensing performance of \gls{FMCW} waveform and alleviate the problem of interference.

\gls{PMCW} radars employ phase codes in their transmitter which are constant modulus and can drive the transmit amplifier at maximum efficiency. 
These waveforms are highly immune to interference as every radar may employ its own unique phase code with small auto- and cross-correlation sidelobes \cite{7485114,7944482}. On the flip side, it suffers from Doppler intolerance which makes it hard to employ in high-speed applications. 
In the case study of the automotive scenarios, the high speed of the targets 
may result in \gls{SNR} reduction if the transmit waveform is not Doppler tolerant.

\subsection{Contribution}
The idea of this framework is to design polyphase sequences that have good Doppler-tolerance properties and sharp autocorrelation functions.
Unlike FMCW, potentially distinct PMCW sequences meeting these requirements exist. This richness in selection of PMCW sequences allows for better interference management than FMCW. In this context, we propose a methodology for generating such a rich set of sequences which can be integrated
into existing optimization-based waveform design tools (like \gls{CAN} \cite{4749273}, \gls{MISL} \cite{7093191}, \gls{MM-PSL} \cite{7362231}, etc.) to provide the designed waveforms with Doppler tolerance properties.   

The main contributions of this work can be summarized as follows:
\begin{itemize}
    \item \textit{Optimization framework}: The problem formulation leads to an objective function which considers the design of phase sequences with polynomial behavior while simultaneously considering other important properties like \gls{ISL} and \gls{PSL}. Although \gls{ISL} and \gls{PSL} minimization has considered earlier, but all these properties are not addressed simultaneously in literature.
    \item \textit{Doppler tolerance}: The optimization problem can handle phase sequences with a degree $Q$ polynomial behavior while building on the literature of $\ell_p$ norm minimization for generic objective subsuming \gls{ISL} and \gls{PSL}. 
    Using \gls{MM} based approach, we design sequences with superior properties as compared to the conventional sequences mentioned in literature. Doppler tolerance property arises in a sequence when its phase changes in a quadratic manner (i.e. $Q=2$). As this is being handled directly in the objective function, the resultant sequence is bound to have better performance for moving targets.
    \item \textit{Flexibility of design - divide and conquer interference}: This approach offers the flexibility of having multiple sub-sequences, designed in parallel while optimizing the ISL and PSL of the entire sequence. 
    Since each sub-sequence has unique polynomial phase coefficients of degree $Q$, length, and code, the entire sequence can avoid interference.
    This provides an additional degree of freedom for waveform design by partitioning the entire sequence into multiple sub-sequences of varying lengths which enables us to tune the optimization variable and generate diverse waveforms whose \gls{AF} is varying from \textit{Thumbtack type} on one extreme to \textit{Doppler-Tolerant} type on the other end.
    The proposed formulation provides the possibility of designing sequences of any length $N$ with piecewise polynomial phase behavior amongst its $L$ sub-sequences each of length $M_l$ where 
    $N = M_1 + M_2 + \cdots + M_L$ if the sub-sequences have different lengths. This approach can be considered for the scenarios in which 
    sequence lengths other than perfect square (unlike Frank) are required.
    \item \textit{Application in automotive scenarios}: A simulation environment is created which generates various automotive scenarios where interference is possible. Using the proposed optimization framework, different sequences have been generated and utilized to interfere with each other. The results of the simulation indicate high interference immunity in sensing. This alleviates the necessity for the requirement of additional sensing and communication protocol for safe operation in dense automotive scenarios where \gls{MI} is possible amongst multiple cars equipped with radar sensors.
\end{itemize}

Finally, by using the proposed method, one can construct many new such polyphase sequences which were not known and/or could not be constructed by the previous formulations in the literature. Thus, the proposed approach is capable of generating unique sequences which in turn would lead to decrement in the interference amongst multiple sensors operating in the nearby region.

\subsection{Organization and Notations}

The rest of this paper is organized as follows. In Section \ref{Sec:ProbFormul}, we formulate the $\ell_p$-norm minimization for designing polynomial phase sequences under unimodular and polynomial phase constraints. 
In Section \ref{Sec:PolyPhaseCodeDesign}, we develop an algorithm based on \gls{MM} framework, where we use a majorizing function to find the optimal solution of the design problem. In Section 
\ref{Sec:perfromanceAnalysis} 
we provide numerical experiments to verify the effectiveness of proposed algorithm. Finally, Section \ref{Sec:Conlcusion} concludes the paper. 

\textit{Notation:} Boldface upper case letters denote matrices, bold-face lower case letters denote column vectors, and italics denote scalars. $\boldsymbol{\mathbb{Z}}$, $\boldsymbol{\mathbb{R}}$ and $\boldsymbol{\mathbb{C}}$ denote the integer, real and complex field, respectively. $\Re(\cdot)$ and $\Im(\cdot)$ denote the real and imaginary part respectively. $\arg(\cdot)$ denotes the phase of a complex number. The superscripts $(\cdot)^T$, $(\cdot)^*$, $(\cdot)^H$ and $(\cdot)^\dagger$ denote transpose, complex conjugate, conjugate transpose, and pseudo-inverse respectively. $X_{i,j}$ denotes the 
$(i, j)^{th}$
element of a matrix and $x_i$ denotes the 
$i^{th}$
element of vector $\mathbf{x}$. 
$\Diag(\boldsymbol{X})$ is a column vector consisting of all the diagonal elements of $\boldsymbol{X}$. $\Diag(\mathbf{x})$ is a diagonal matrix formed with $\mathbf{x}$ as its
principal diagonal. $vec(\boldsymbol{X})$ is a column vector consisting of all
the columns of $\boldsymbol{X}$ stacked. 

\section{Problem Formulation}\label{Sec:ProbFormul}

Let $\{x_n\}_{n=1}^{N}$ be the transmitted complex unit-modulus radar code sequence of length $N$.
The aperiodic autocorrelation of the transmitting waveform at lag $k$ (e.g. matched filter output at the \gls{PMCW} radar receiver) is defined as
\begin{equation}\label{eq:rAP}
\begin{aligned}
r_k &= \sum_{n=1}^{N-k} x_n x_{n+k}^*=r_{-k}^*, ~~~ k \in [0,\ldots, N-1].
\end{aligned}
\end{equation}
The \gls{ISL} and \gls{PSL} can be mathematically defined by
\begin{equation}\label{eq:ISL}
    \text{ISL} = \sum_{k=1}^{N-1}|r_k|^2,
\end{equation}
\begin{equation}\label{eq:PSL1}
    \text{PSL} = \maximum_{k=1,2,\ldots,N-1}|r_k|.
\end{equation}
It is clear that the  \gls{ISL} metric is the squared
$\ell_2$-norm of the autocorrelation sidelobes. 
Further, the $\ell_\infty$-norm of autocorrelation sidelobes of a sequence is the \gls{PSL} metric.
These can, in fact, be generalized by considering the $\ell_p$ norm, $p\geq 1$ which offers additional flexibility in design while subsuming ISL and PSL.
In general $\ell_p$-norm metric of the autocorrelation sidelobes is defined as
\begin{equation}\label{eq:PSL2}
    \left(\sum_{k=1}^{N-1} |r_k|^p \right)^{1/p}, ~~2\leq p < \infty.
\end{equation}
Sequences are designed to minimize the various $\ell_p$ norm metric in literature \cite{7093191,7362231,7547360,8168352,raei2021design}. It is known that in general longer the code, better are the \gls{ISL}/\gls{PSL} \cite{7362231}, 
but the system suffers from high design complexity.
One possible solution for reducing complexity is to design long sequences with multiple segments, which is considered in this paper.
Let the sequence $\{x_n\}_{n=1}^{N}$
be partitioned into $L$ sub-sequences each having length of $M_l$, where $l \in \{1, 2, \ldots, L\}$
such that every sub-sequence of it, say 
\begin{equation}
    \widetilde{\mathbf{x}}_l = [x_{\{1,l\}}, x_{\{2,l\}}, \cdots, x_{\{M_l,l\}}]^T \in \mathbb{C}^{M_l},
\end{equation}
has a polynomial phase, which can be expressed as
\begin{equation}
    \arg(x_{\{m,l\}}) = \sum_{q=0}^{Q} a_{\{q,l\}} m^q,
\end{equation}
where $a_{\{q,l\}}$ is the $Q^{th}$ degree polynomial coefficient for the phase of the  $l$-th sub-sequence with  $q \in \{0, 1, 2, \ldots, Q\}$, and 
$ l = 1,\ldots, L$. 
Further,
\begin{equation}
    \begin{aligned}
        \widetilde{\mathbf{X}} = [\widetilde{\mathbf{x}}_1, \widetilde{\mathbf{x}}_2, \cdots, \widetilde{\mathbf{x}}_l] \in \mathbb{C}^{N},
    \end{aligned}
\end{equation}
is a collection of all the sub-sequences where $\{x_n\}_{n=1}^{N} = \text{vec(}\widetilde{\mathbf{X}}\text{)}$.
It can be observed that the length of each sub-sequence can be arbitrarily chosen.
The problem of interest 
is to design the code vector $\{x_n\}_{n=1}^{N}$ with
a generic polynomial phase of a degree $Q$ in its sub sequences while having impulse-like autocorrelation function.
Therefore, by considering $\sum_{k=1}^{N-1}|r_k|^p$ as the objective function, the optimization problem can be compactly written as
\begin{equation}\label{eq:PSLOptC2}
{\cal{P}}_{1}
\begin{dcases}
\minimize_{a_{\{q,l\}}}& \sum_{k=1}^{N-1}|r_k|^p \\
\subjectto 
            & \arg(x_{\{m,l\}}) = \sum_{q=0}^{Q} a_{\{q,l\}} m^q,\\
            & |x_{\{m,l\}}| = 1 , 
\end{dcases}
\end{equation}
where $m = 1,\ldots, M_l$, and  $l = 1,\ldots, L$.
 
In the following section, we  propose an efficient algorithm to design 
a set of sub-sequences with polynomial phase relationship of degree $Q$ amongst its sub-sequences
 based on \gls{MM} framework.

The \gls{MM} (majorization-minimization) - is an approach to solve optimization problems that are difficult to solve directly. The principle behind the MM method is to transform a difficult problem into a series of simple problems. Suppose, we want to minimize $f(\mathbf{x})$ over $\boldsymbol{\chi} \subseteq \mathbb{C}$. Instead of minimizing the cost function $f(\mathbf{x})$ directly, the MM approach optimizes the sequence of approximate objective functions that majorize $f(\mathbf{x})$.
More specifically, starting from a feasible point $\mathbf{x}^{(0)}$, the algorithm produces a sequence ${\mathbf{x}^{(i)}}$ according to the following update rule
\begin{equation}\label{eq:MM1}
    \mathbf{x}^{(i+1)} \in \displaystyle{\argmin_{\mathbf{x} \in \boldsymbol{\chi}}}~ u(\mathbf{x}, \mathbf{x}^{(i)}),
\end{equation}
where $\mathbf{x}^{(i)}$ is the point generated by the algorithm at iteration $i$, and $u(\mathbf{x}, \mathbf{x}^{(i)})$ is the majorization function of $f(\mathbf{x})$ at $\mathbf{x}^{(i)}$. Formally, the function $u(\mathbf{x},\mathbf{x}^{(i)})$ is said to majorize the function $f(\mathbf{x})$ at the point $\mathbf{x}^{(i)}$ if
\begin{equation}\label{eq:MM2}
    \begin{aligned}
    u(\mathbf{x}, \mathbf{x}^{(i)}) \geq& f(\mathbf{x}), \forall \mathbf{x} \in \boldsymbol{\chi},\\
    u(\mathbf{x}^{(i)}, \mathbf{x}^{(i)}) =& f(\mathbf{x}^{(i)}),\\
    \nabla u(\mathbf{x}^{(i)}, \mathbf{x}^{(i)}) = & \nabla f(\mathbf{x}^{(i)}).
    \end{aligned}
\end{equation}
In other words, function $u(\mathbf{x}, \mathbf{x}^{(i)})$ is an upper bound of $f(\mathbf{x})$ over $\boldsymbol{\chi}$ and coincides with $f(\mathbf{x})$ at $\mathbf{x}^{(i)}$.
To summarize, in order to minimize $f(\mathbf{x})$ over $\boldsymbol{\chi} \subseteq \mathbb{C}^n$, the main steps of the majorization-minimization scheme are:
\begin{enumerate}
    \item Find a feasible point $\mathbf{x}^{(0)}$ and set $i=0$.
    \item Construct a function $u(\mathbf{x}, \mathbf{x}^{(i)})$ that majorizes $f(\mathbf{x})$ at $\mathbf{x}^{(i)}$ and is easier to optimize.
    \item Let $\mathbf{x}^{(i+1)} \in \displaystyle{\argmin_{\mathbf{x} \in \boldsymbol{\chi}}}~ u(\mathbf{x}, \mathbf{x}^{(i)})$.
    \item If some convergence criterion is met, exit; otherwise, set $i = i+1$ and go to step (2).
\end{enumerate}
It is easy to show that with this scheme, the objective value is monotonically decreasing at every iteration, i.e.,
\begin{equation}\label{eq:MM3}
    f(\mathbf{x}^{(i+1)}) \leq u(\mathbf{x}^{(i+1)}, \mathbf{x}^{(i)}) \leq u(\mathbf{x}^{(i)},\mathbf{x}^{(i)}) = f(\mathbf{x}^{(i)}).
\end{equation}
The first inequality and the third equality follow from the properties of the majorization function, namely \eqref{eq:MM2} and the second inequality follows from \eqref{eq:MM1}. The monotonicity makes \gls{MM} algorithms very stable in practice.
For more details on \gls{MM} framework, refer \cite{7547360, hunter2004tutorial, 10.5555/3086980} and references therein.

\begin{figure}[t!]
    \centering
    \includegraphics[width=.95\linewidth]{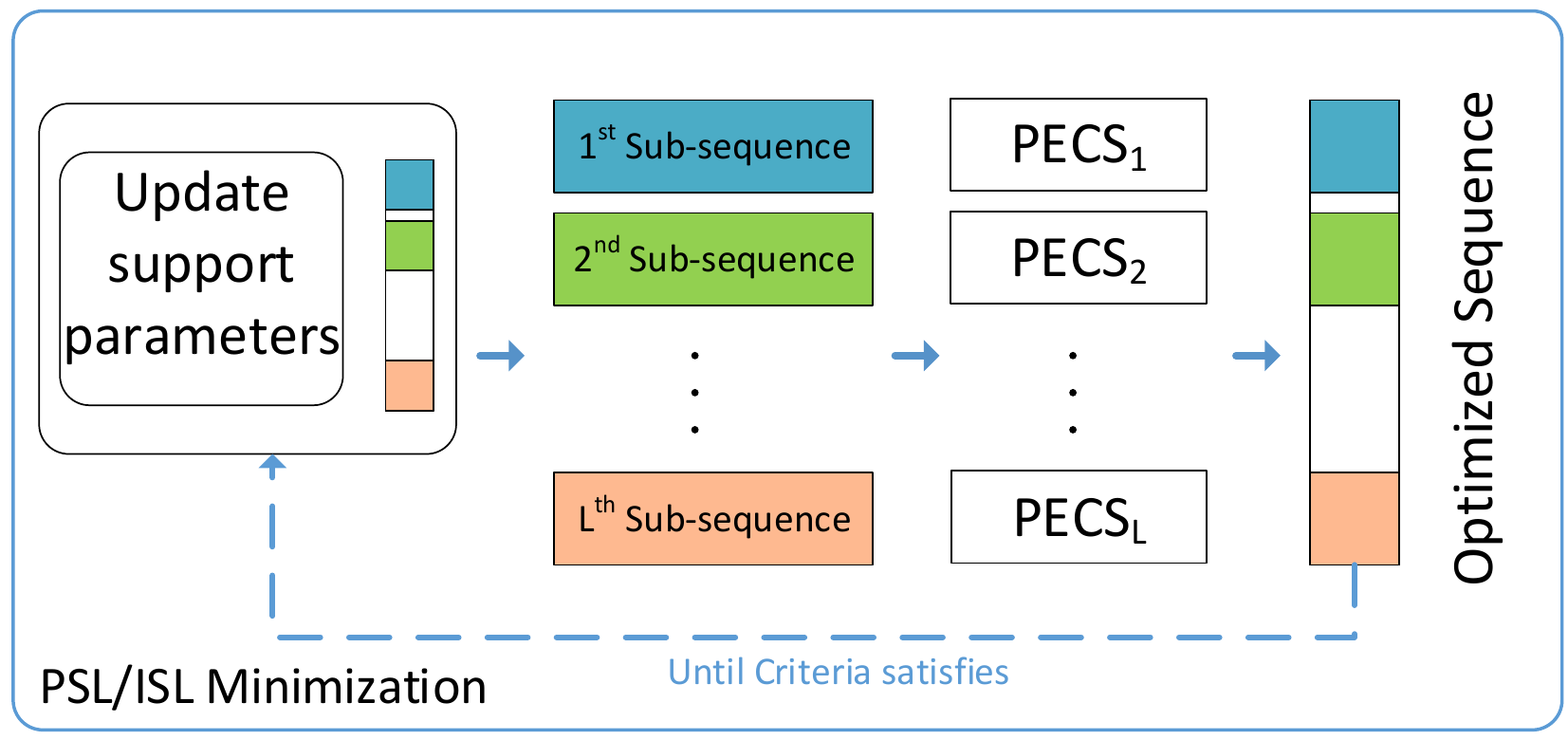}
    \caption{Workflow of PECS for PSL/ISL Minimization.}
    \label{fig:PECS_MISL}
\end{figure}

\section{Proposed Method}\label{Sec:PolyPhaseCodeDesign}

The optimization problem mentioned in \eqref{eq:PSLOptC2} is hard to solve since each $\mathbf{r}_k$ is quadratically related to $\{x_n\}_{n=1}^{N}$ and each $\{x_n\}_{n=1}^{N}$ is non-linearly related to $a_{\{q,l\}}$. Moreover, the $\ell_p$ norm of autocorrelation sidelobes becomes difficult to solve using classical optimization approach. As a result \gls{MM} is considered and after several majorization steps (refer Appendix-\ref{Ap:MPSL_Dervn}), simplifies to the following optimization problem
\begin{equation}\label{eq:PSLfinalOpt}
{\cal{P}}_{2}
\begin{dcases}
    \minimize_{a_{\{q,l\}}}& ~~ || \mathbf{x} - \mathbf{y}||_2\\
    \subjectto & \arg(x_{\{m,l\}}) = \sum_{q=0}^{Q} a_{\{q,l\}} m^q,\\
               & |x_{\{m,l\}}| = 1 , 
\end{dcases}
\end{equation}
where
\begin{equation*}
    \mathbf{y} = \left( \lambda_{\text{max}}(\boldsymbol{L})N + \lambda_u \right) \mathbf{x}^{(i)} - \widetilde{\boldsymbol{R}}\mathbf{x}^{(i)},
\end{equation*}
with $\lambda_{\text{max}},\boldsymbol{L}, N, \lambda_u \ignore{\mathbf{x}^{(i)}}$ and $\widetilde{\boldsymbol{R}}$ defined in Appendix \ref{Ap:MPSL_Dervn}. Note that in \cite{7362231} the polynomial phase constraint was not considered. Since the objective in \eqref{eq:PSLfinalOpt} is separable in the sequence variables, the minimization problem can now be split into $L$ sub-problems (each of which can be solved in parallel).
Let us define, $\boldsymbol{\rho} = |\mathbf{y}|$, and $\boldsymbol{\psi} = \arg(\mathbf{y})$, where $\rho_m$ and $\psi_m$ are the magnitude and phase of every entry of $\mathbf{y}$, respectively. 
Also, for ease of notation let us assume that the polynomial phase coefficients and sub-sequence length of the $l$-th sub-sequence, say $a_{\{q,l\}}$ and $M_l$ are indicated as $\widetilde{a}_{q}$ and $\widetilde{M}$ respectively. 
Thus, dropping the subscript-$l$, each of the sub-problem can be further defined as
\begin{equation}\label{eq:ISLOptC6}
{\cal{P}}_{3}
\begin{dcases}
\minimize_{\widetilde{a}_{q}}& \sum_{m=1}^{\widetilde{M}}\left| e^{j(\sum_{q=0}^{Q} \widetilde{a}_{q} m^q)} - \rho_m e^{j\psi_m}\right|^2,
\end{dcases}
\end{equation}
where we have considered the unimodular and polynomial phase constraints of Problem ${\cal{P}}_{2}$ directly in the definition of the code entries in Problem ${\cal{P}}_{3}$.
Further, the above problem can be simplified as
\begin{equation}\label{eq:ISLOptC7}
\begin{dcases}
\minimize_{\widetilde{a}_{q}}&-\left[\sum_{m=1}^{\widetilde{M}}  {\rho}_m \cos \left(\sum_{q=0}^{Q} \widetilde{a}_{q} m^q - {\psi}_m  \right) \right].
\end{dcases}
\end{equation}
The ideal step would be to minimize the majorized function in \eqref{eq:ISLOptC7} for $\widetilde{a}_{q}$
given the previous value of $\widetilde{a}_{q}^{(i)}$, the objective function can be upper-bounded (at a given $\widetilde{a}_{q}^{(i)}$). 
However, since the optimization variables are in the argument of the cosine function in the objective of \ref{eq:ISLOptC7}, the solution to this problem is not straight-forward. 
Hence, we resort to a second MM step. Towards this,
let us define\footnote{$\theta_m$ depends on the optimization variables $\widetilde{a}_{q} $.}
\begin{equation*}\label{eq:App2_1_d}
\theta_m =  \sum_{q=0}^{Q} \widetilde{a}_{q} m^q - {\psi}_m,   
\end{equation*}
a majorizer ($g(\theta_m,\theta_m^{(i)})$) of the function 
$f(\theta_m) = -\rho_m \cos(\theta_m)$
can be obtained by
\begin{equation}  \label{eq:App2_2}
    \begin{aligned}
        g(\theta_m,&\theta_m^{(i)})  = -\rho_m \cos(\theta_m^{(i)}) + (\theta_m - \theta_m^{(i)})\rho_m \sin(\theta_m^{(i)}) +  \\
        & \frac{1}{2} (\theta_m - \theta_m^{(i)})^2 \rho_m \cos(\theta_m^{(i)}) \geq - \rho_m cos(\theta_m),
    \end{aligned}
\end{equation}
where 
 $\theta_m$ is the variable and 
$\theta_m^{(i)}$ 
is the phase value of the last iteration. 
This follows from exploiting the fact that if a function is continuously differentiable with a Lipschitz continuous gradient, then second order Taylor expansion can be used as a majorizer \cite{7547360}.
Using the aforementioned majorizer function, at $i$-th iteration of \gls{MM} algorithm, 
the optimization problem
\begin{equation} \label{eq:TaylorMajorizer}
{\cal{P}}_{4}
\begin{dcases}
\minimize_{\widetilde{a}_{q}}  & \sum_{m=1}^{\widetilde{M}} \Bigg[ -\rho_m \cos(\theta_m^{(i)}) + \\
&\left(\theta_m - \theta_m^{(i)}\right)\rho_m \sin(\theta_m^{(i)})\\
& + \frac{1}{2} \left( \theta_m - \theta_m^{(i)}\right)^2 \rho_m\cos(\theta_m^{(i)}) \Bigg]
\end{dcases}
\end{equation}
\begin{table}[t!]
    \caption{Supporting parameters for Algorithm \ref{Algo:PseudoCode_PECS_LpNorm}
    }
    \centering
         \begin{tabular}{|M{1.5em}|M{4em}|M{21em}|}
            \hline
            \textbf{S.No}   &    \textbf{Parameter}    & \textbf{Relation}     \\\hline\hline
            1 & $\boldsymbol{F}$ & $2N \times 2N$ FFT Matrix with $F_{m,n} = e^{-j\frac{2mn\pi}{2N}}$\\\hline
            2 & $f$ & $\boldsymbol{F}[\mathbf{x}^{(i)T}, \mathbf{0}_{1\times N}]$ \\ \hline
            3 & $\mathbf{r}$ & $\frac{1}{2N} \boldsymbol{F}^H |\mathbf{f}|^2$  \\ \hline
            4 & $t$ & $||\mathbf{r}_{2:N}||_p$ \\ \hline
            5 & $a_k$ & $\frac{1 +(p-1)\left( \frac{|r_{k+1}|}{t} \right)^p - p\left( \frac{|r_{k+1}|}{t} \right)^{p-1} }{(t - |r_{k+1}|)^2}$ \\ \hline
            6 & $\hat{w}_k$ & $\frac{p}{2t^2} \left(\frac{|r_{k+1}|}{t}\right)^{p-2}, k = 1,\ldots, N-1$ \\ \hline
            7 & $\Tilde{\mathbf{c}}$ & $ \mathbf{r} \circ [ \hat{w}_1,\dots,\hat{w}_{N-1},0,\hat{w}_{N-1},\ldots,\hat{w}_1]^T$ \\ \hline
            8 & $\widetilde{\boldsymbol{\mu}}$ & $ \boldsymbol{F}\Tilde{\mathbf{c}}$ \\ \hline
            9 & $\lambda_L$ & $\max_k \{\Tilde{\alpha}_k(N-k) | k = 1,\ldots,N\}$ \\ \hline
            10 & $\lambda_u$ & $\frac{1}{2}(\max_{(1\leq \Tilde{i} \leq N)} \Tilde{\mu}_{2\Tilde{i}} + \max_{(1\leq \Tilde{i} \leq N)} \Tilde{\mu}_{2\Tilde{i}-1})$ \\ \hline
            \end{tabular}
          \label{Table:SuprtAlgoPECS}
\end{table}
\begin{algorithm}[t!]
\caption{PECS}
\label{Algo:PseudoCode_PECS_LpNorm}
\begin{algorithmic}[1]
\Require{Seed sequence $\mathbf{x}^{(0)}$, $N$, $\widetilde{M}$, $L$ and $p$ }
\Ensure{$\mathbf{x}$} 
\State Set $i = 0$, initialize $\mathbf{x}^{(0)}$.
\While{stopping criterion is not met}
\State Calculate $\boldsymbol{F},\widetilde{\boldsymbol{\mu}}, \mathbf{f}, \lambda_L, \lambda_u$ from \tablename{~\ref{Table:SuprtAlgoPECS}}
\State $\mathbf{y} = \mathbf{x}^{(i)} - \frac{\boldsymbol{F}_{:,1:N}(\widetilde{\boldsymbol{\mu}} \circ \mathbf{f})}{2N(\lambda_L N + \lambda_u)}$
\State $\boldsymbol{\psi} = \arg(\boldsymbol{y})$ $|~\boldsymbol{\psi} = [\widetilde{\boldsymbol{\psi}}_1^T, \cdots, \widetilde{\boldsymbol{\psi}}_L^T]^T$, $\widetilde{\boldsymbol{\psi}}_l \in \mathbb{R}^{\widetilde{M}}$
\State $\boldsymbol{\rho} = |\mathbf{y}|$ $|~\boldsymbol{\rho} = [\widetilde{\boldsymbol{\rho}}_1^T, \cdots, \widetilde{\boldsymbol{\rho}}_L^T]^T$, $\widetilde{\boldsymbol{\rho}}_l \in \mathbb{R}^{\widetilde{M}}$
\For{$l\leftarrow 1 ~to ~L$}
\State $\widetilde{\boldsymbol{\psi}}_l = [\psi_1, \cdots, \psi_{\widetilde{M}}]^T$
\State $\widetilde{\boldsymbol{\rho}}_l = [\rho_1, \cdots, \rho_{\widetilde{M}}]^T$
\State ${\theta}_m^{(i)} = \sum_{q=0}^{Q} \widetilde{a}_{q} m^q - {\psi}_m $, ~$m = 1,\ldots,\widetilde{M}$
\State $\widetilde{b}_m = - \rho_m \cos{\theta_m^{(i)}}\left( \psi_m + \theta_m^{(i)} \right) + \rho_m \sin{\theta_m^{(i)}}$
\State $\mathbf{\eta} = [1, 2, 3, \cdots, \widetilde{M}]^T \in \mathbb{Z}^{\widetilde{M}}$
\State $\widetilde{\boldsymbol{A}} = [\mathbf{\eta}^0, \mathbf{\eta}^1, \cdots, \mathbf{\eta}^{Q}] \in \mathbb{Z}^{\widetilde{M} \times Q},$
\State $\mathbf{z} = [\widetilde{a}_0, \widetilde{a}_1, \cdots, \widetilde{a}_{Q}]^T \in \mathbb{R}^Q$
\State $\widetilde{\mathbf{b}} = [\widetilde{b}_1, \widetilde{b}_2, \cdots, \widetilde{b}_{\widetilde{M}}]^T \in \mathbb{R}^{\widetilde{M}}$
\State $\mathbf{z}^{\star} = \widetilde{\boldsymbol{A}}^{(\dagger)
}\widetilde{\mathbf{b}}$
\State $\widetilde{\mathbf{x}}_l = e^{(j(\widetilde{\boldsymbol{A}}\mathbf{z}^{\star}))}$
\EndFor
\State $\widetilde{\mathbf{X}} = [\widetilde{\mathbf{x}}_1, \widetilde{\mathbf{x}}_2, \cdots, \widetilde{\mathbf{x}}_L  ]$
\State $\mathbf{x}^{(i+1)} = \text{vec}({\widetilde{\mathbf{X}}}) $
\State $i \leftarrow i + 1$
\EndWhile \textbf{return} $\mathbf{x}^{(i+1)}$
\end{algorithmic}
\end{algorithm}
The objective function in \eqref{eq:TaylorMajorizer} can be rewritten into perfect square form and the constant terms independent to the optimization variable $\widetilde{a}_{q}$ can be ignored.
Thus, a surrogate optimization problem deduced from \eqref{eq:TaylorMajorizer} is given below
\begin{equation} \label{eq:FinalMajorizer}
{\cal{P}}_{5}
\begin{dcases}
\minimize_{\widetilde{a}_{q}} ~ & ~ \sum_{m=1}^{\widetilde{M}} \Bigg[ -\rho_m \cos(\theta_m^{(i)}) \left(\sum_{q=0}^{Q} \widetilde{a}_{q} m^q\right)  + \widetilde{b}_m \Bigg]^2
\end{dcases}
\end{equation}
where $\widetilde{b}_m = -\rho_m \cos(\theta_m^{(i)})\left( \psi_m + \theta_m^{(i)} \right) + \rho_m \sin(\theta_m^{(i)})$.

Now, considering a generic sub-sequence index $l$ we define
\begin{equation*}
    \bfeta = [1, 2, 3, \cdots, \widetilde{M}]^T \in \mathbb{Z}^{\widetilde{M}},
\end{equation*}
$\bfeta^q$ implying each element of $\bfeta$ is raised to the power of $q$, $q = 0, 1, \ldots, Q$.
Further,
\begin{equation}\label{eq:LstSqr_Init}
    \begin{aligned}
        \bfgamma &= \rho_m \cos(\theta_m^{(i)}) \odot [1,\cdots, 1]^T \in \mathbb{R}^{\widetilde{M}},\\
        \widetilde{\boldsymbol{A}} &= \Diag{(\bfgamma)} [\bfeta^0, \bfeta^1, \cdots, \bfeta^{Q}] \in \mathbb{Z}^{\widetilde{M} \times Q},\\
        \mathbf{z} &= [\widetilde{a}_0, \widetilde{a}_1, \cdots, \widetilde{a}_{Q}]^T \in \mathbb{R}^Q,\\
        \widetilde{\mathbf{b}} &= [\widetilde{b}_1, \widetilde{b}_2, \cdots, \widetilde{b}_{\widetilde{M}}]^T \in \mathbb{R}^{\widetilde{M}},
    \end{aligned}
\end{equation}
the optimization problem in 
\eqref{eq:FinalMajorizer} can be rewritten as
\begin{equation}\label{eq:LstSqr}
\begin{dcases}
    \minimize_{\mathbf{z}}~&~|| \widetilde{\boldsymbol{A}}\mathbf{z} - \widetilde{\mathbf{b}} ||_2^2,
\end{dcases}
\end{equation}
which is the standard \gls{LS} problem. 
As a result, the optimal $\mathbf{z}^{\star} = \widetilde{\boldsymbol{A}}^{(\dagger)
}\widetilde{\mathbf{b}} =[\widetilde{a}_0^{\star}, \widetilde{a}_1^{\star}, \cdots, \widetilde{a}_{Q}^{\star}]^T $ would be calculated\footnote{We can use ``lsqr'' in \textit{Sparse Matrices Toolbox} of MATLAB 2021a to solve \eqref{eq:LstSqr}.} and the optimal sequence will be synthesized. 

Using the aforementioned setup for a generic sub-sequence index $l$, we calculate all the  $\widetilde{\mathbf{x}}_l$s pertaining to different sub-sequences and derive $\widetilde{\mathbf{X}}$.
The sequence for the next iteration is derived by vectorizing $\widetilde{\mathbf{X}}$ where vector length is again $N$. 
The algorithm successively improves the objective and an optimal value of $\mathbf{x}$ is achieved.
Details of the implementation for the proposed method in the form of pseudo code are summarized in Algorithm \ref{Algo:PseudoCode_PECS_LpNorm}\footnote{MATLAB codes of Algorithm \ref{Algo:PseudoCode_PECS_LpNorm} can be shared per request.} and would be referred further as \gls{PECS}.

\begin{remark}
\textit{Computational complexity}\\
Assuming $L$ sub-sequences are processed in parallel, the computational load of Algorithm - \ref{Algo:PseudoCode_PECS_LpNorm} is dependent on deriving (i) the supporting parameters: $\mathbf{f}$, $\mathbf{r}$, $t$, $\mathbf{a}$ and $\mathbf{\hat{w}}$ mentioned in Table \ref{Table:SuprtAlgoPECS} and (ii) the least squares operation in every iteration of the algorithm. 
In (i), the order of computational complexity is $\boldsymbol{O}(2N)$ real additions/subtractions, $\boldsymbol{O}(Np)$ real multiplications, $\boldsymbol{O}(N)$ real divisions and $\boldsymbol{O}(N\log_2 N)$ for \gls{FFT}.
In (ii), assume $M_1 = M_2 =\cdots =  M_l = M$ for simplicity, therefore, the complexity of least squares operation: {$\boldsymbol{O}(M^2 Q)$ + $\boldsymbol{O}(Q^2 M)$ + $\boldsymbol{O}(QM)$} real matrix multiplications and $\boldsymbol{O}(Q^{3})$ real matrix inversion  \cite{10.5555/3174304.3175337} \cite{GoluVanl96}.
Therefore, the overall computational complexity is $\boldsymbol{O}(M^2 Q)$ (provided $M > Q$ which is true in general). 
In case the $L$ sub-sequences are processed sequentially, the complexity is $\boldsymbol{O}(M^2 LQ)$.
\end{remark}
\subsection*{Extension of other methods to PECS}
In the previous section, by applying a constraint of $Q$-th degree polynomial phase variation on the sub-sequences, we have addressed the problem of 
minimizing the autocorrelation sidelobes to obtain optimal \gls{ISL}/\gls{PSL}
of the complete sequence using $\ell_p$ norm minimization with the method \gls{MM-PSL}.
\subsubsection{Extension of MISL}\label{ap:ISL_Minmzn}
In \cite{4749273}, the \gls{ISL} metric minimization is addressed with a different approach. As discussed previously in \eqref{eq:ISL}, it is just the squared $\ell_2$-norm of the autocorrelation sidelobes.
Therefore, the \gls{ISL} minimization problem under piece-wise polynomial phase constraint of degree $Q$ can be written as follows
\begin{equation}\label{eq:ISLOptC1}
{\cal{M}}_{1}
\begin{dcases}
\minimize_{a_{\{q,l\}}}& \sum_{k=1}^{N-1}|r_k|^2 \\
\subjectto 
            & \arg(x_{\{m,l\}}) = \sum_{q=0}^{Q} a_{\{q,l\}} m^q,\\
            & |x_{\{m,l\}}| = 1 , ~~ m = 1,\ldots, M, \\
            & ~~~~~~~~~~~~~~~~~~ l = 1,\ldots, L.
\end{dcases}
\end{equation}
where $a_{\{q,l\}}$ indicate the coefficients of $l$-th segment of the optimized sequence whose phase varies in accordance to the degree of the polynomial $Q$.
It has been shown in \cite{4749273} that the \gls{ISL} metric of the aperiodic autocorrelations can be equivalently expressed in the frequency domain as
\begin{equation}\label{eq:ISLOptC2}
    \text{ISL} = \frac{1}{4N} \sum_{g=1}^{2N} \left[ \left| \sum_{n=1}^{N} x_n e^{-j\omega_g(n-1)} \right|^2 - N\right]^2 \\
\end{equation}
where $\omega_g = \frac{2\pi}{2N}(g-1)$, $g=1,...,2N$. 
%
Let us define $\mathbf{x}=[x_1, x_2, \ldots , x_N]^T, ~\mathbf{b}_g=[1, e^{j\omega_g},.... ,e^{j\omega_g(N-1)}]^T, \textrm{where  } g = 1, \ldots, 2N$.\\
Therefore, writing (\ref{eq:ISLOptC2}) in a compact form
\begin{equation}\label{eq:ISLOptC3}
\text{ISL} = \sum_{g=1}^{2N} \left( \mathbf{b}_{g}^{H} \mathbf{x} \mathbf{x}^H \mathbf{b}_g\right)^2 
\end{equation}
%

The \gls{ISL} in \eqref{eq:ISLOptC3} is  quartic with respect to $\mathbf{x}$ and its minimization is still difficult. 
The \gls{MM} based algorithm (\gls{MISL}) developed in \cite{7093191} computes a minimizer of (\ref{eq:ISLOptC3}). So given any sequence $\mathbf{x}$, the surrogate minimization problem in \gls{MISL} algorithm is given by
\begin{equation}\label{eq:ISLOptC4}
{\cal{M}}_{2}
\begin{dcases}
\minimize_{a_{\{q,l\}}}& \Re(\mathbf{x}^H (\Grave{\boldsymbol{A}} - 2N^2\mathbf{x}^{(i)} (\mathbf{x}^{(i)^H})) \mathbf{x}^{(i)}) \\
\subjectto & \arg(x_{\{m,l\}}) = \sum_{q=0}^{Q} a_{\{q,l\}} m^q\\
           & |x_{\{m,l\}}| = 1.
\end{dcases}
\end{equation}
where $\boldsymbol{A} = [b_1,\ldots,b_{2N}]$, $\mathbf{f}^{(i)}=|\boldsymbol{A}^H\mathbf{x}^{(i)}|^2$, $f_{max}^{(i)}=\text{max}_f\{f_g^{(i)}: g = 1,\ldots,2N\}$,
$\Grave{\boldsymbol{A}} = \mathbf{A} (\Diag(\mathbf{f}^{(i)}) - f_{\text{max}}^{(i)}\mathbf{I})\mathbf{A}^H$. 
The problem in (\ref{eq:ISLOptC4}) is majorized once again and the surrogate minimization problem is given as
\begin{equation}\label{eq:ISLOptC5}
{\cal{M}}_{3}
\begin{dcases}
\minimize_{a_{\{q,l\}}}& \Vert \mathbf{x} - \mathbf{y} \Vert_2 \\
\subjectto & \arg(x_{\{m,l\}}) = \sum_{q=0}^{Q} a_{\{q,l\}} m^q\\
           & |x_{\{m,l\}}| = 1.
\end{dcases}
\end{equation}
where $ \mathbf{y} = -\mathbf{A} (\Diag(\mathbf{f}^{(i)}) - f_{\text{max}}^{(i)}\mathbf{I})\mathbf{A}^H\mathbf{x}^{(i)}$.
Once the optimization problem in \eqref{eq:ISLOptC5} is achieved, i.e ${\cal{M}}_{3}$, it is exactly equal to the problem in \eqref{eq:PSLfinalOpt}, i.e. ${\cal{P}}_{2}$  and hence its solution\ignore{the solution for ${\cal{P}}_{2}$} can be pursued further. The details of the implementation can be found in Algorithm \ref{alg:euclid} and \ref{alg:PseudoCodeMISL}.
\begin{algorithm}[!htbp]
\caption{{PECS} subroutine}\label{alg:euclid}
\begin{algorithmic}[1]
\Procedure{PECS}{$\mathbf{y}^{(i)}, L, M_l,{a}_{q,l}^{(i)}$}
\State $\boldsymbol{\psi} = \arg(\boldsymbol{y})$ $|~\boldsymbol{\psi} = [\widetilde{\boldsymbol{\psi}}_1^T, \cdots,
\widetilde{\boldsymbol{\psi}}_L^T]^T$, $\widetilde{\boldsymbol{\psi}}_l \in \mathbb{R}^{M_l}$
\State $\boldsymbol{\rho} = |\mathbf{y}|$ $|~\boldsymbol{\rho} = [\widetilde{\boldsymbol{\rho}}_1^T, \cdots, 
\widetilde{\boldsymbol{\rho}}_L^T]^T$, $\widetilde{\boldsymbol{\rho}}_l \in \mathbb{R}^{M_l}$
\For{$l\leftarrow 1 ~to ~L$} 
\State $\widetilde{\boldsymbol{\psi}}_l = [\psi_1, \cdots, \psi_{M_l}]^T$
\State $\widetilde{\boldsymbol{\rho}}_l = [\rho_1, \cdots, \rho_{M_l}]^T$
\State ${\theta}_m^{(i)} = \sum_{q=0}^{Q} \widetilde{a}_{q} m^q - {\psi}_m $, ~$m = 1,\ldots,M_l$
\State $\widetilde{b}_m = - \rho_m \cos{\theta_m^{(i)}}\left( \psi_m + \theta_m^{(i)} \right) + \rho_m \sin{\theta_m^{(i)}}$
\State $\mathbf{\eta} = [1, 2, 3, \cdots, M_l]^T \in \mathbb{Z}^{M_l}$
\State $\widetilde{\boldsymbol{A}} = [\mathbf{\eta}^0, \mathbf{\eta}^1, \cdots, \mathbf{\eta}^{Q}] \in \mathbb{Z}^{M_l \times Q},$
\State $\mathbf{z} = [\widetilde{a}_0, \widetilde{a}_1, \cdots, \widetilde{a}_{Q}]^T \in \mathbb{R}^Q$
\State $\widetilde{\mathbf{b}} = [\widetilde{b}_1, \widetilde{b}_2, \cdots, \widetilde{b}_{M_l}]^T \in \mathbb{R}^{M_l}$
obtain $\mathbf{z}^{\star}$ by \State $\displaystyle\minimize_{\mathbf{z}}~~~|| \widetilde{\boldsymbol{A}}\mathbf{z} - \widetilde{\mathbf{b}} ||_2^2$ 

\State $\widetilde{\mathbf{x}}_l = e^{(j(\widetilde{\boldsymbol{A}}\mathbf{z}^{\star}))}$
\EndFor
\State $\widetilde{\mathbf{X}} = [\widetilde{\mathbf{x}}_1, \widetilde{\mathbf{x}}_2, \cdots, \widetilde{\mathbf{x}}_L  ]$
\State $\mathbf{x}^{(i+1)} = \text{vec}({\widetilde{\mathbf{X}}}) $
\State $i \leftarrow i + 1$
\textbf{return} $\mathbf{x}^{(i+1)}$
\EndProcedure
\end{algorithmic}
\end{algorithm}
\begin{algorithm}[!htbp]
\caption{Optimal sequence with minimum {ISL} and polynomial phase parameters $a_{q,l}$ using {MISL}}\label{alg:PseudoCodeMISL}
\begin{algorithmic}[1]
\Require{$N$ and $M$ }
\Ensure{$\mathbf{x}^{(i+1)}$} 
\State Set $i =0$, initialize $\mathbf{x}^{(0)}$
\While{\textrm{Stopping criterion is not met}}
\State $\mathbf{f} = \left| \mathbf{A}^H \mathbf{x}^{(i)}\right|^2$
\State $f_{\text{max}} = \max \left( \mathbf{f} \right)$
\State $\mathbf{y}^{(i)} = -\mathbf{A} \left( \Diag(\mathbf{f}) - f_{\text{max}}\mathbf{I} - N^2\mathbf{I} \right)\mathbf{A}^H\mathbf{x}^{(i)}$
\State $\mathbf{x}^{(i+1)} = \text{PECS}(\mathbf{y}^{(i)}, L, M_l, a_{\{q,l\}}^{(i)})$
\EndWhile
\State \textbf{return} $\mathbf{x}^{(i+1)}$
\end{algorithmic}
\end{algorithm}
\subsubsection{Extension of CAN}
In addition to the above mentioned procedure using \gls{MM-PSL}, the optimization problem in (\ref{eq:ISLOptC1}) can also be solved using \gls{CAN} method \cite{4749273}. As opposed to the approach pursued in \cite{7093191} of directly minimizing a quartic function, in \cite{4749273} the solution of the objective function in (\ref{eq:ISLOptC1}) is assumed to be ``almost equivalent'' to minimizing a quadratic function
\begin{equation}\label{eq:ISL_CAN}
\minimize_{\{x_n\}_{n=1}^{N}; \{\psi_g\}_{g=1}^{2N}} \sum_{g=1}^{2N} \left| \sum_{n=1}^{N} x_n e^{-j\omega_g n} - \sqrt{N} e^{j \psi_g} \right|^2. \\
\end{equation}
It can be written in a more compact form (to within a multiplicative constant)
\begin{equation}\label{eq:ISL_CAN_cmpct}
    || \boldsymbol{A}^* \mathbf{\bar{x}} - \mathbf{v}||^2
\end{equation}
where $\mathbf{a}_g^* = [e^{-j\omega_g}, \cdots, e^{-j 2N\omega_g}]$ and 
$A^*$ is the following unitary $2N \times 2N$ \gls{DFT} matrix
\begin{equation}
    \boldsymbol{A}^* = \frac{1}{\sqrt{2N}}
    \begin{bmatrix}
    \mathbf{a}_1^* \\ \vdots \\ \mathbf{a}_{2N}^*,
    \end{bmatrix}
\end{equation}
$\mathbf{\bar{x}}$ is the sequence $\{x_n\}_{n=1}^N$ padded with $N$ zeros, i.e.
$\mathbf{\bar{x}} = [x_1,\cdots, x_N, 0, \cdots, 0]_{2N\times 1}^T$
and
$\mathbf{v} = \frac{1}{\sqrt{2}}[e^{j\psi_1}, \cdots, e^{j\psi_{2N}}]^T$.
For given $\{ x_n\}$, CAN minimizes \eqref{eq:ISL_CAN_cmpct} by alternating the optimization between $\mathbf{\bar{x}}$ and $\mathbf{v}$. Let $\mathbf{\bar{x}}^{(i)} = [x_1^{(i)},\cdots, x_N^{(i)}, 0 , \cdots, 0]_{2N\times1}^T$, and let $D_i$ represent the value of $||\boldsymbol{A}^* \mathbf{\bar{x}}^{(i)} - \mathbf{v}^{(i)}||$ at iteration $i$. Then we have $D_{i-1} \geq D_{i}$. Further in the $i^{th}$ iteration, the objective can be minimized using the technique proposed for solving  \eqref{eq:PSLfinalOpt} by assuming
\begin{equation}
    \begin{aligned}
        \mathbf{x} &= \mathbf{\bar{x}}^{(i)},\\
        \mathbf{y} &= e^{j\arg(\mathbf{d})}
    \end{aligned}
\end{equation}
where $\mathbf{d} = \boldsymbol{A}\mathbf{v}$ denotes the inverse \gls{FFT} of $\mathbf{v}$.
The details of the implementation can be found in Algorithm  \ref{Pseudo-Code2}.
\begin{algorithm}[!htbp]
\caption{Optimal sequence with minimum \gls{ISL} and polynomial phase parameters $a_{q,l}$ using \gls{CAN}}\label{Pseudo-Code2}
\begin{algorithmic}[1]
\Require{$N$ and $M$ }
\Ensure{$\mathbf{x}^{(i+1)}$} 
\State Set $i =0$, initialize $\mathbf{x}^{(0)}$.
\While{\textrm{Stopping criterion is not met}}
\State $\mathbf{f} = \mathbf{A}^H \mathbf{x}^{(i)}$
\State $v_g = e^{j(\arg(f_g))},  ~g = 1,\ldots,2N$
\State $\mathbf{d} = \mathbf{A} \mathbf{v}$
\State $y_n^{(i+1)} = e^{j (\arg(d_n))},  ~n = 1,\ldots,N$
\State $\widetilde{\mathbf{X}} = \text{PECS}(\mathbf{y}^{(i)}, L, M_l, a_{\{q,l\}}^{(i)})$
\State $\mathbf{x}^{(i+1)} = \text{vec}({\widetilde{\mathbf{X}}}) $
\EndWhile
\State \textbf{return} $\mathbf{x}^{(i+1)}$
\end{algorithmic}
\end{algorithm}

\section{Performance Analysis}\label{Sec:perfromanceAnalysis}
In this section, we assess the performance of the 
proposed \gls{PECS} algorithm\footnote{Source code is available at "https://github.com/robin-amar/ResearchPapers/blob/eb83e644c164ac998bac7820268ee5929563583e/ PECS.m"}  
and  compare it with prior work in the literature.
We then emphasize its potential to design sequences for various automotive radar applications while considering the effects of Doppler and 
interference.

\subsection{\texorpdfstring{$\ell_p$}{} Norm Minimization}

%
At first, we evaluate the performance of the proposed Algorithm-\ref{Algo:PseudoCode_PECS_LpNorm} in terms of $\ell_p$ norm minimization
by several examples.
For the initialization, we chose a random seed sequence and $Q=2$.

\begin{figure}[t!]
    \centering
    \begin{subfigure}{.5\textwidth}
        \centering
        \includegraphics[width=.95\linewidth]{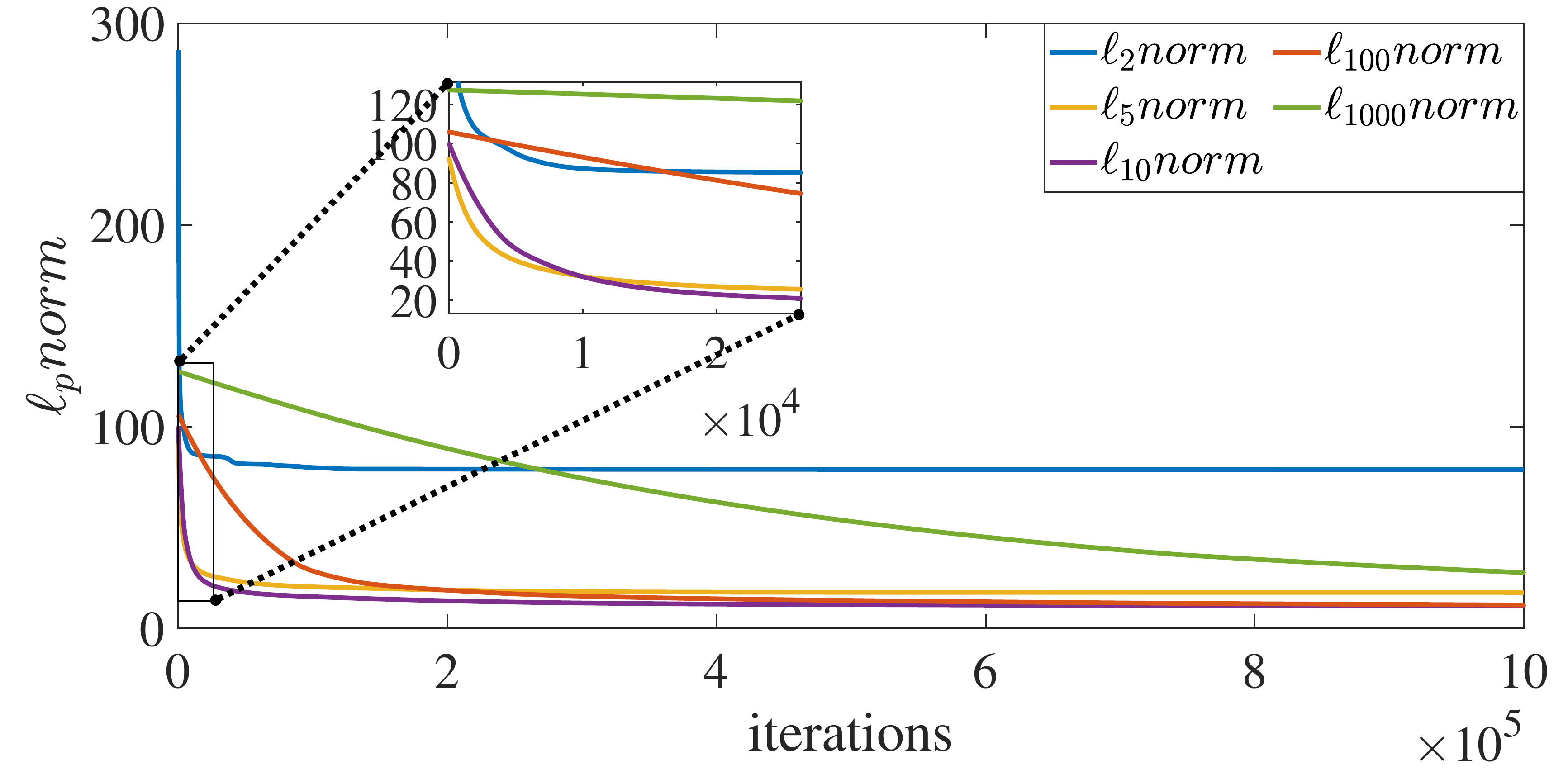}
        \caption{Objective convergence}
    \end{subfigure}
    \begin{subfigure}{.5\textwidth}
        \centering
        \includegraphics[width=.95\linewidth]{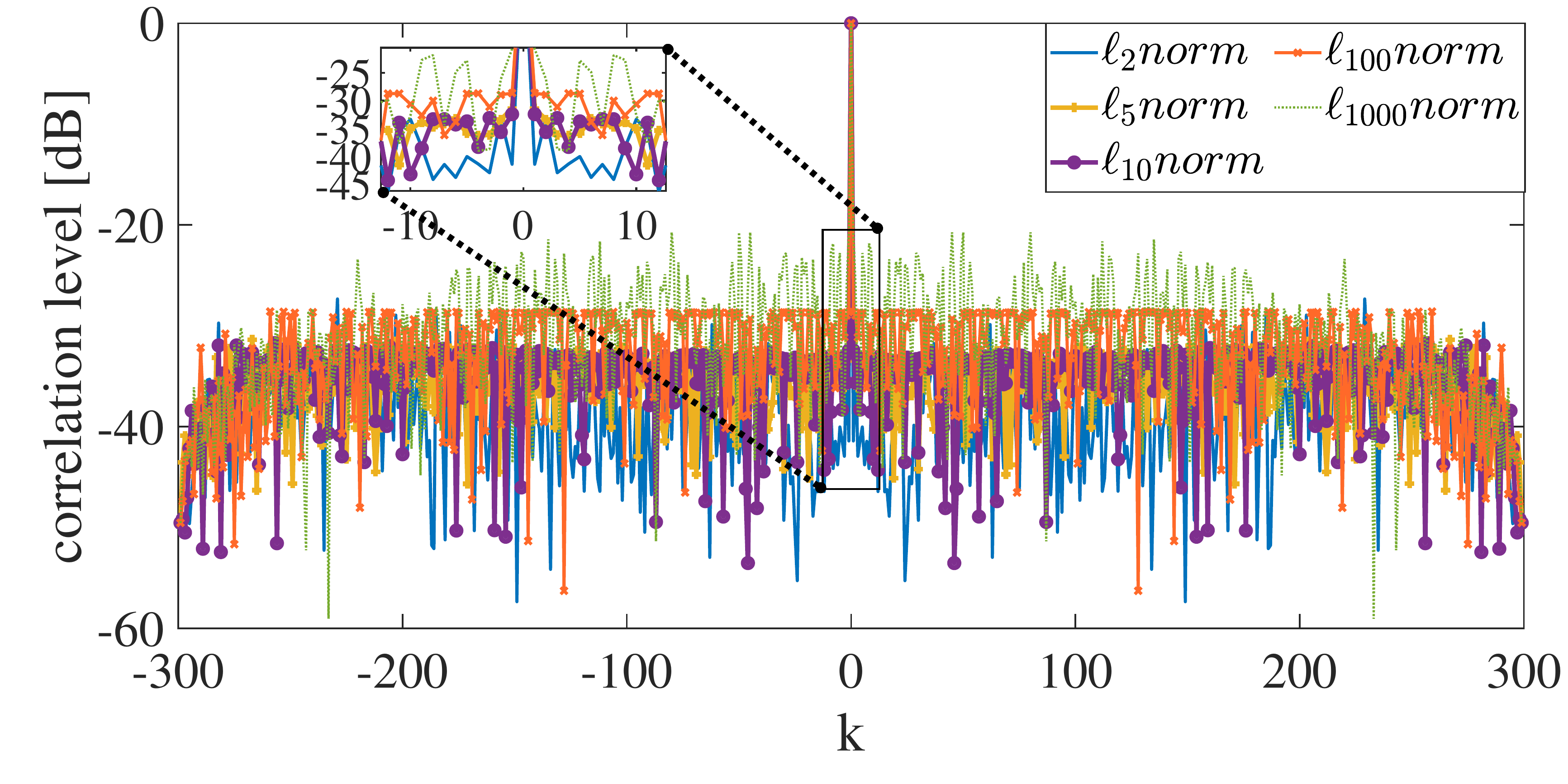}
        \caption{Autocorrelation response}
    \end{subfigure}
    \caption{$\ell_p$ norm convergence and Autocorrelation comparison with varying $p$ in $\ell_p$ norm for a sequence with input parameters $N = 300, M = 5, Q = 2$ and iterations $= 10^6$.}
    \label{fig:PSLOptmznCmprsn}
\end{figure}

\figurename{~\ref{fig:PSLOptmznCmprsn}} shows the convergence behaviour of the proposed algorithm when the simulation is mandatorily run for $10^6$ iterations. 
We chose different values of $p$ (i.e. $2$, $5$, $10$, $100$ and $1e3$) as an input parameter of the Algorithm-\ref{Algo:PseudoCode_PECS_LpNorm}, these allow to trade-off between good \gls{PSL}  and \gls{ISL}.
For this figure, we keep the values of sequence length, sub-sequence length and 
polynomial degree fixed, by setting 
$N = 300$, $M = 5$, and $Q=2$. 
Nevertheless, we observed similar behaviour in the convergence for the different values of $N$, $M$, and $Q$.

As evident from the \figurename{~\ref{fig:PSLOptmznCmprsn}}, the objective function is reduced rapidly for $p=2$ and this rate reduces with the objective saturating
after $10^5$ iterations. 
Whereas, by increasing the value of $p$ to $5, 10$ and $100$, we achieve similar convergence rate as observed earlier. 
Uniquely, while computing the $\ell_{1000}$ norm\footnote{Computationally,
we cannot use $p = \infty$, but by setting $p$ to a tractable value, e.g., $p \geq 10$, we find that the peak sidelobe
is effectively minimized.}, the objective converges slowly and continues to decrease until $10^6$ iterations.

Further, while analysing the autocorrelation sidelobes in \figurename{~\ref{fig:PSLOptmznCmprsn}} for the same set of input parameters of $p, N, M$ and maximum number of iterations, we numerically observe that the lowest \gls{PSL} values\footnote{$
    \text{PSL}_{dB} \triangleq 10\log_{10}(\text{PSL}),~~
    \text{ISL}_{dB} \triangleq 10\log_{10}(\text{ISL}).$} 
are observed for $\ell_{10}$ norm and other \gls{PSL} values are higher for $p \neq 10$.

\begin{figure}[t!]
    \centering
    \begin{subfigure}{.5\textwidth}
        \centering
        \includegraphics[width=.95\linewidth]{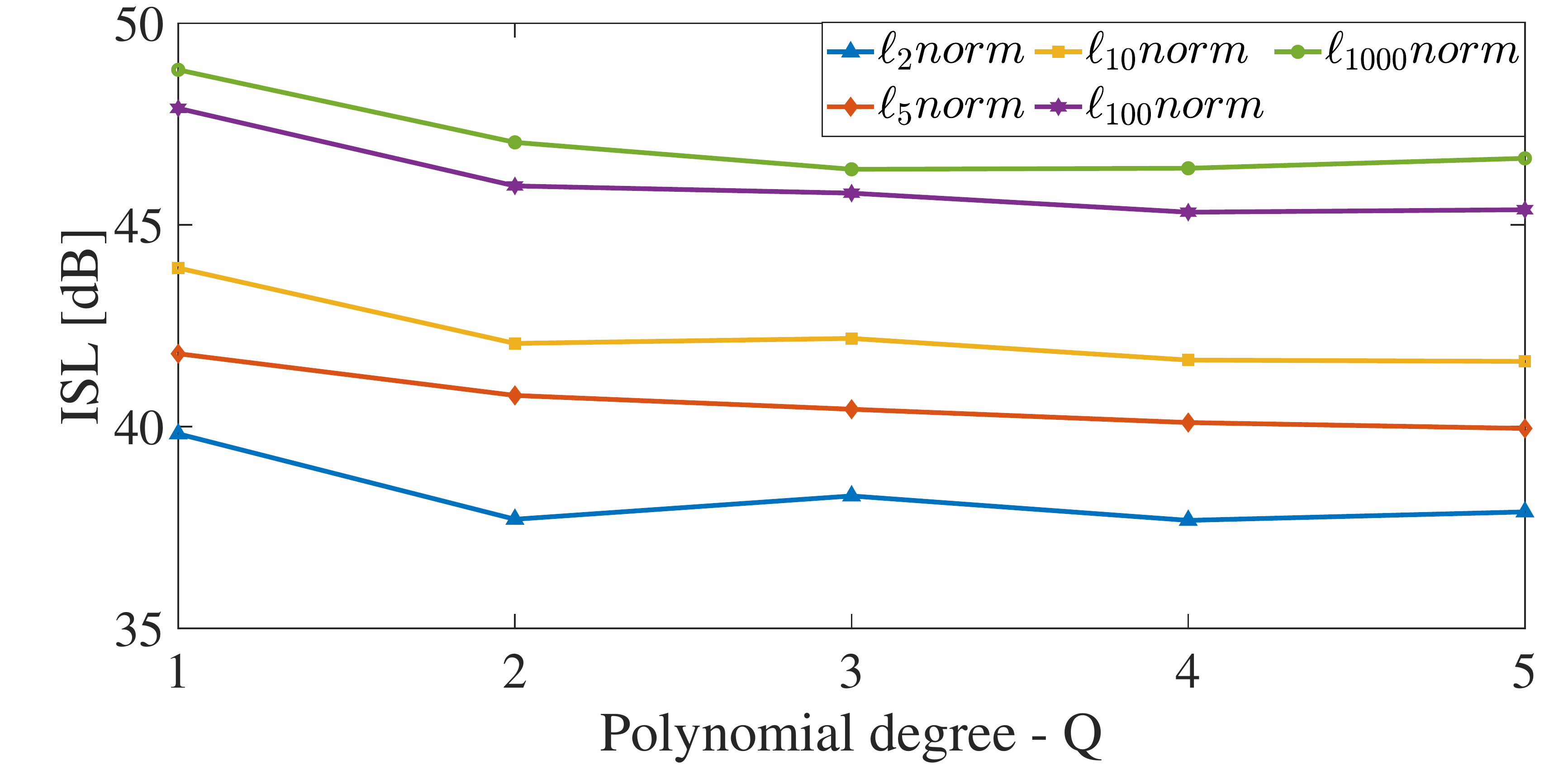}
        \caption{ISL vs Q}
    \end{subfigure}
    \begin{subfigure}{.5\textwidth}
        \centering
        \includegraphics[width=.95\linewidth]{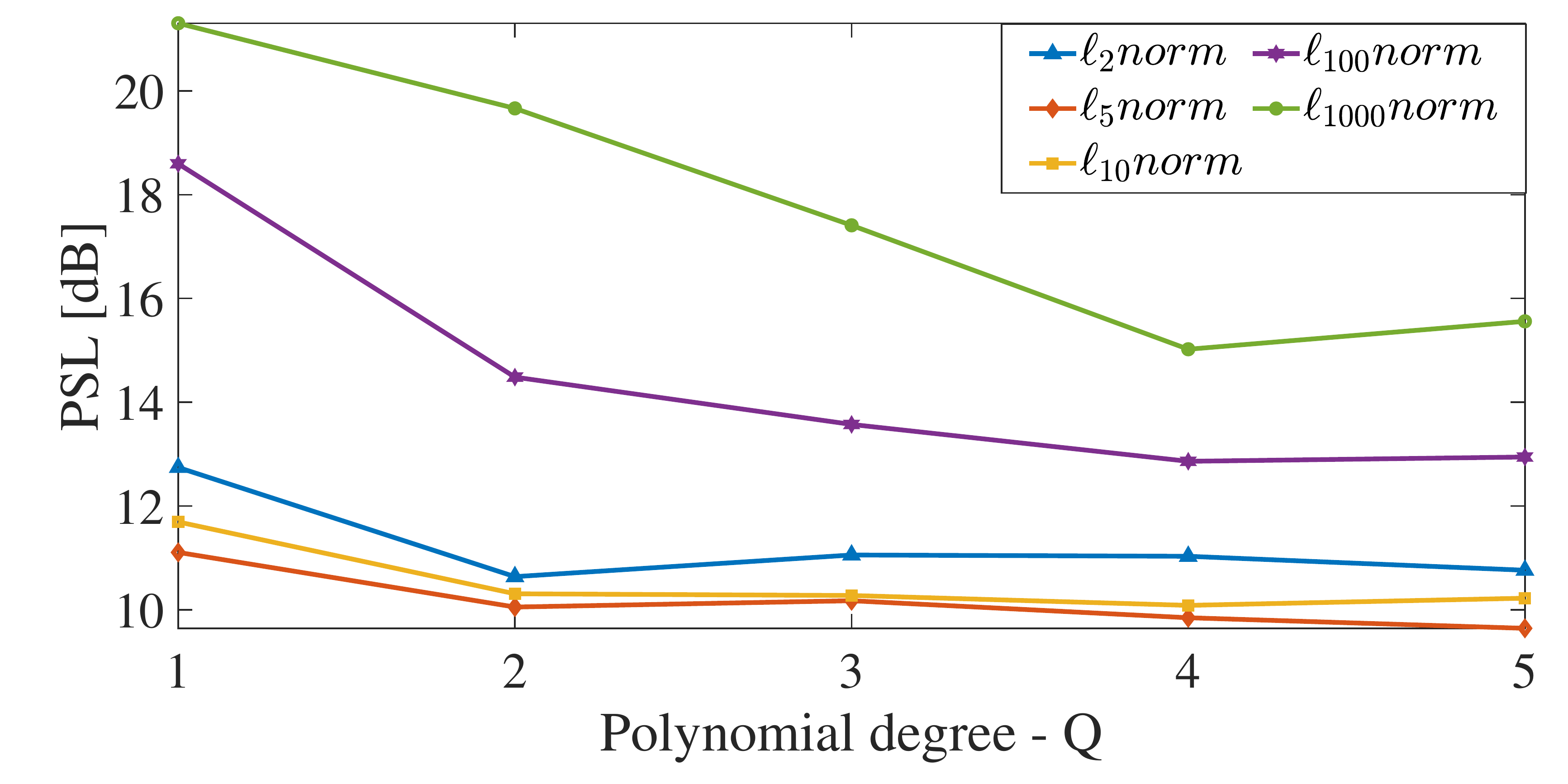}
        \caption{PSL vs Q}
    \end{subfigure}
    \caption{ISL and PSL variation with increasing Q.}
    \label{fig:ISL_PSL_vs_Q}
\end{figure}

In \figurename{~\ref{fig:ISL_PSL_vs_Q}}, we assess the relationship of Polynomial phase of degree $Q$ as a tuning parameter with \gls{PSL} and \gls{ISL}. The parameter $Q$, can be considered as another degree of freedom available for the design problem. Other input parameters are kept fixed (i.e. $N=300$ and $M=5$) and same seed sequence is fed to the algorithm. As the value of $Q$ is increased, we observe a decrement in the optimal \gls{PSL} and \gls{ISL} values generated from \gls{PECS} for different norms (i.e. $p = 2,5, 10, 100$ and $10^3$). Therefore, the choice of the input parameters would vary depending upon the application.

\subsection{Shaping the Ambiguity Function}

In Table (\ref{Table:PhsAmbigFn_Cmprsn}), we show the capability and effectiveness of the \gls{PECS} algorithm to shape the \gls{AF} by changing the input parameters. For the plots shown here, the input parameters are $N=300$, varying sub-sequence lengths of $M = 5, 50, 150$ and $300$, $Q=2$ and $p$ limited to $2, 10$ and $100$. In the unwrapped phase plots of the sequences, the quadratic nature of the sub-sequence is retained for all values of $M$ and $p$. As evident, the \gls{AF} achieves the thumbtack type shape for $M=5$ keeping quadratic behaviour in its phase and as the value of $M$ increases, it starts evolving into a ridge-type shape. For $M=N =300$ (i.e. only one sub-sequence, $L=1$), it achieves a perfect ridge-shaped \gls{AF}. Further, the sharpness of the ridge-shape is observed as the value of $p$ increases (i.e. $p = 10, 100$).


\begin{table*}[tbh]
    \caption{Unwrapped phase and Ambiguity Function comparison for $N=300$, $Q=2$, $M \in [5, 50, 150,300]$ and $p \in [2, 10,100]$ in $\ell_p$ norm.}
    \centering
    \begin{center}
            \begin{tabular}{|M{1em}|M{20em}|M{20em}|M{20em}|}
            \hline
            &\textbf{$\boldsymbol{\ell}_2$ norm}   &    \textbf{$\boldsymbol{\ell}_{10}$ norm}    & \textbf{$\boldsymbol{\ell}_{100}$ norm}     \\\hline\hline
            \rotatebox{90}{Unwrapped Phase}&\includegraphics[width=5cm,height=3cm]{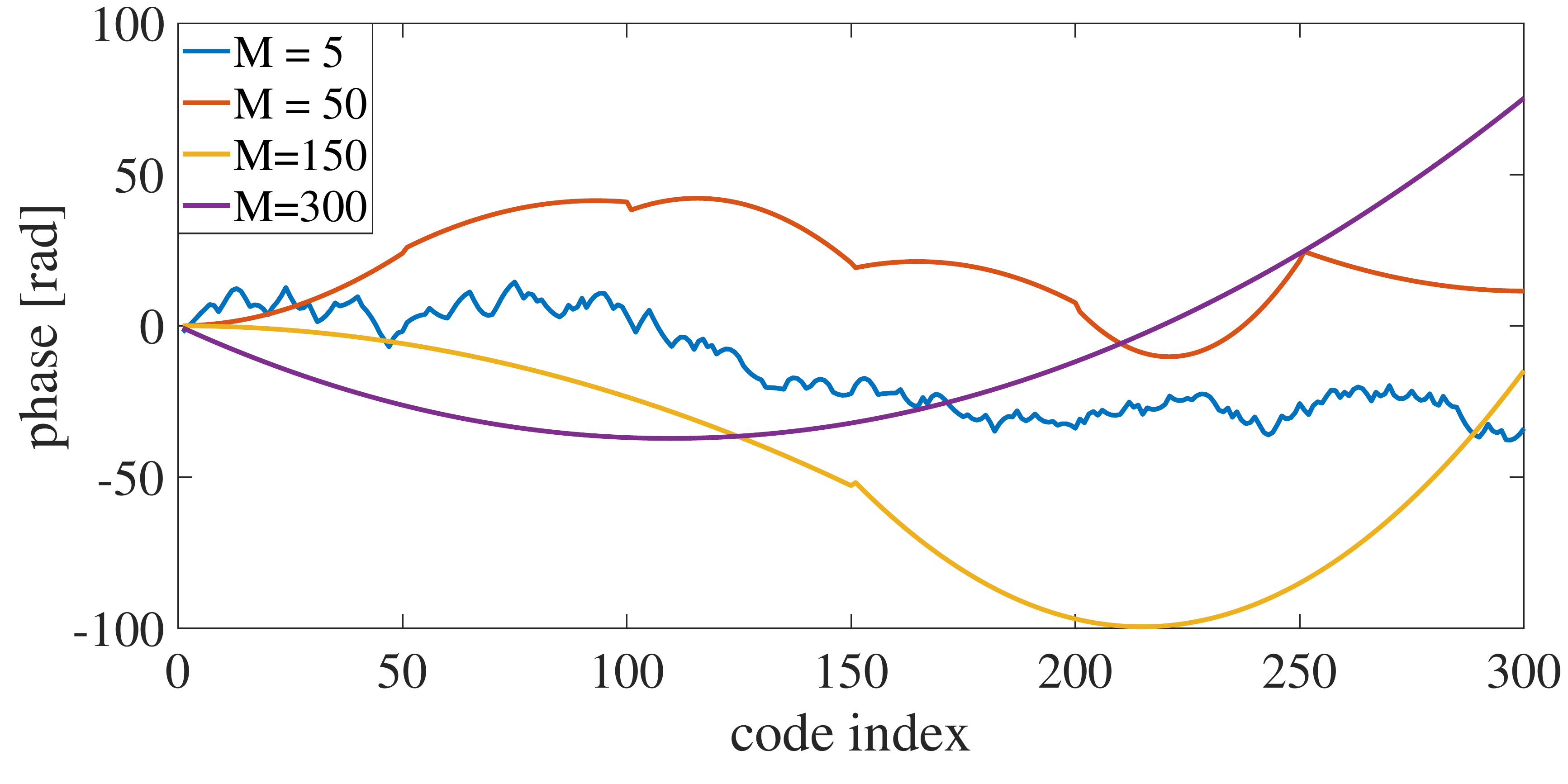} & \includegraphics[width=5cm,height=3cm]{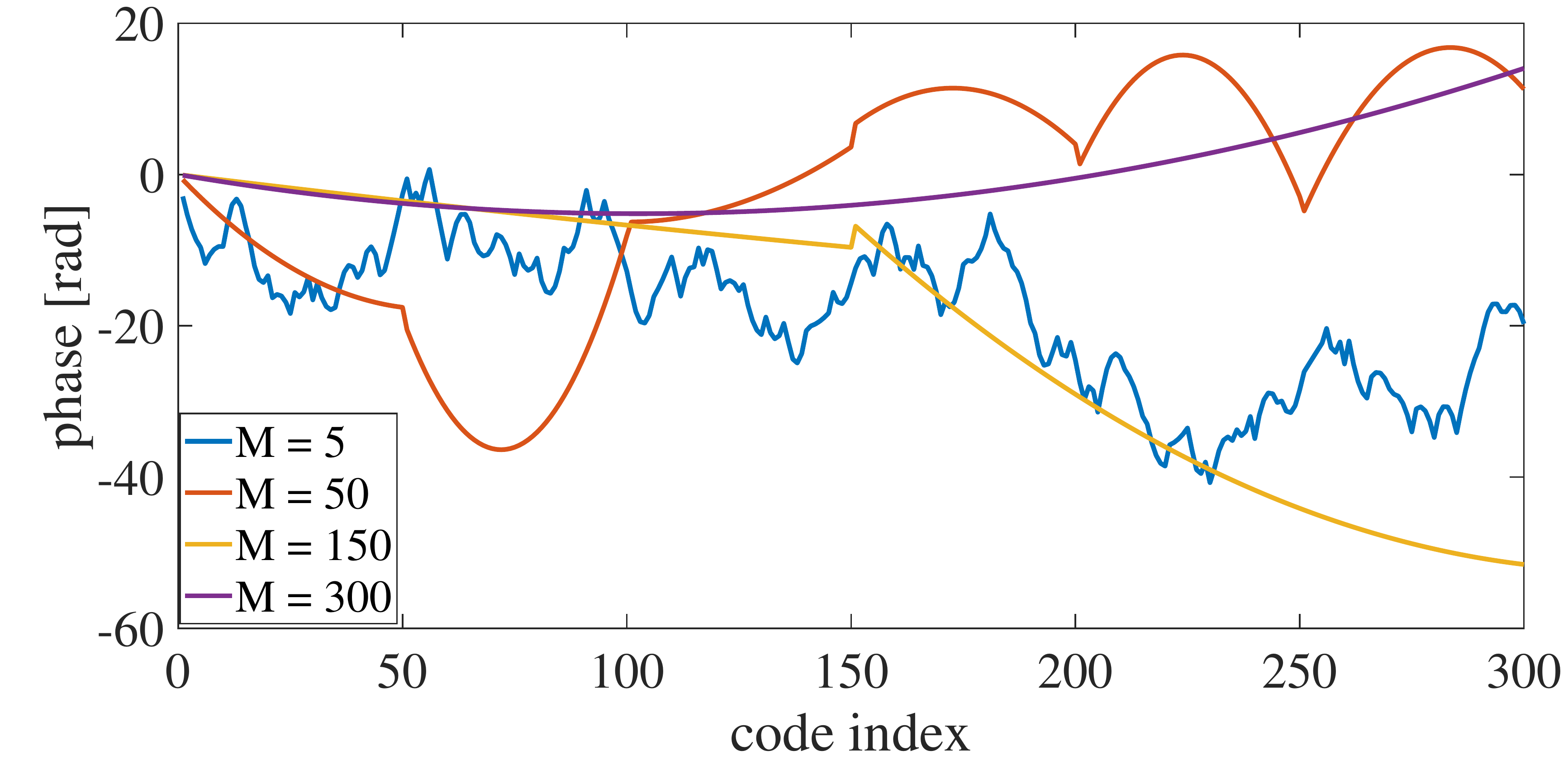} &\includegraphics[width=5cm,height=3cm]{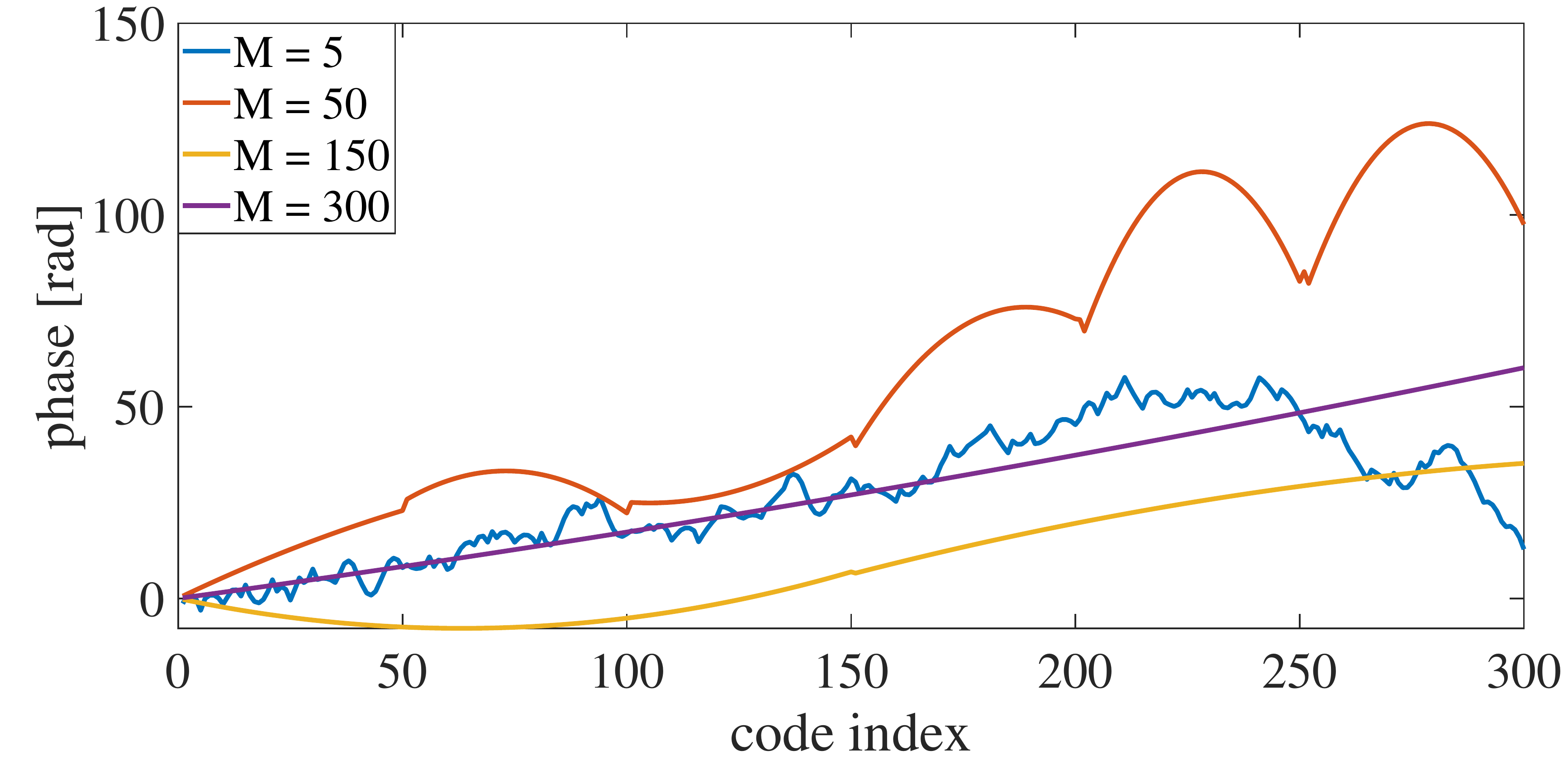}\\ \hline
            \rotatebox{90}{AF $\rightarrow M = 5$} &  \includegraphics[width=5cm,height=3cm]{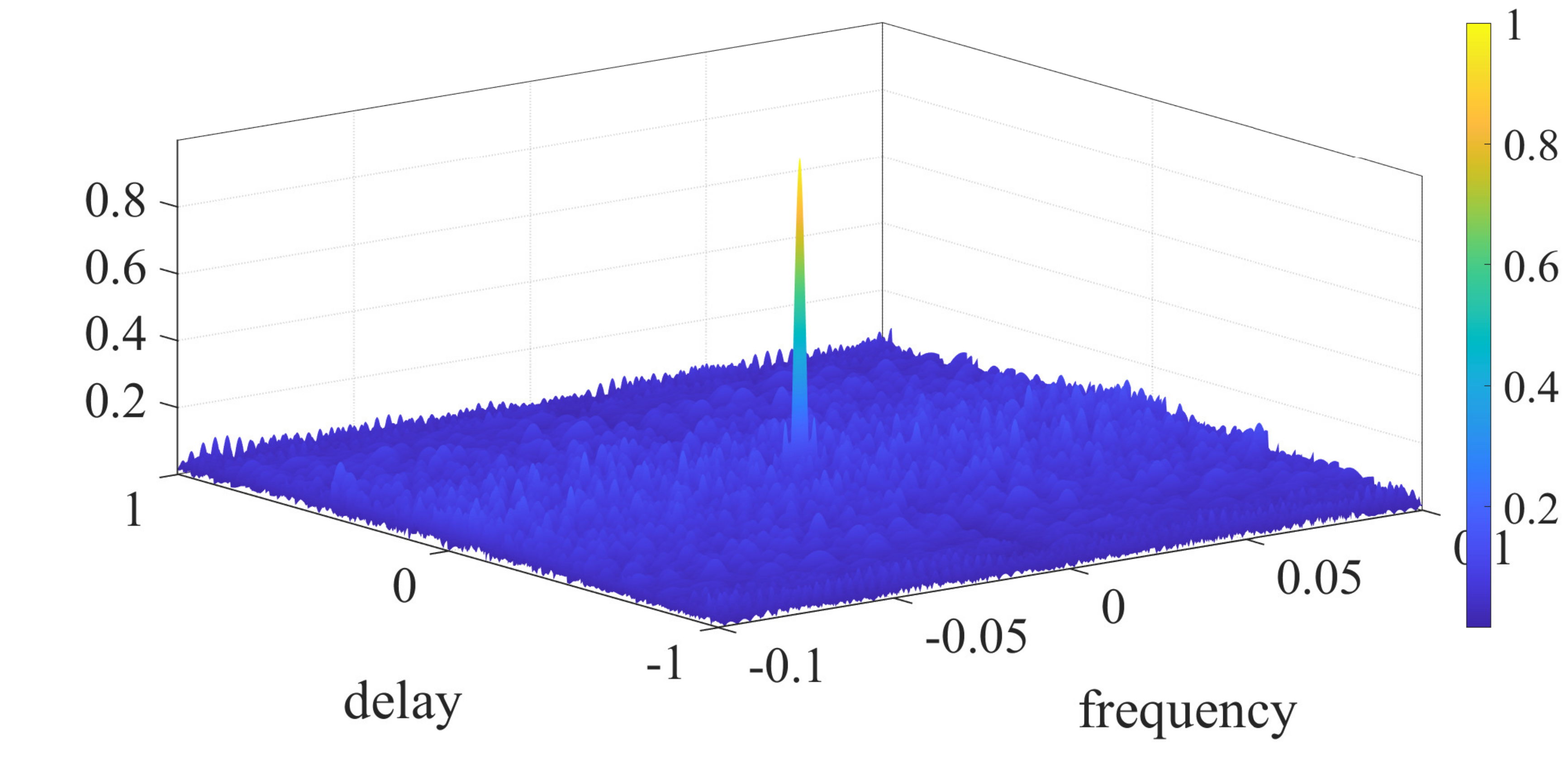} &  \includegraphics[width=5cm,height=3cm]{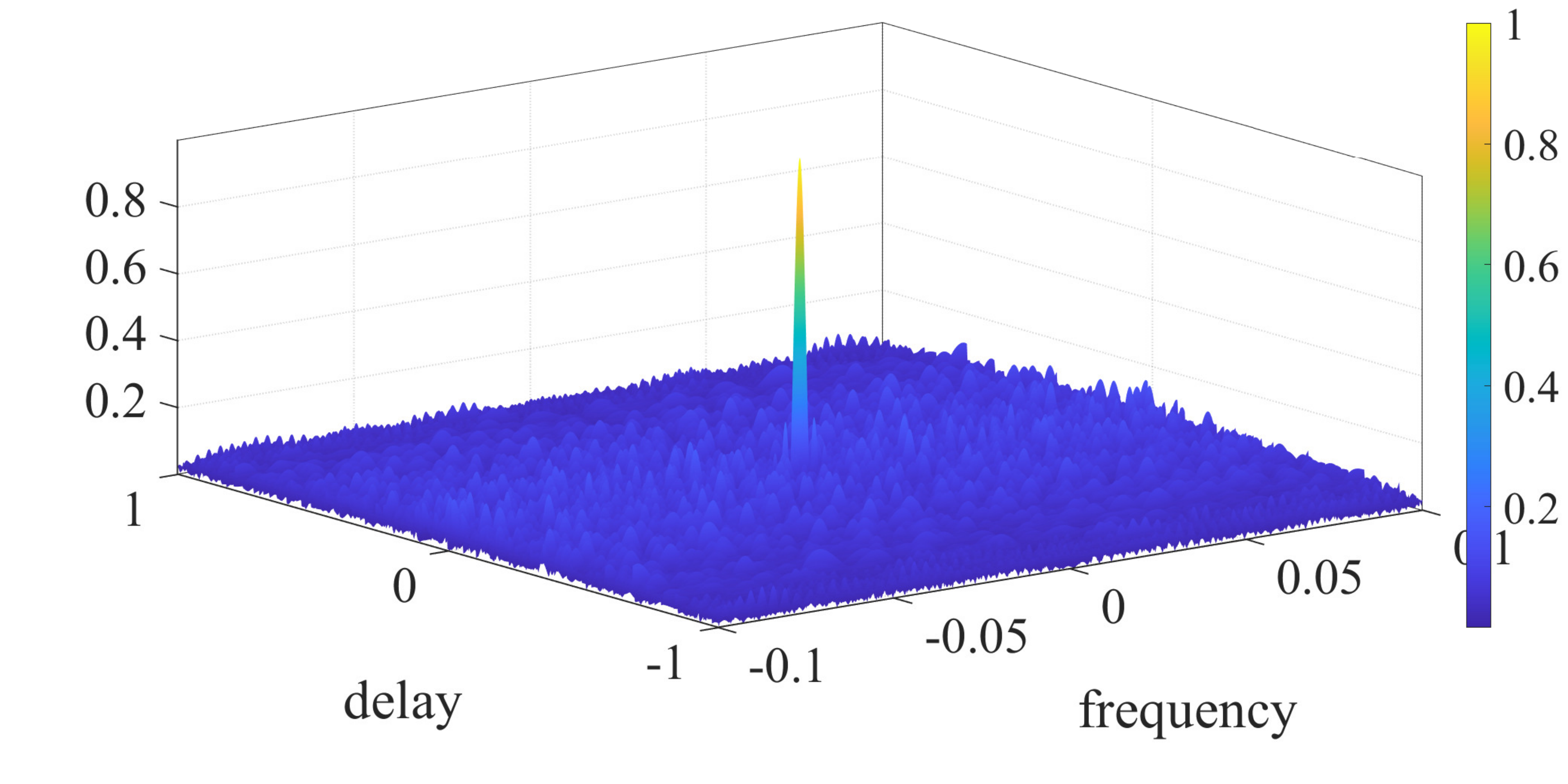} & \includegraphics[width=5cm,height=3cm]{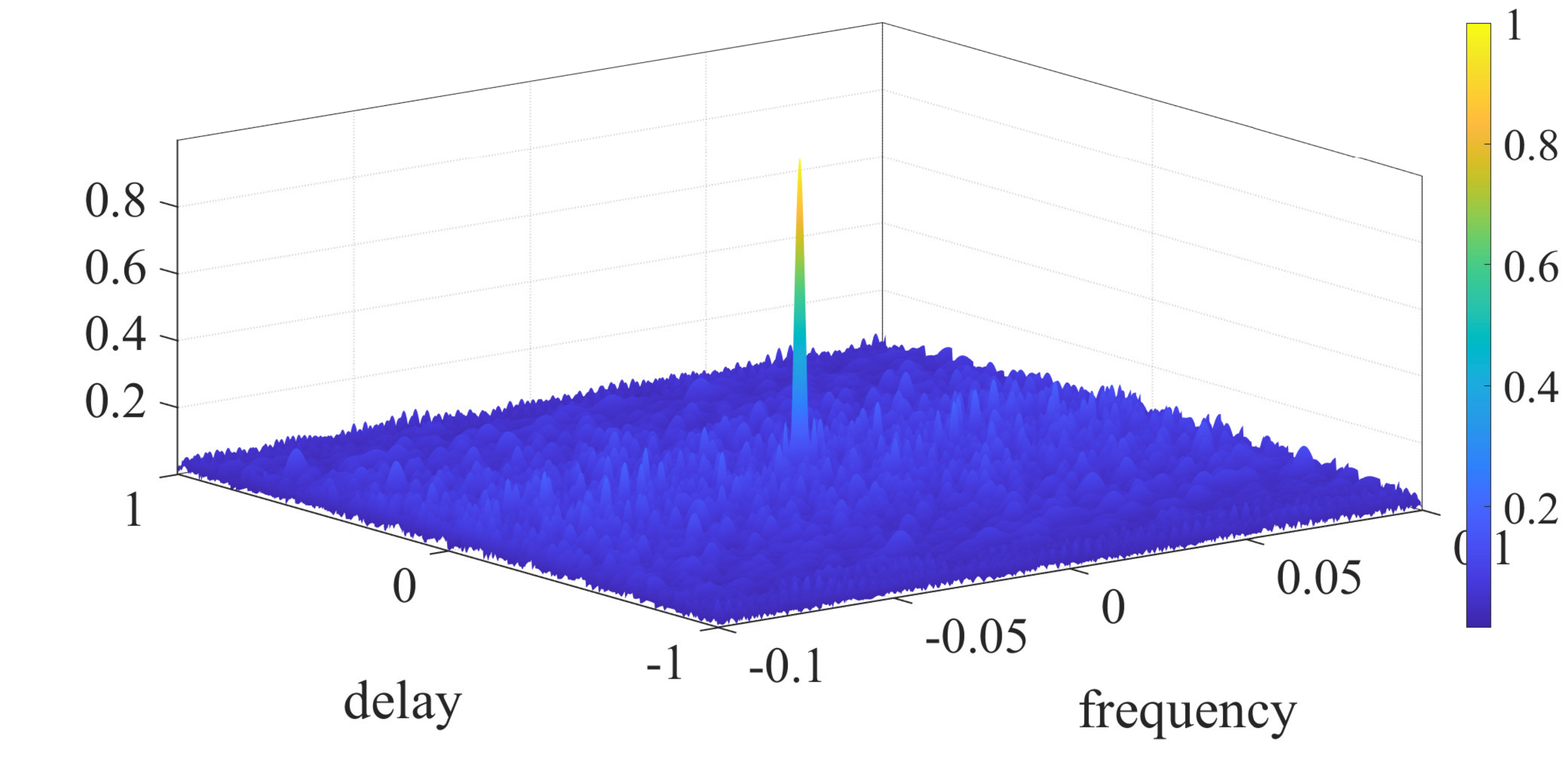}  \\ \hline
            \rotatebox{90}{AF $\rightarrow M = 50$} &  \includegraphics[width=5cm,height=3cm]{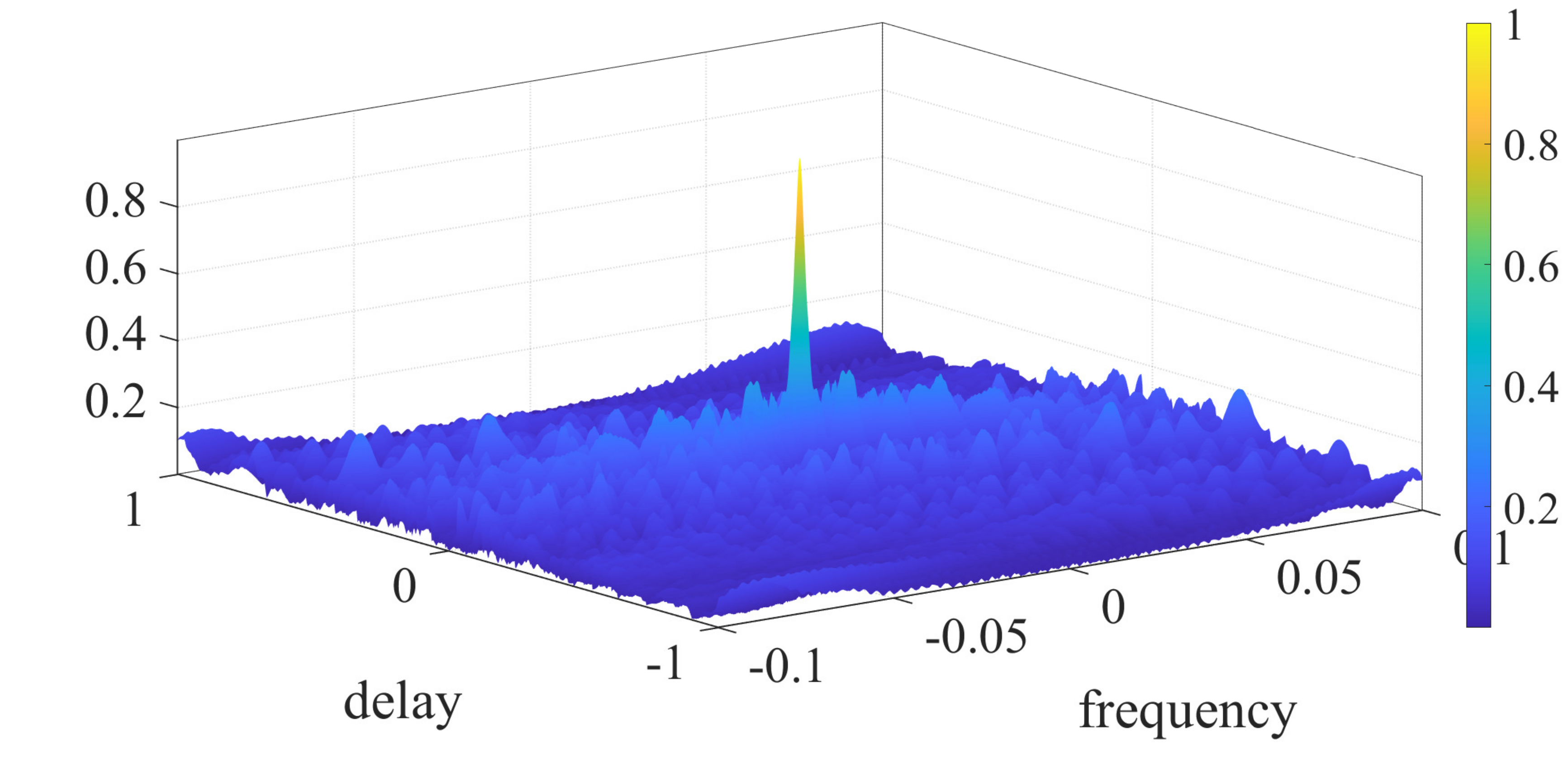} &  \includegraphics[width=5cm,height=3cm]{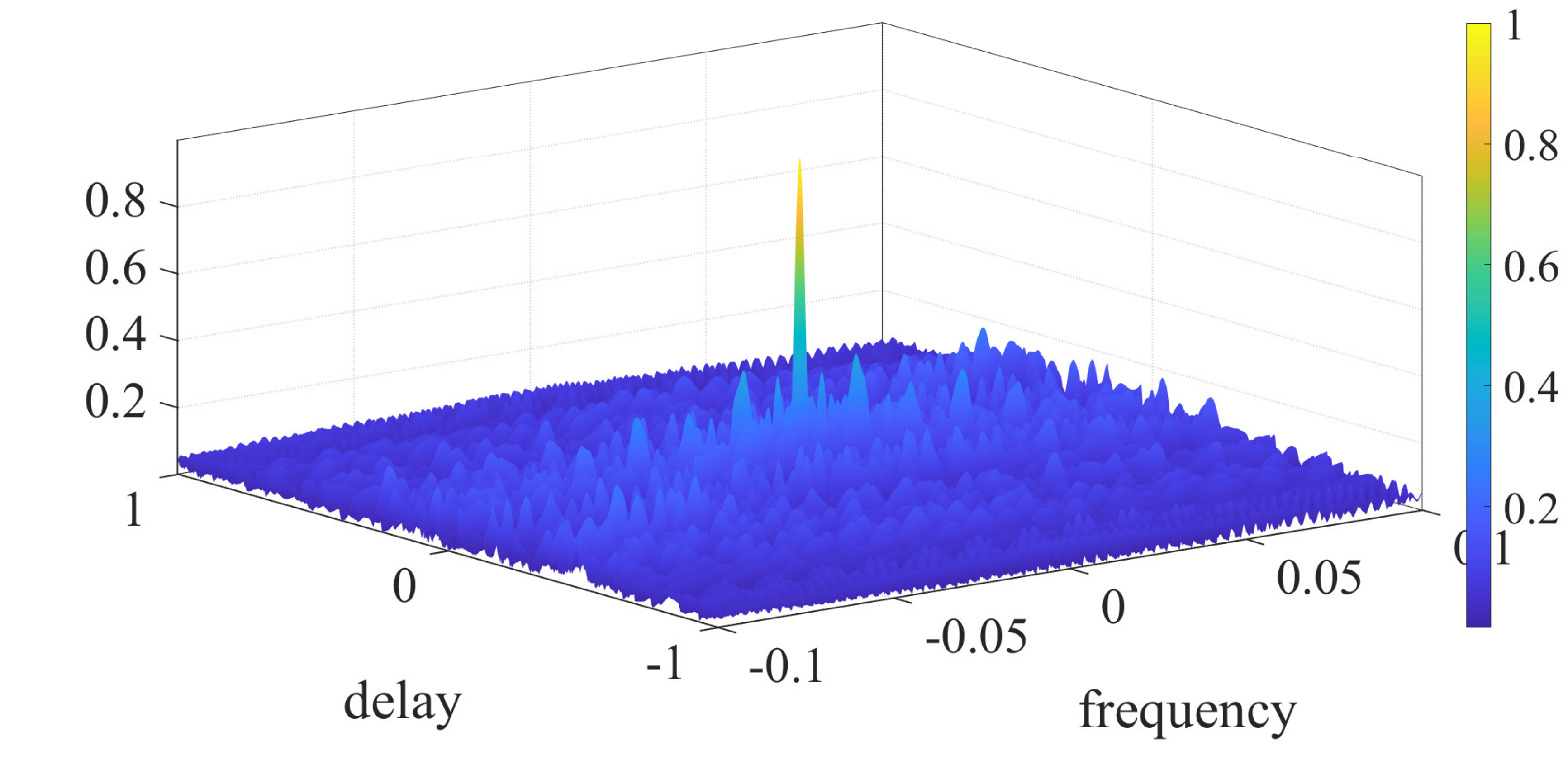} & \includegraphics[width=5cm,height=3cm]{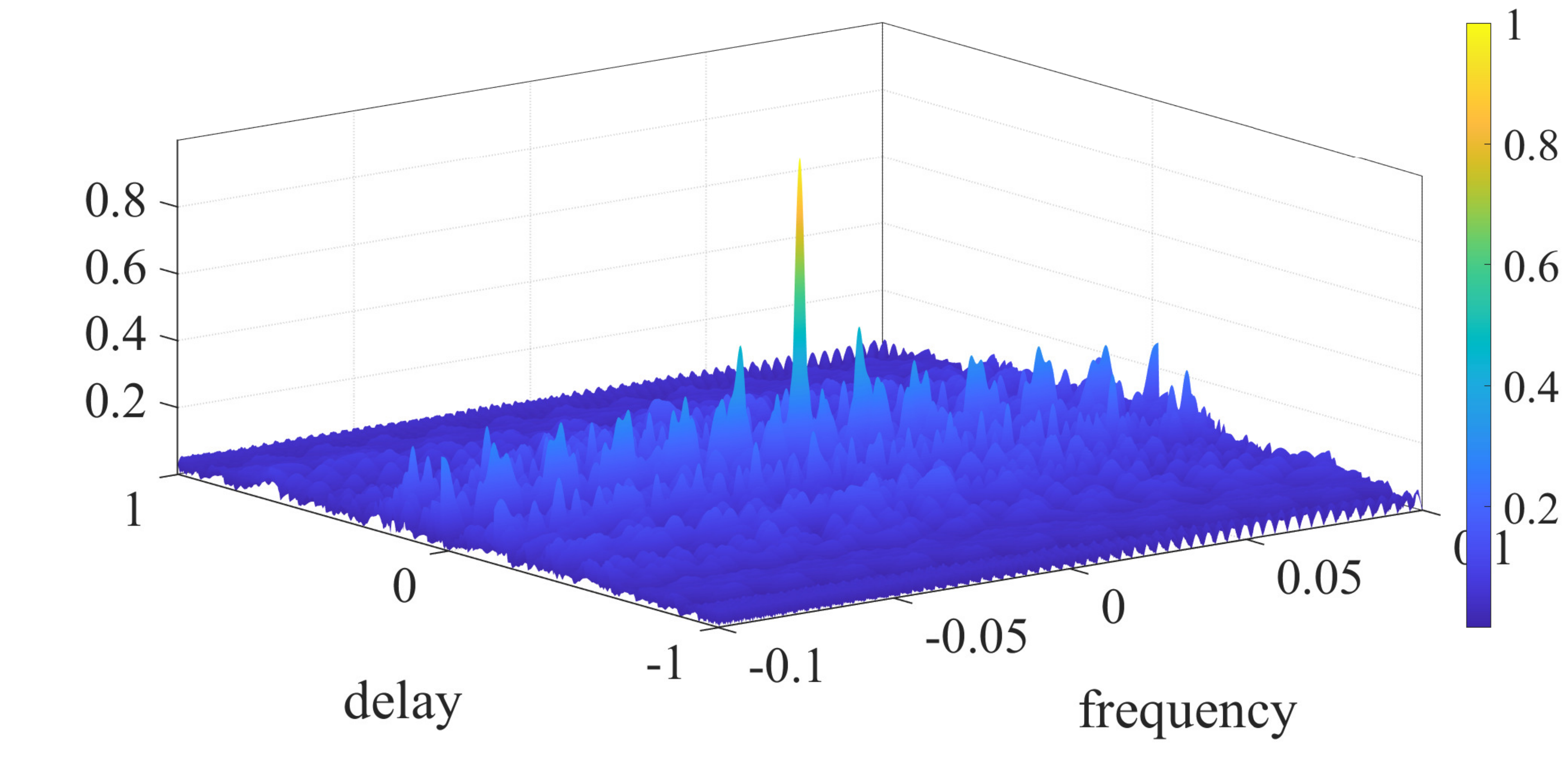}  \\ \hline
            \rotatebox{90}{AF $\rightarrow M = 150$} &  \includegraphics[width=5cm,height=3cm]{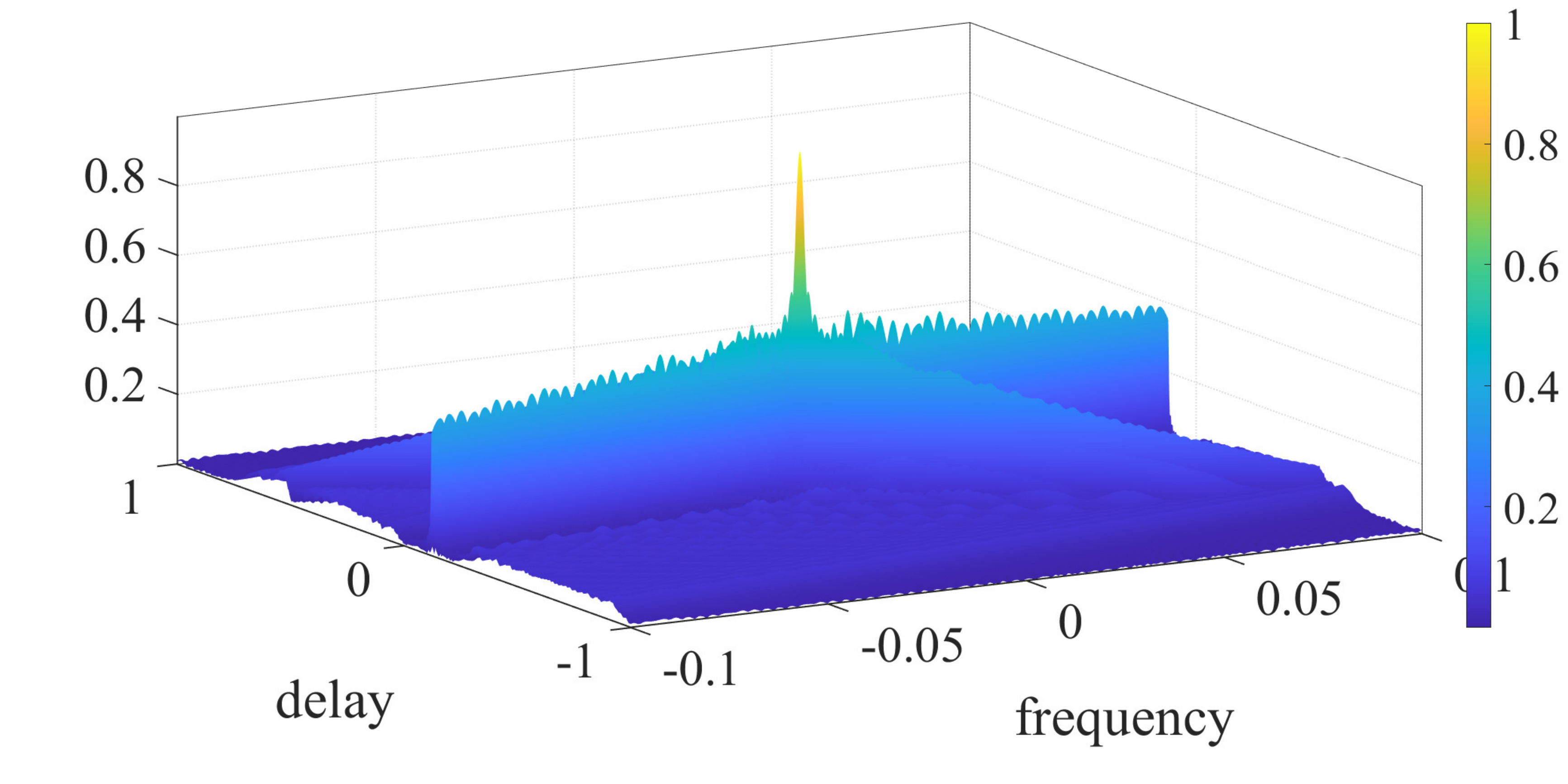} &  \includegraphics[width=5cm,height=3cm]{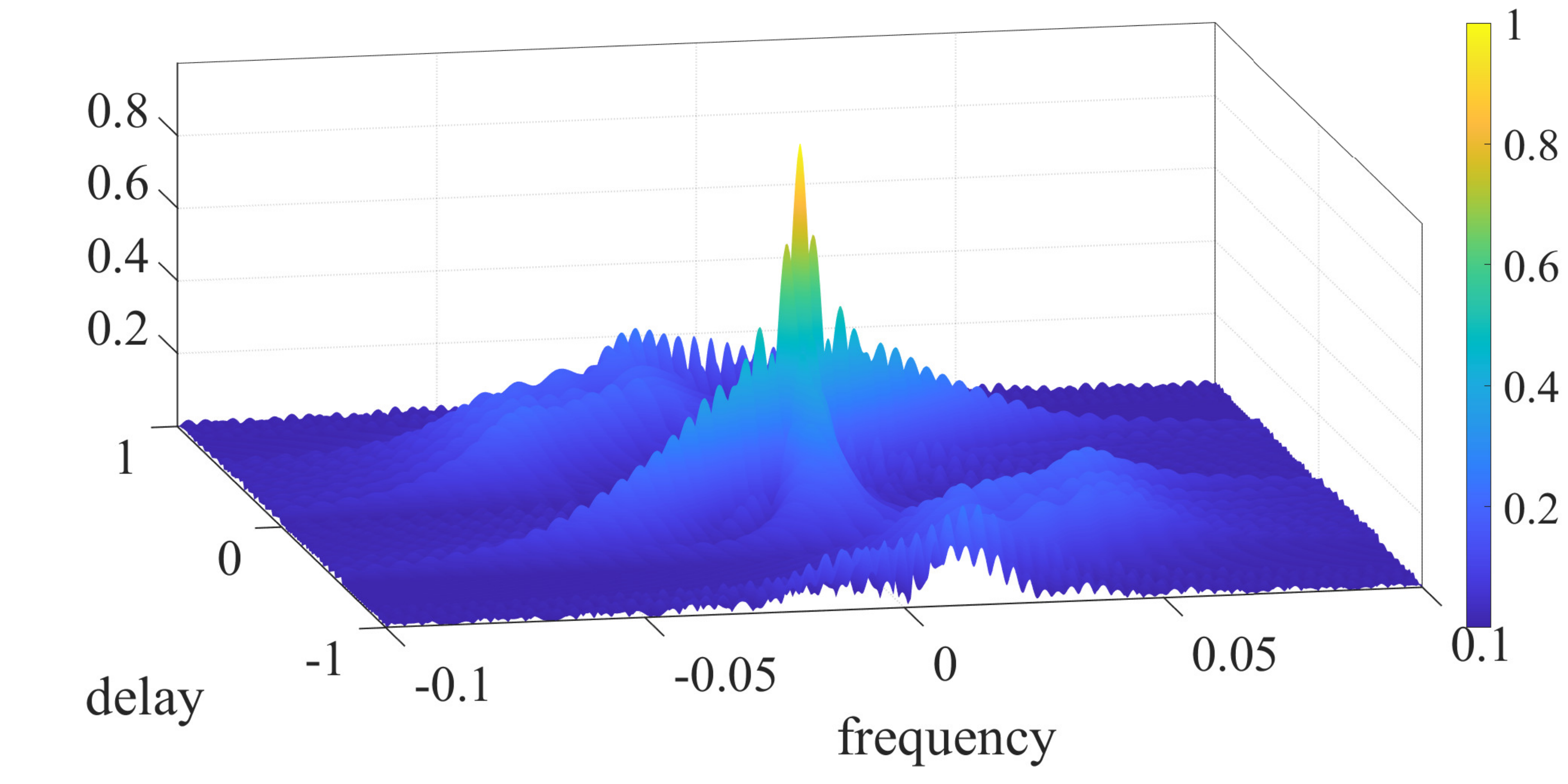} & \includegraphics[width=5cm,height=3cm]{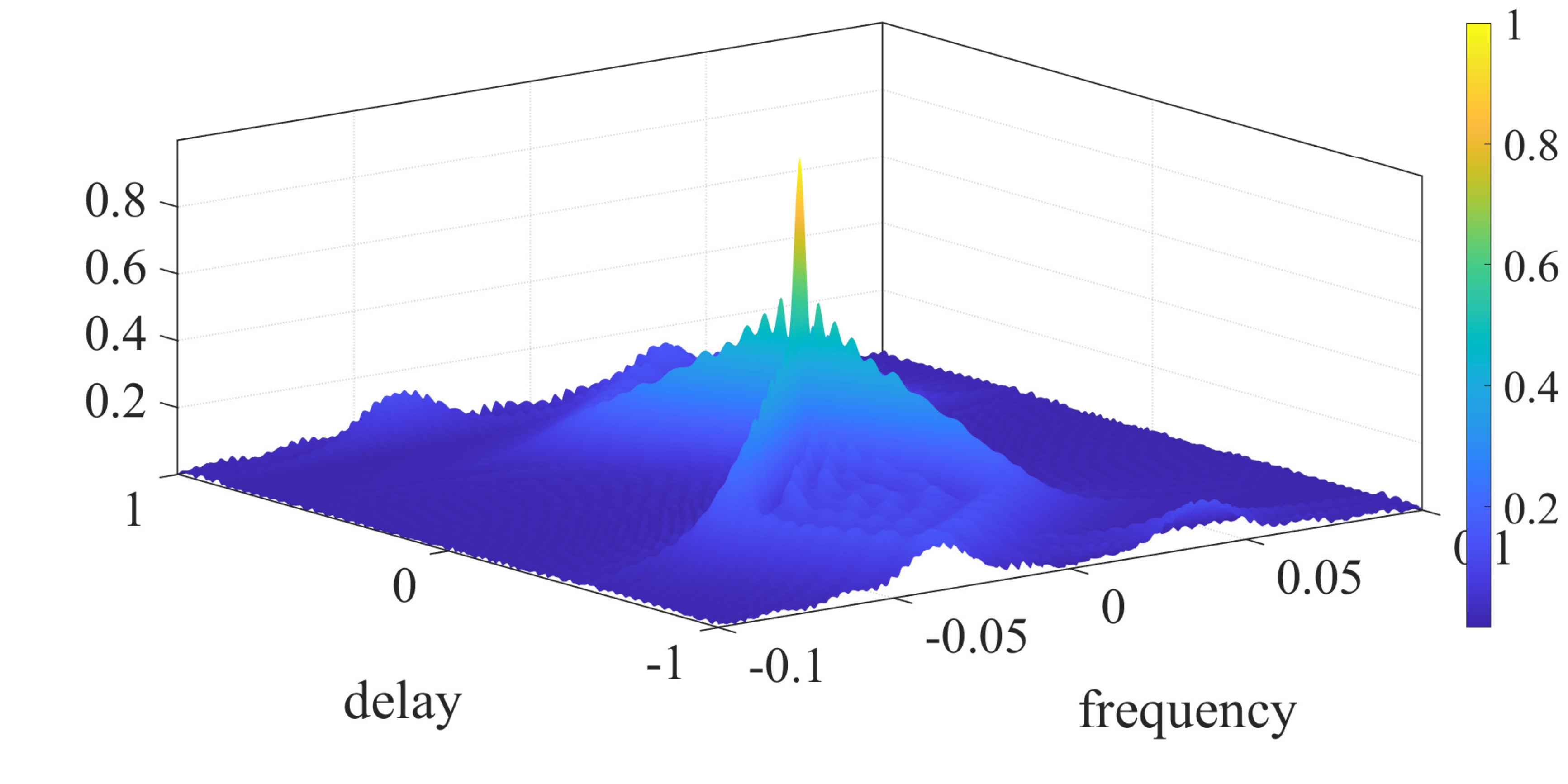}  \\ \hline
           \rotatebox{90}{AF $\rightarrow M = 300$} & \includegraphics[width=4.5cm,height=3cm]{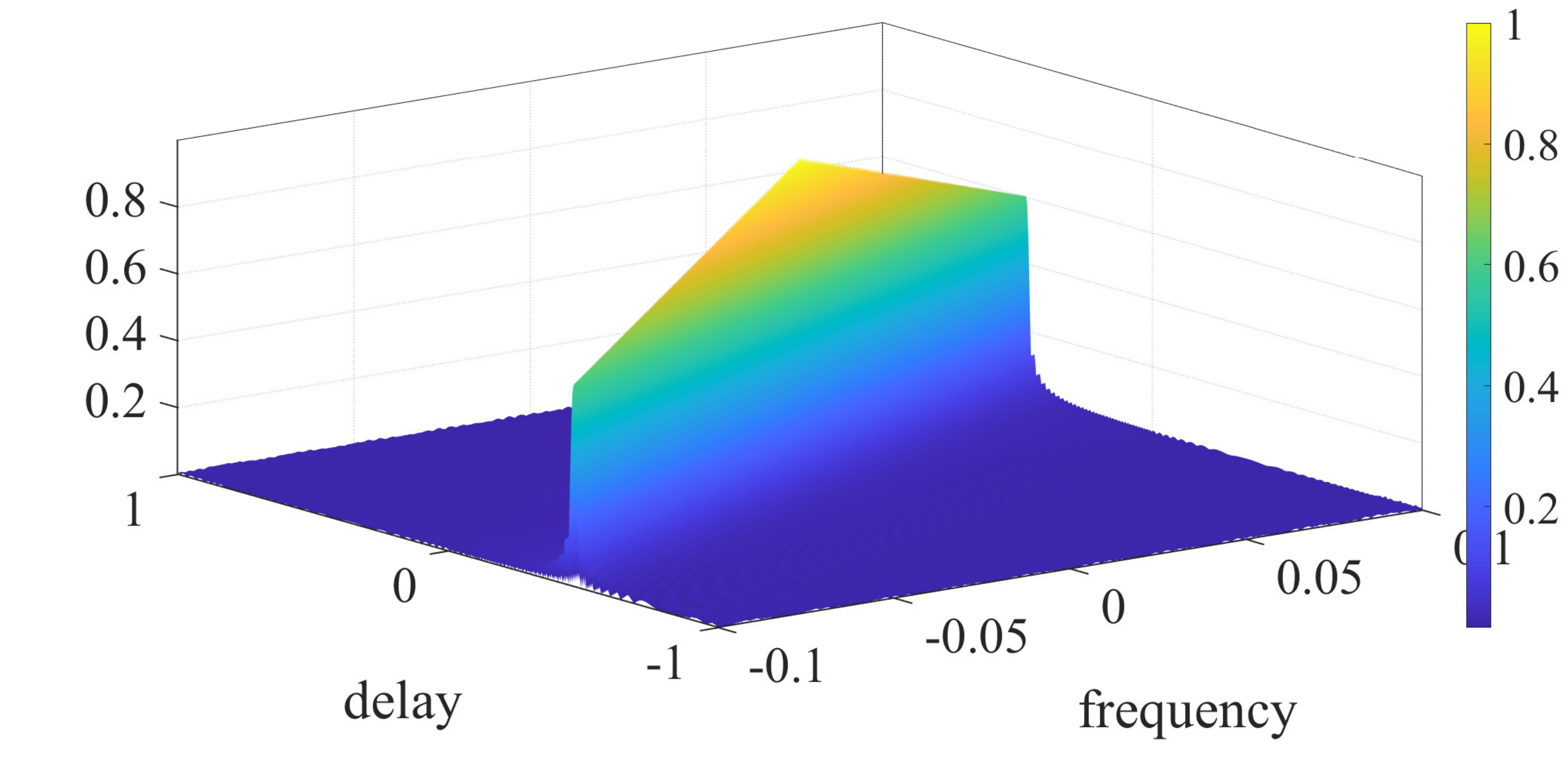}      &  \includegraphics[width=5cm,height=3cm]{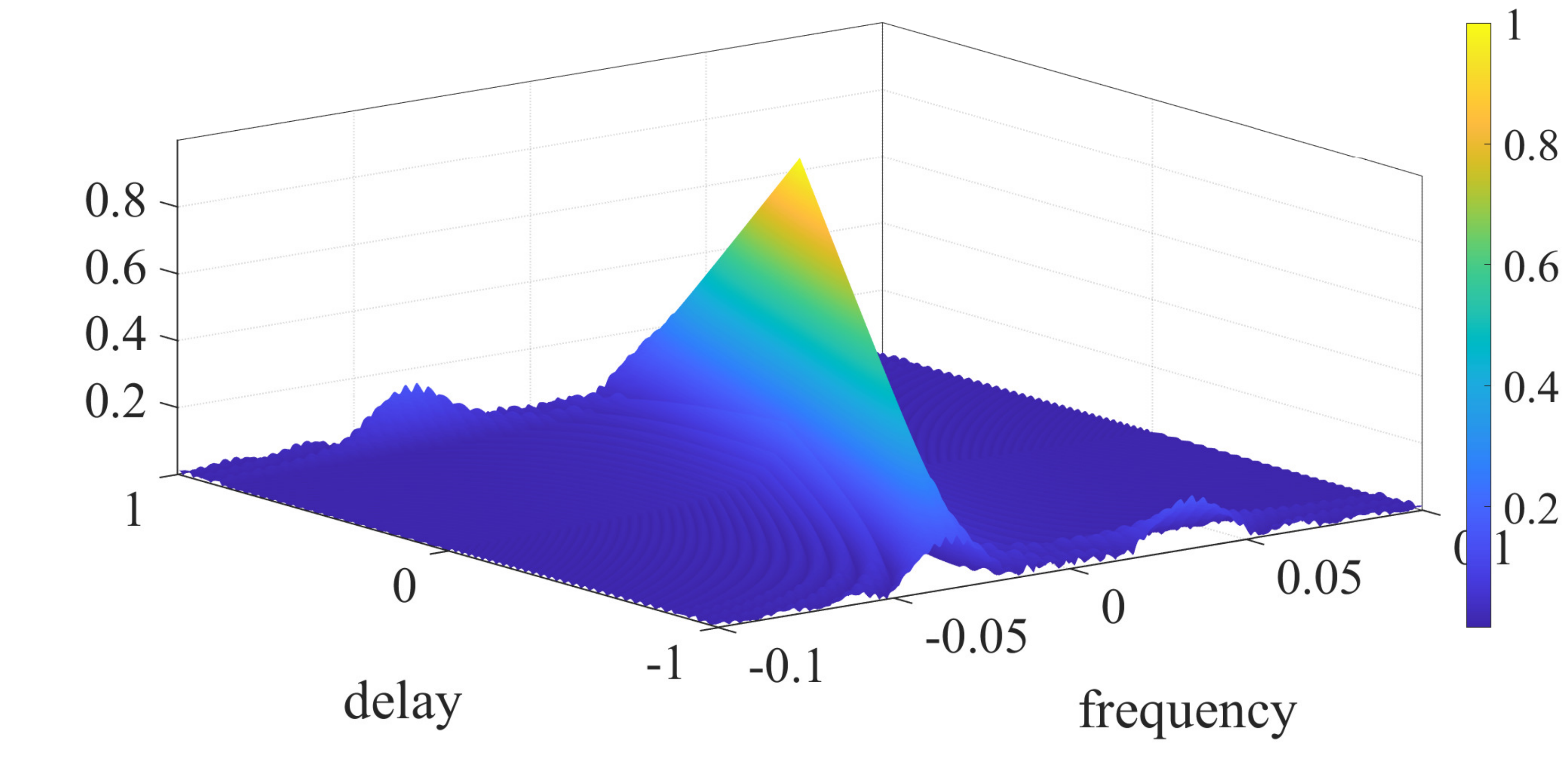}    & \includegraphics[width=5cm,height=3cm]{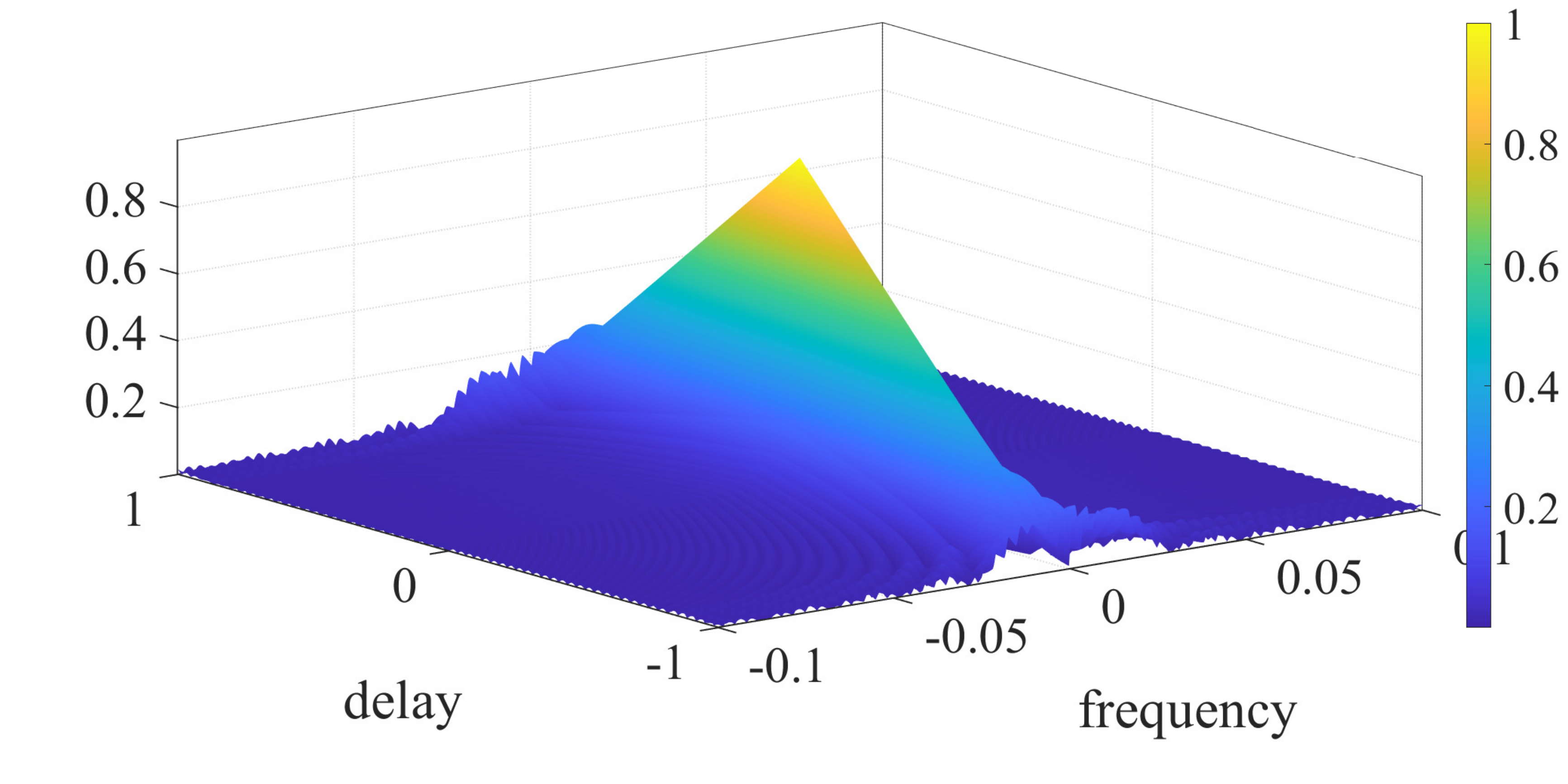} \\ \hline
            \end{tabular}\label{Table:PhsAmbigFn_Cmprsn}
    \end{center}
\end{table*}
\subsection{Doppler Tolerance Evaluation}
\begin{figure}[tbh]
    \centering
    \begin{subfigure}{.5\textwidth}
        \centering
        \includegraphics[width=.95\linewidth]{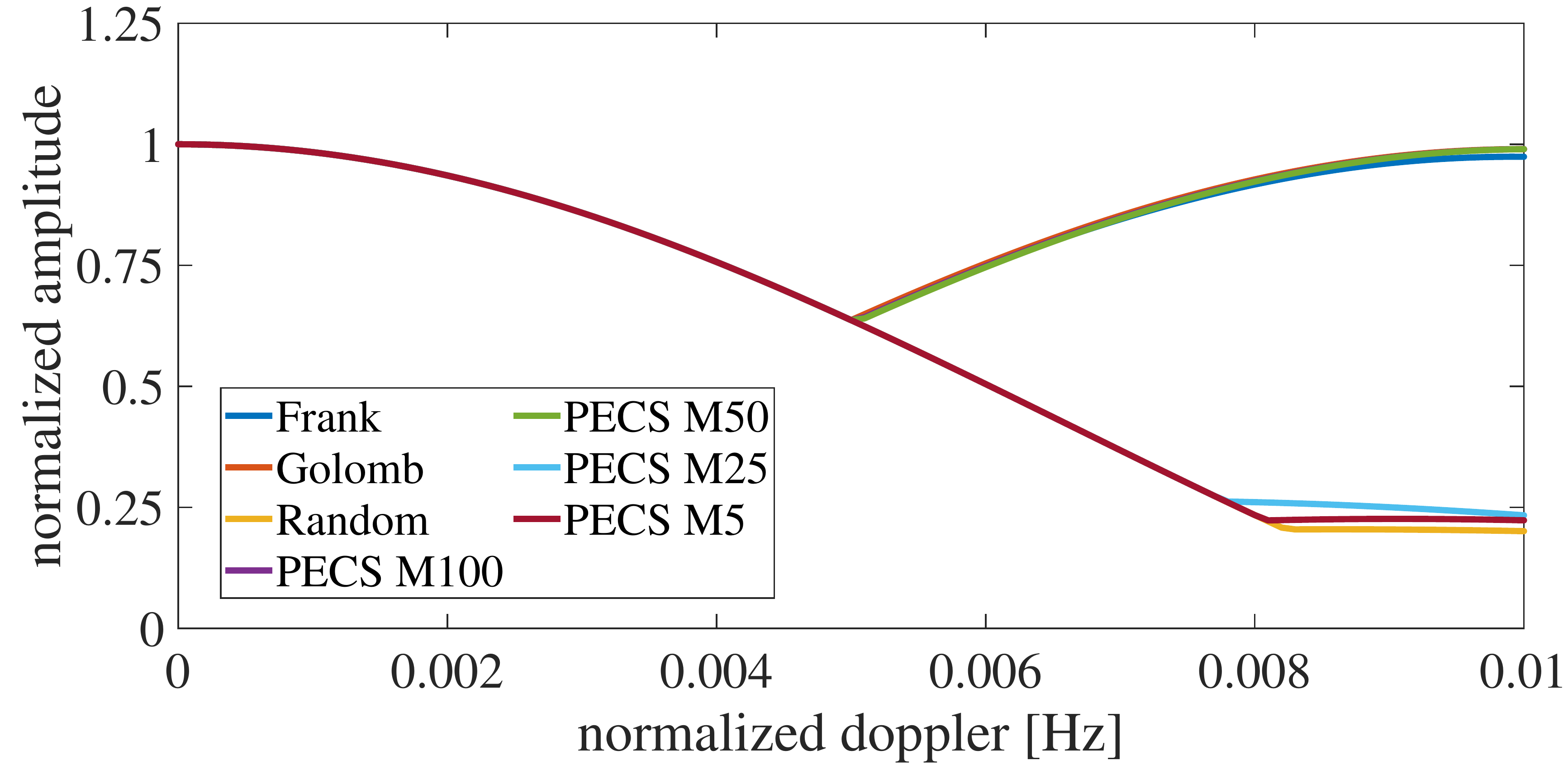}
        \vspace{-0.15in}
        \caption{Autocorrelation response comparison with Doppler shift}
        \label{fig:DplrTlrnce}
        \vspace{0.1in}
    \end{subfigure}
    \begin{subfigure}{.5\textwidth}

        \includegraphics[width=.95\linewidth]{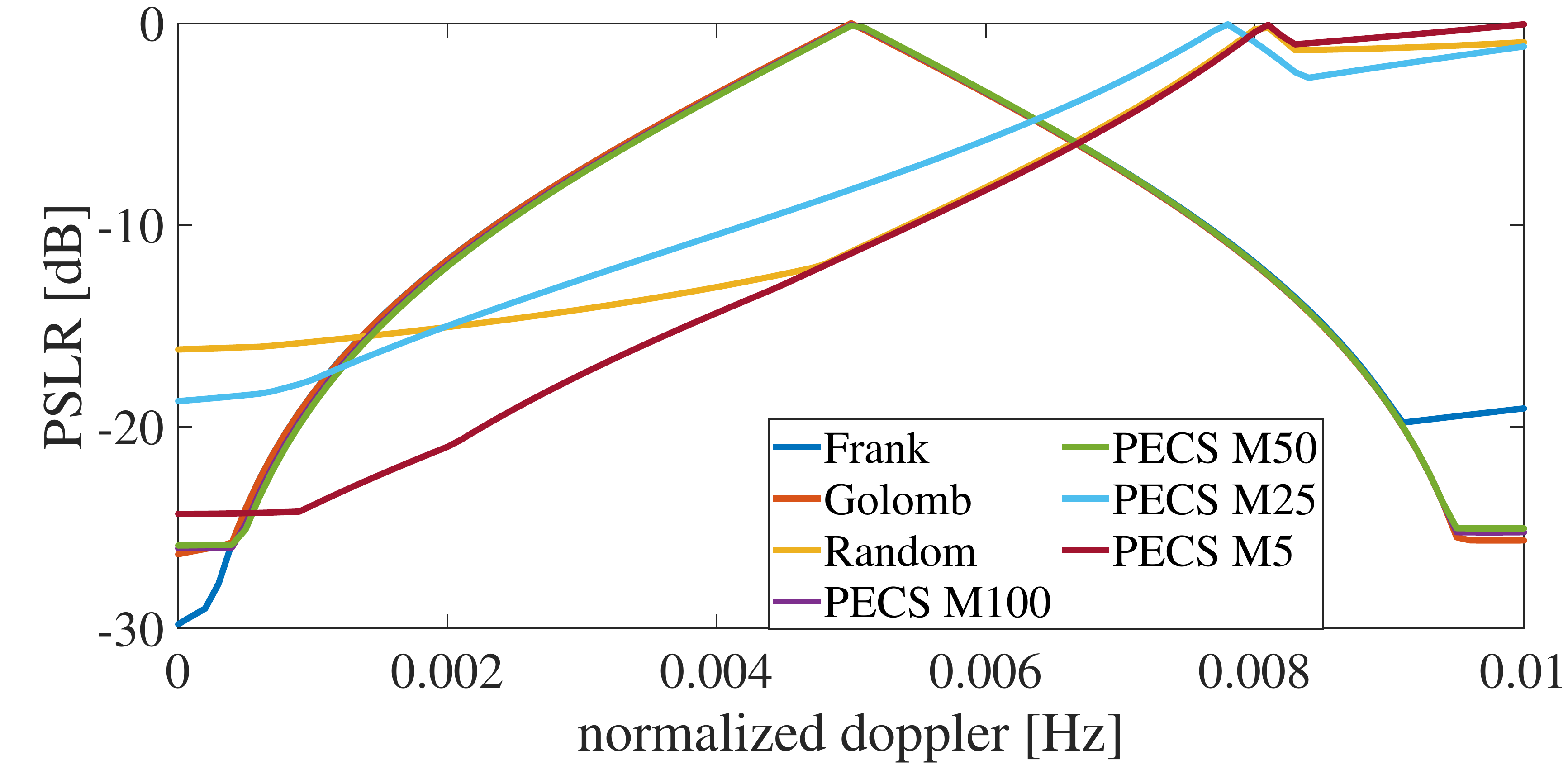}
        \caption{PSLR}
        \label{fig:PSLR_DplrTlrnce}
    \end{subfigure}
    \begin{subfigure}{.5\textwidth}

        \includegraphics[width=.95\linewidth]{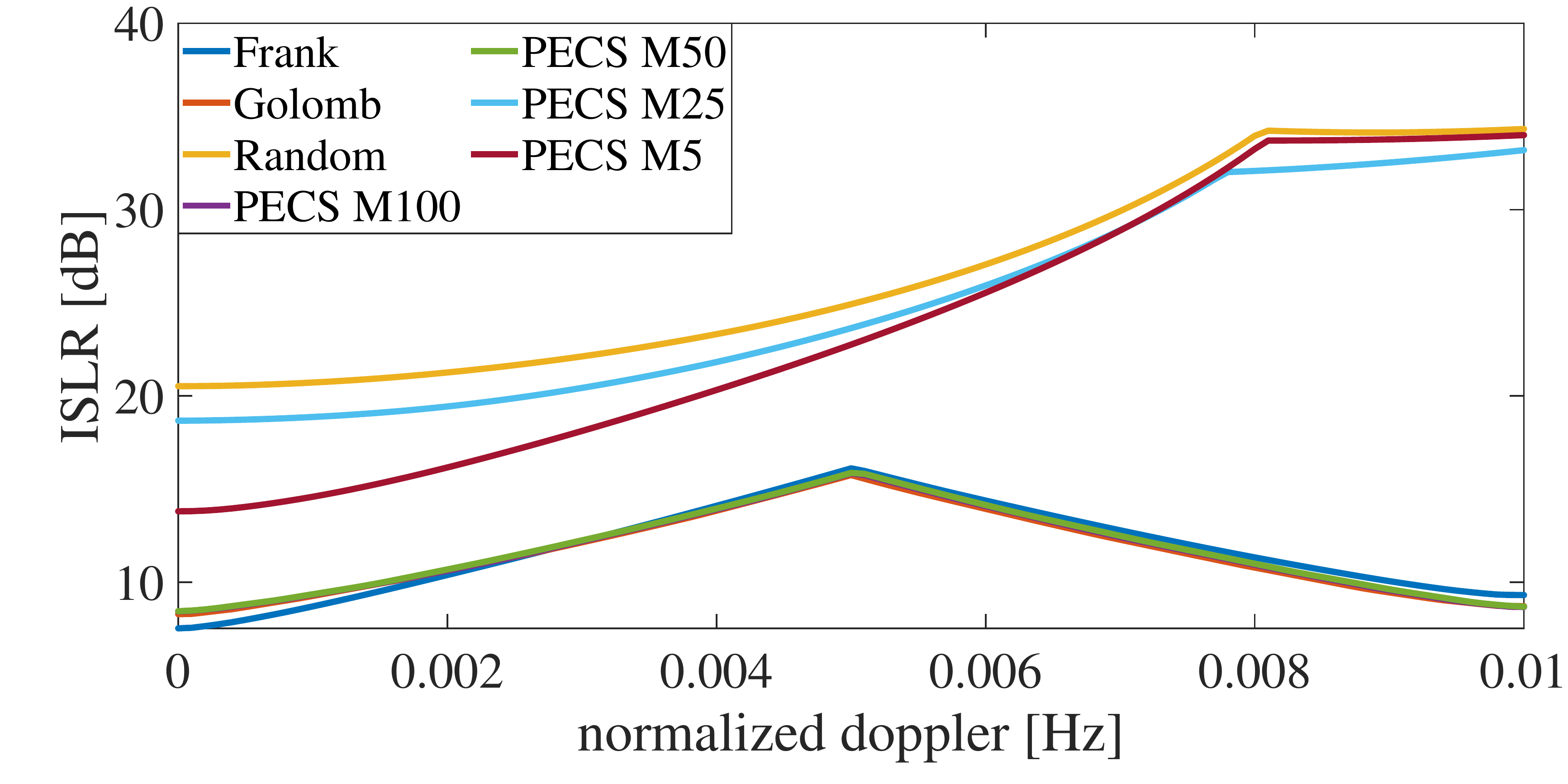}
        \caption{ISLR}
        \label{fig:ISLR_DplrTlrnce}
    \end{subfigure}
    \caption{Comparison of various sequences for evaluating the Doppler tolerance}
    \label{fig:DopplerTolerance}
\end{figure}

In \figurename{~\ref{fig:DopplerTolerance}}, we compare the Doppler tolerance characteristics of sequences: Frank, Golomb, Random and PECS with length $N=100$. 
In the automotive case study considered in this paper, 
the maximum expected Doppler frequency  for a 
radar operating at $79$GHz  from  the targets cruising at the highest absolute relative velocity of $400$kmph is roughly
$58$KHz. By considering range resolution of $1$m (or equivalently $150$MHz bandwidth), this Doppler frequency is less than $1\%$ of the transmit signal bandwidth. 
Therefore for a single chirp time period the normalized Doppler frequency range of interest shown in the figure is limited to 0.01Hz.
In \gls{PECS} algorithm, the input parameters are: polynomial degree, $Q=2$ and the sub-sequence length, $M$ varies with a relation: $\{5 \leq M \leq 100\}$ and the variations in the peak amplitude of the autocorrelation as a function of normalized Doppler are shown in \figurename{~\ref{fig:DplrTlrnce}}. 
As is evident from the plot, the peak of the matched filter decays for all the sequences as the Doppler shift increases due to target motion. But, a sharp decay of the peak value is observed in the case of Random sequence as compared to a Golomb sequence where the decay is rather linear in manner. In relation to these when the sequence is generated using PECS, and $M = N = 100$, its response is better than Frank sequence and similar to Golomb sequence. Further, as we decrease $M$, we observe that the Doppler shift impacts the peak and the response starts resembling the Random sequence for $M=5$. Similar conclusions can be drawn from the \figurename{~\ref{fig:ISLR_DplrTlrnce}} where \gls{ISLR} values of various sequences are observed. Further, the correlation level, \gls{PSLR} and the \gls{ISLR} variation with normalized Doppler is shown in \figurename{~\ref{fig:PSLR_DplrTlrnce}} and \figurename{~\ref{fig:ISLR_DplrTlrnce}} respectively. 
\begin{equation}
    \begin{aligned}
        \text{correlation} &\text{~level} \triangleq \big| \frac{r_k}{r_0} \big|, \\
        \text{PSLR}_{dB} &\triangleq 20\log_{10}\frac{\text{PSL}}{\text{max}\{ |r_k|\}_{k=0}^{k=N-1}},\\
        \text{ISLR}_{dB} &\triangleq 20\log_{10}\frac{\text{ISL}}{\text{max}\{ |r_k|\}_{k=0}^{k=N-1}}
    \end{aligned}
\end{equation}
On the other hand, as the number of sub-sequences increase in the \gls{PECS} sequences (e.g "PECS M25" and "PECS M5") the \gls{PSL} begins to rise and eventually lowers the \gls{PSLR}. This may impact the detection probability of the target.
\subsection{Comparison with the Counterparts}
\subsubsection{Performance}
\begin{figure}[tbh]
    \centering
    \begin{subfigure}{.5\textwidth}
        \centering
        \includegraphics[width=.95\linewidth]{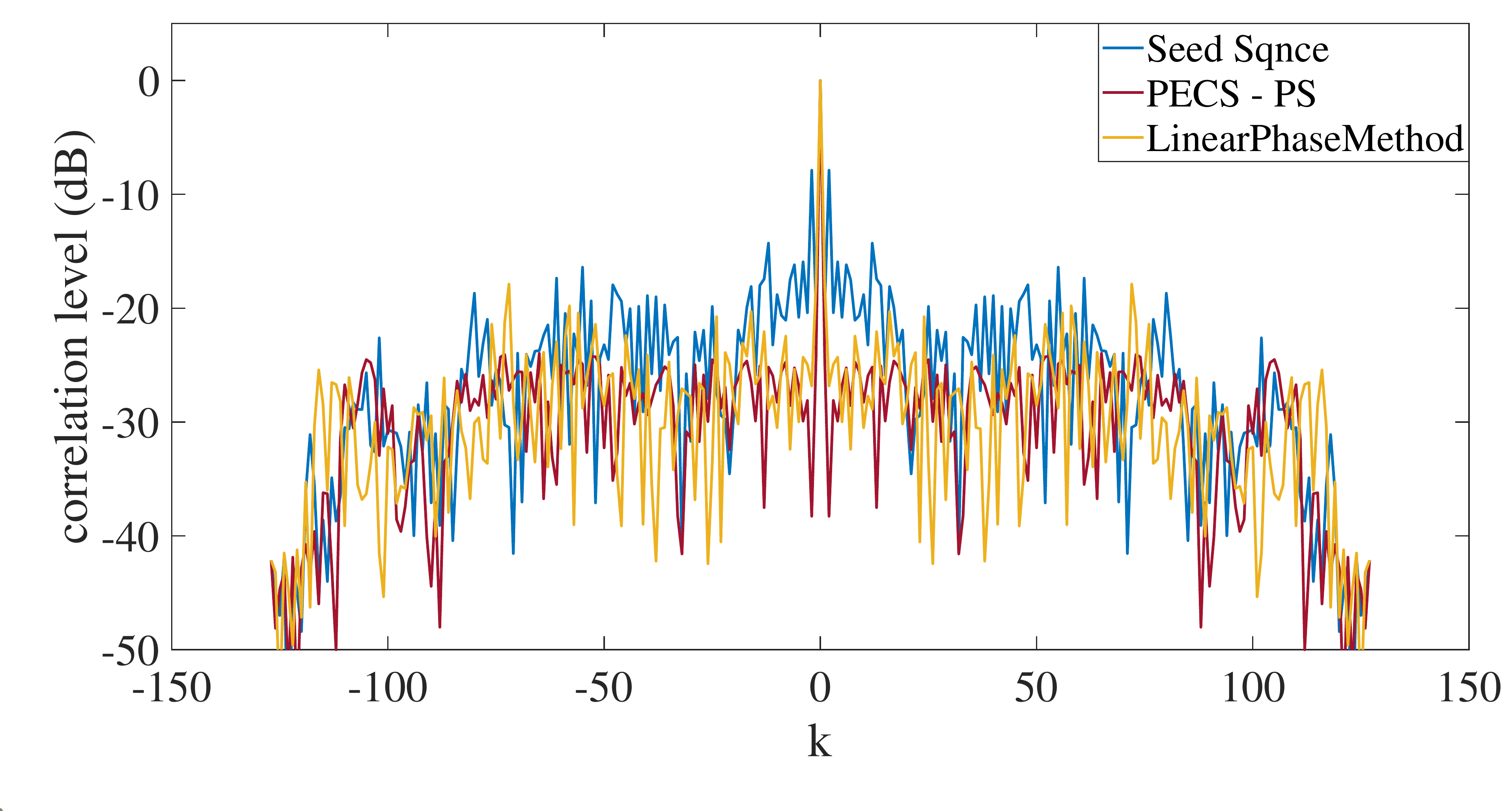}
        \vspace{-0.15in}
        \caption{Autocorrelation response comparison}
        \label{fig:MjtbaAtoCrln}
        \vspace{0.1in}
    \end{subfigure}
    \begin{subfigure}{.5\textwidth}

        \includegraphics[width=.95\linewidth]{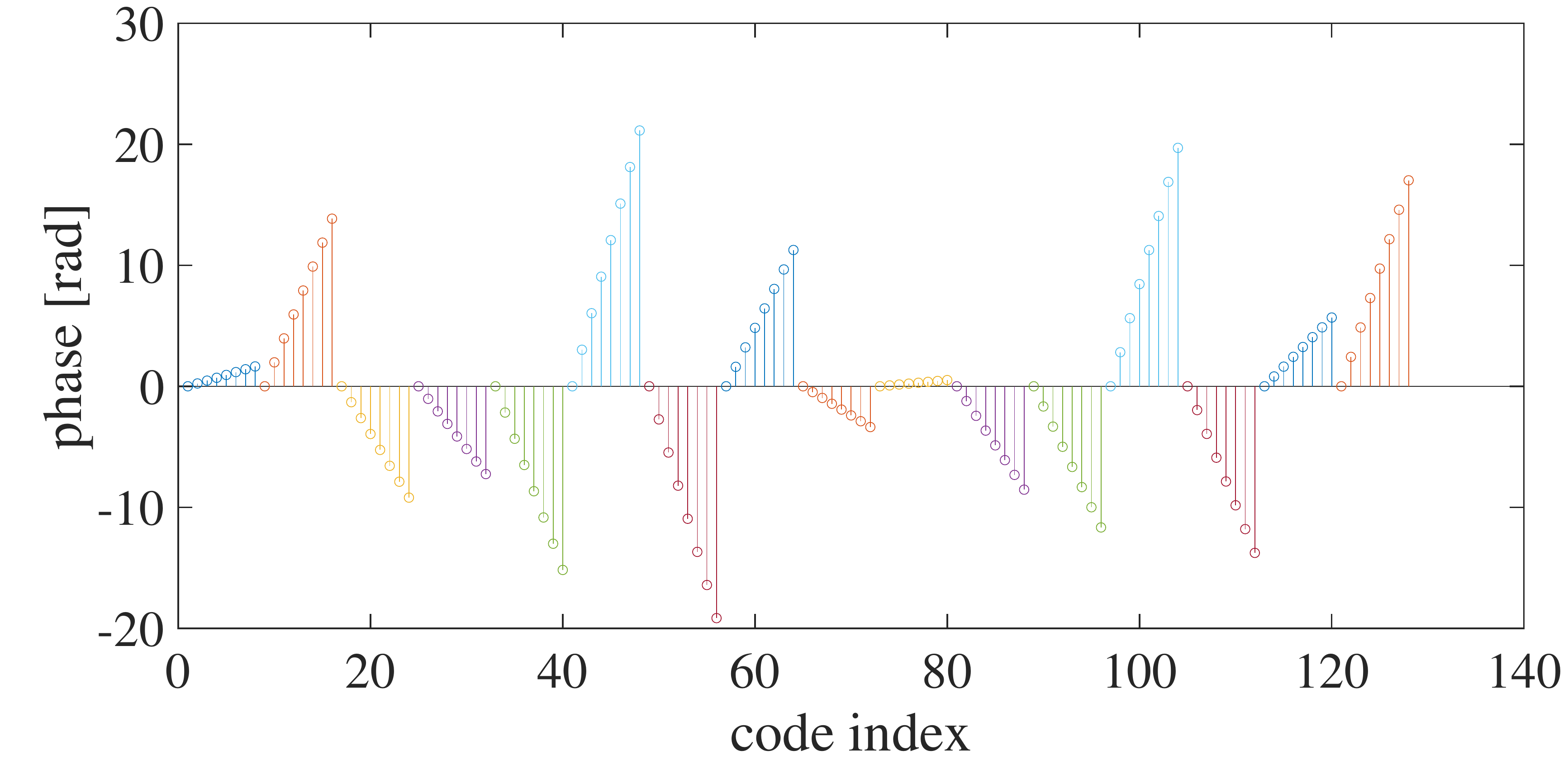}
        \caption{Unwrapped phase: Linear Phase Method}
        \label{fig:MjtbaUnwrpPhs}
    \end{subfigure}
    \begin{subfigure}{.5\textwidth}

        \includegraphics[width=.95\linewidth]{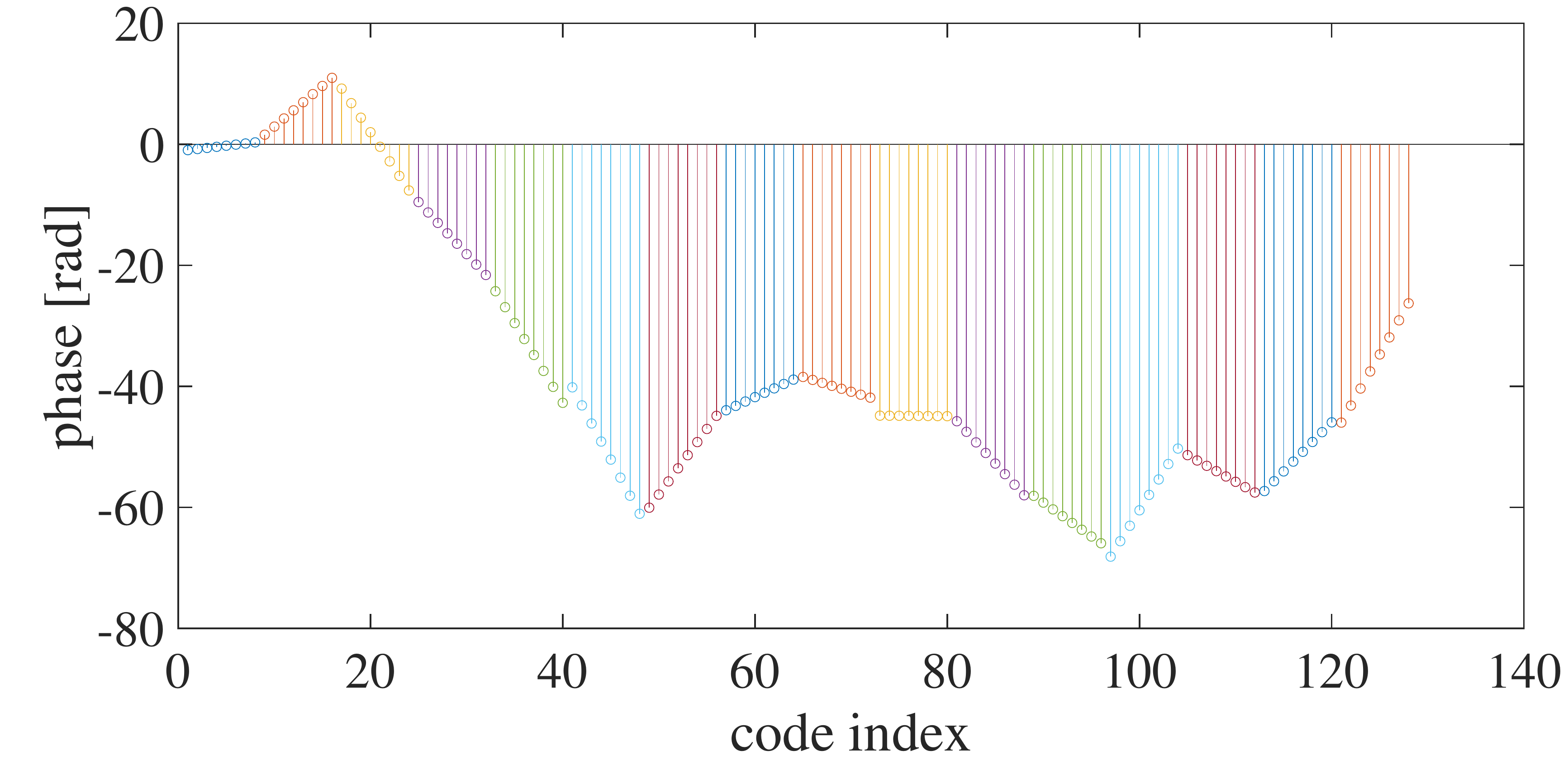}
        \caption{Unwrapped phase: PECS Method}
        \label{fig:PECSUnwrpPhs}
    \end{subfigure}
    \caption{Comparison of Linear Phase Method and PECS to design linear polyphase sequence with good autocorrelation properties.}
    \label{fig:PerfCmprsnMjtba}
\end{figure}

In \cite{6952459}, an approach was presented for designing polyphase sequences with piecewise linearity and impulse like autocorrelation properties (further referred as "LinearPhaseMethod"). 
In order to compare the performance of the LinearPhaseMethod with the proposed \gls{PECS}, we use both the algorithms to design a piecewise linear polyphase sequence of length $N = 128$ with sub-sequence length $8$ and thereby the total number of sub-sequences are $16$. Normalized autocorrelation of the optimized sequences from both the approaches is shown in \figurename{~\ref{fig:MjtbaAtoCrln}} and it shows lower \gls{PSL} values of the autocorrelation for PECS as compared to LinearPhaseMethod. 
Unwrapped phase of the optimized sequences are shown in \figurename{~\ref{fig:MjtbaUnwrpPhs}}\footnote{the phase unwrapping operation can be expressed mathematically as 
\begin{equation*}\label{eq:unwrap}
\mathbf{x}_{U} = \ignore{\textcolor{Red}{\cal{U}}} {\cal{F}}[\mathbf{x}_{W}] = \text{arg}(\mathbf{x}_{W}) + 2 k \pi
\end{equation*}
where \ignore{\textcolor{Red}{$\cal{U}$}}${\cal{F}}$ is the phase unwrapping operation, k is an integer, $\mathbf{x}_{W}$ and $\mathbf{x}_{U}$ are the wrapped and unwrapped phase sequence, respectively}.
Linear Phase Method generates an optimal sequence whose \gls{ISL} and \gls{PSL} values are $36.47$dB and $12.18$dB respectively whereas \gls{PECS} results in an optimal sequence whose \gls{ISL} and \gls{PSL} values are $32.87$dB and $9.09$dB. Therefore, better results are obtained using \gls{PECS} approach. 

\begin{figure}[!htbp]
    \centering
    \begin{subfigure}{.5\textwidth}
        \centering
        \includegraphics[width=.95\linewidth]{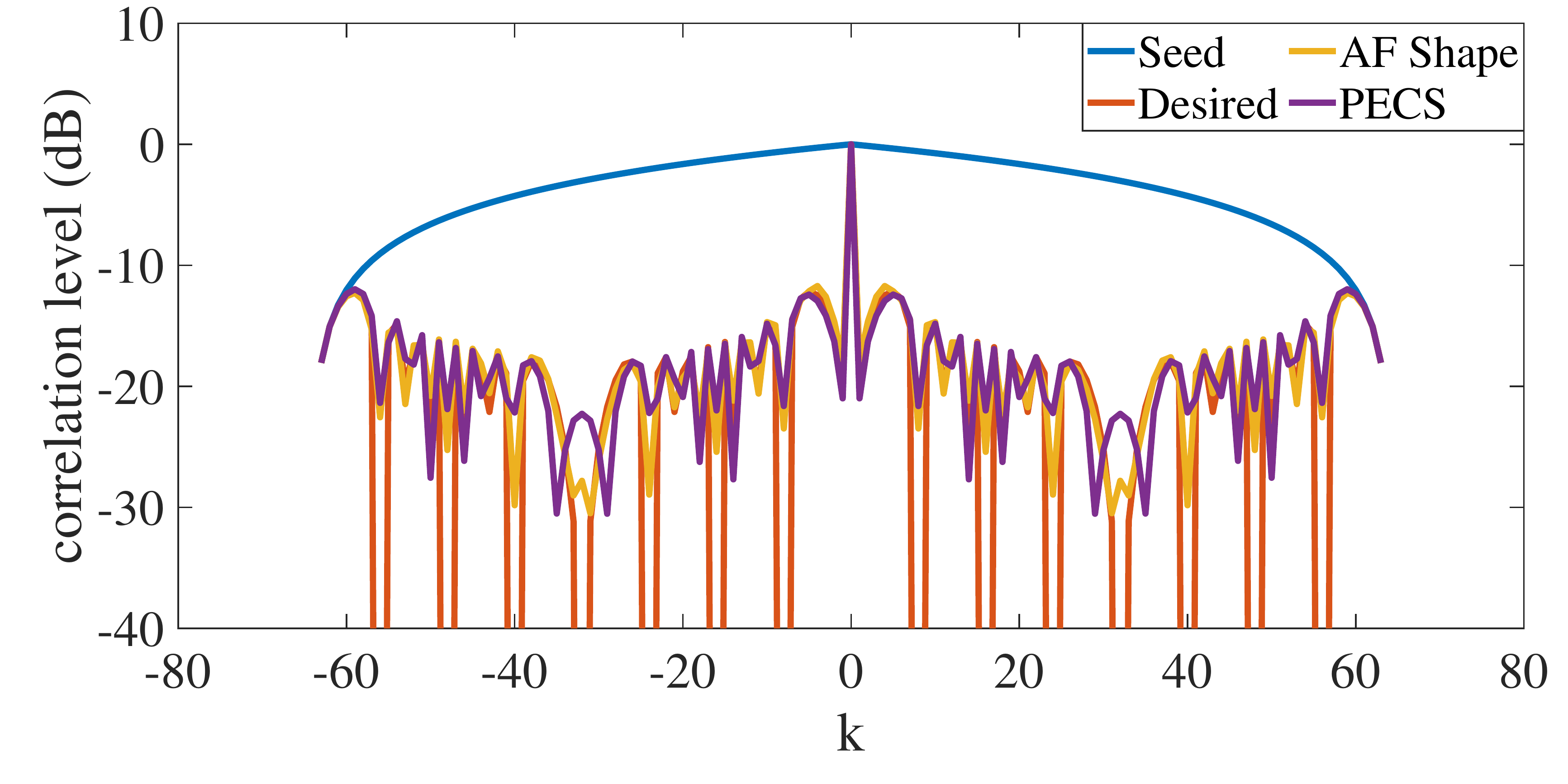}
        \vspace{-0.1in}
        \caption{Autocorrelation response comparison}
        \vspace{0.1in}
    \end{subfigure}
    \begin{subfigure}{.5\textwidth}
        \centering
        \includegraphics[width=.95\linewidth]{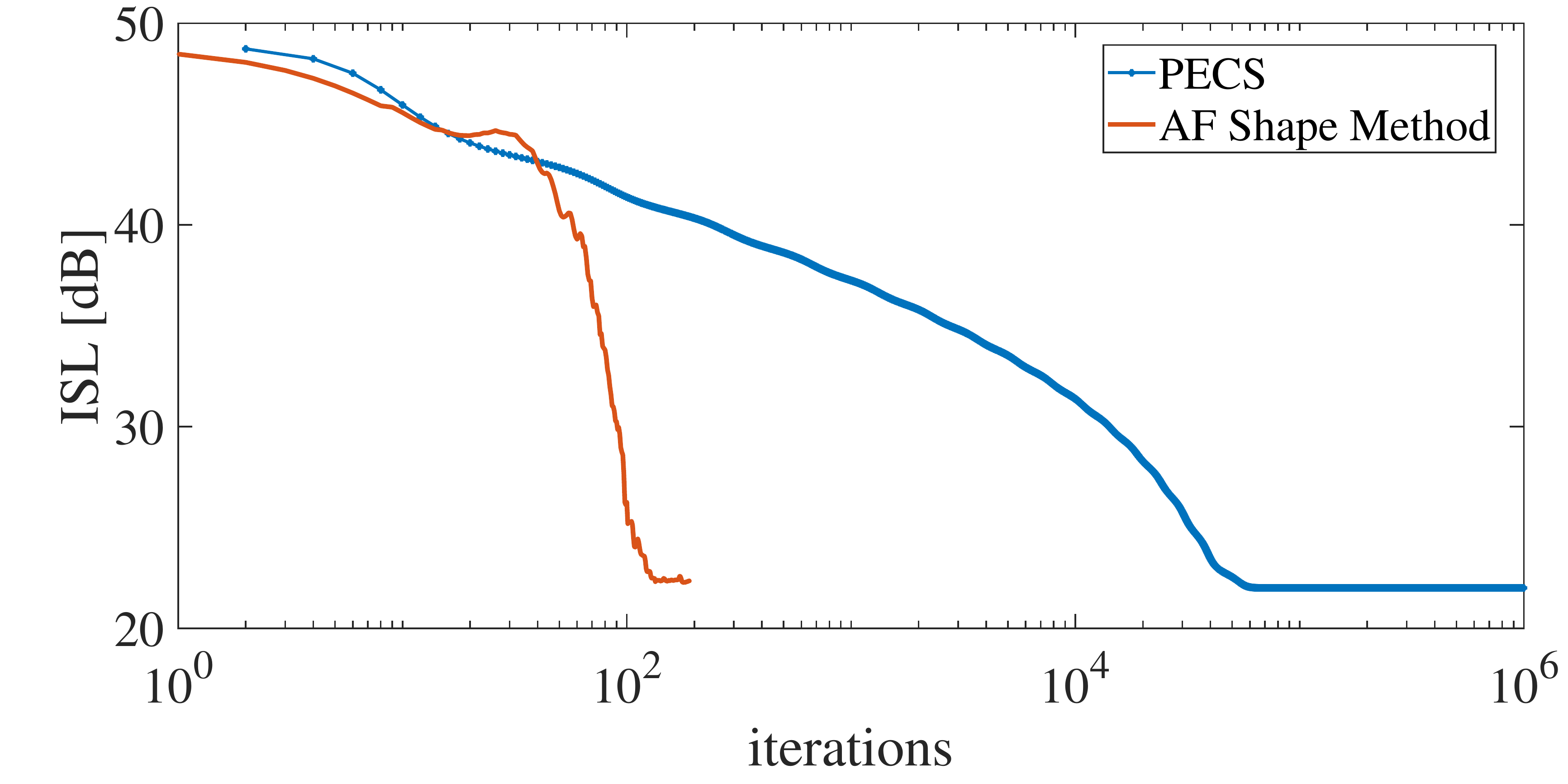}
        \caption{ISL Convergence: PECS Method}
        \label{fig:PECS_ISLCnvrgnce}
    \end{subfigure}
    \caption{Performance comparison of AF Shape method and PECS algorithms.}
    \label{fig:PerfCmprsn1}
\end{figure}

In \cite{8682216}, an approach to shape the \gls{AF} of a given sequence w.r.t. a desired sequence was proposed (later referred as "AF shape method"). Here, we consider an example where the two approaches (i.e. AF shaping method and PECS) strive to achieve the desired \gls{AF} of a Golomb sequence of length $N=64$. The performance of the two approaches would be assessed by comparing the autocorrelation responses and ISL/PSL values of the optimal sequences. 
Both the algorithms are fed with the same seed sequence and the convergence criterion is kept the same for better comparison.
As evident from the \figurename{~\ref{fig:PerfCmprsn1}}, the autocorrelation function of the optimal sequence derived from \gls{PECS} shows improvement as compared to the optimal sequence of benchmark approach. The initial \gls{ISL} of the seed sequence was $49.30$dB and the desired Golomb sequence was $22.050$dB. After the optimization was performed, the optimal \gls{ISL} using AF shaping approach was $22.345$dB and using \gls{PECS} was $22.002$dB. In addition, the ridge shape of the \gls{AF} generated using both the approaches is equally matched to the desired \gls{AF} of Golomb sequence. Noteworthy point over here is that the monotonic convergence is absent in the AF shape method as it optimizes using the \gls{CD} approach whereas in the \gls{PECS} method, monotonic convergence is achieved. Further, PECS has the capability of achieving better ISL values than the Golomb sequence as it aims to minimize the objective in \eqref{eq:PSLOptC2} and its proof can be seen from the optimal ISL value quoted above (i.e. $0.048$dB improvement w.r.t. ISL of Golomb sequence).

\subsubsection{Run-time}
To calculate run-time of the algorithm 
we used a PC with the following specifications: $2.6$GHz i$9-11950$H CPU and $32$GB RAM. 
No acceleration schemes (i.e. Parallel Computing Toolbox in MATLAB) are used to generate the results and are evaluated from purely sequential processing.

For a sequence length of $N=300$, computational time was derived by varying two input parameters: sub-sequence length $M = 5, 50, 150 ~\&~ 300$ and $Q = 2, 3, 4, 5 ~\&~ 6$. The results mentioned in \tablename{~\ref{Table:ComputationalTime}} indicate that the computation time increases in proportion to the increasing values of $Q$ keeping $M$ fixed. On the other hand, computation time decreases as we keep $Q$ fixed and increase $M$.
\begin{table}[!htbp]
    \caption{PECS runtime - sequence length $N = 300$}
    \centering
    \begin{center}
            \begin{tabular}{|M{4em}|M{4em}|M{4em}|M{4em}|M{4em}|}
            \hline
            & \textbf{M$ = 5$} & \textbf{M$=50$} & \textbf{M$=150$} & \textbf{M$=300$}     \\\hline\hline
            \textbf{Q$=2$} & $429.25$s & $84.26$s  & $52.53$s  & $45.15$s\\ \hline
            \textbf{Q$=3$} & $521.41$s & $88.32$s  & $55.03$s  & $47.42$s\\ \hline
            \textbf{Q$=4$} & $562.04$s & $97.20$s  & $64.55$s  & $56.54$s\\ \hline
            \textbf{Q$=5$} & $573.72$s & $108.37$s & $73.47$s  & $65.47$s\\ \hline
            \textbf{Q$=6$} & $620.90$s & $118.82$s & $83.39$s  & $74.06$s\\ \hline
            \end{tabular}\label{Table:ComputationalTime}
    \end{center}
\end{table}

\subsection{Automotive Scenario}

\begin{figure}[t!]
    \centering
    \begin{subfigure}{.5\textwidth}
        \centering
        \includegraphics[width=.99\linewidth]{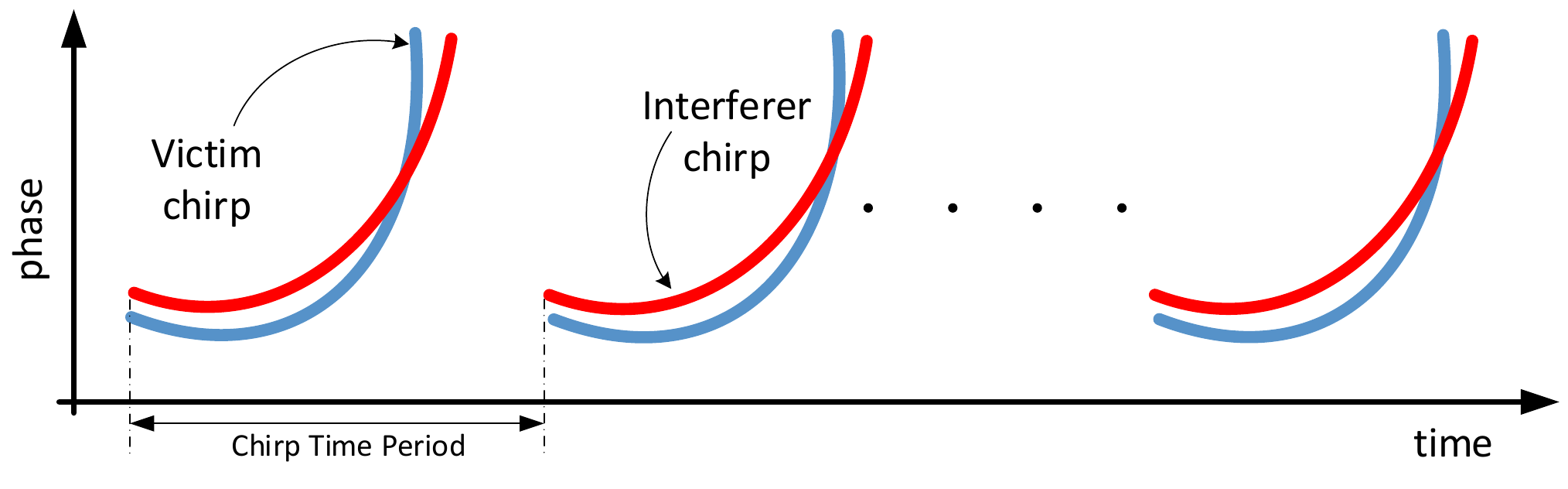}
    \caption{Quadratic phase FMCW waveform}
    \label{fig:PhaseCmprsn_FMCW}
    \end{subfigure}
    \begin{subfigure}{.5\textwidth}
        \centering
        \includegraphics[width=.99\linewidth]{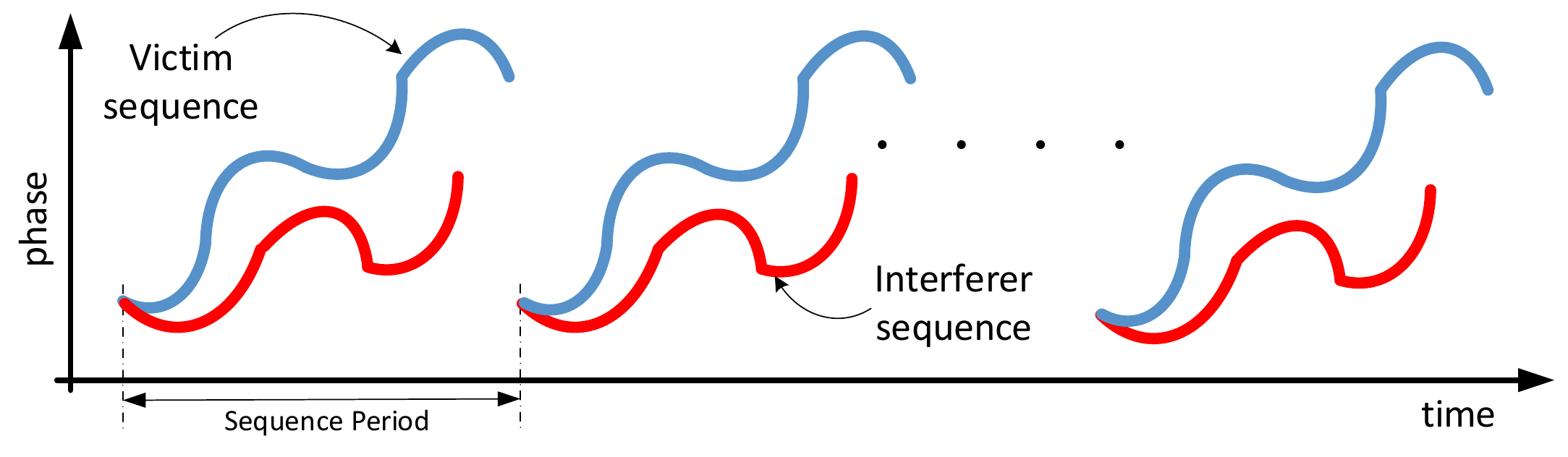}
    \caption{PECS based PMCW waveform}
    \label{fig:PhaseCmprsn_PMCW}
    \end{subfigure}
    \caption{Phase variation comparison of FMCW and PMCW waveform.}
    \label{fig:PhaseCmprsn_FMCW_PMCW}
\end{figure}

\begin{figure*}[t!]
    \centering
    \begin{subfigure}{0.24\textwidth}
        \centering
        \includegraphics[width=0.99\textwidth]{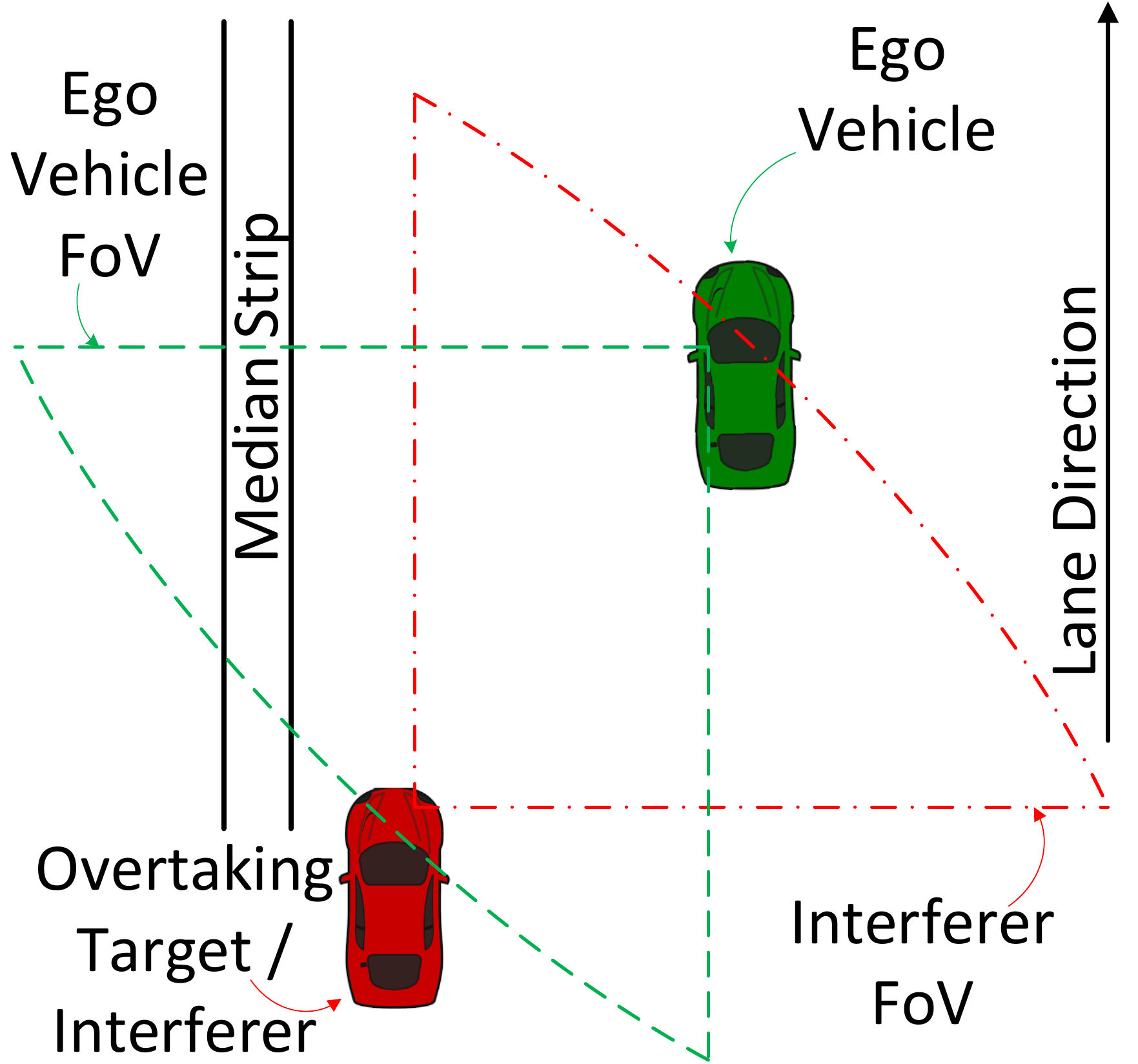}
        \caption{Blind Spot Detection}
        \label{fig:BSD}
    \end{subfigure}
    \begin{subfigure}{0.24\textwidth}
        \centering
        \includegraphics[width=0.99\textwidth]{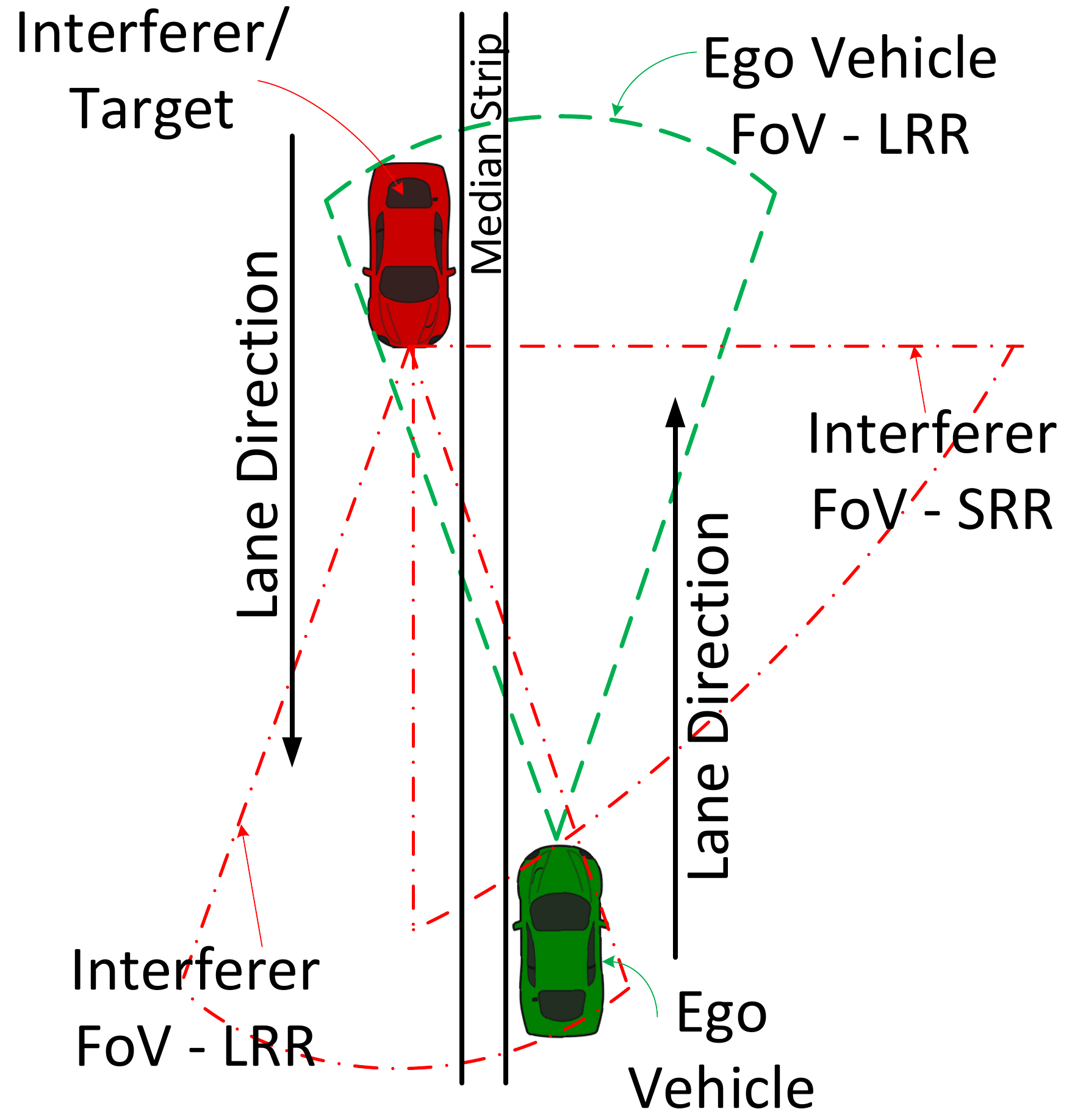}
        \caption{Highway Scenario}
        \label{fig:LRR_SRR}
    \end{subfigure}
    \begin{subfigure}{0.24\textwidth}
        \centering
        \includegraphics[width=0.9\textwidth]{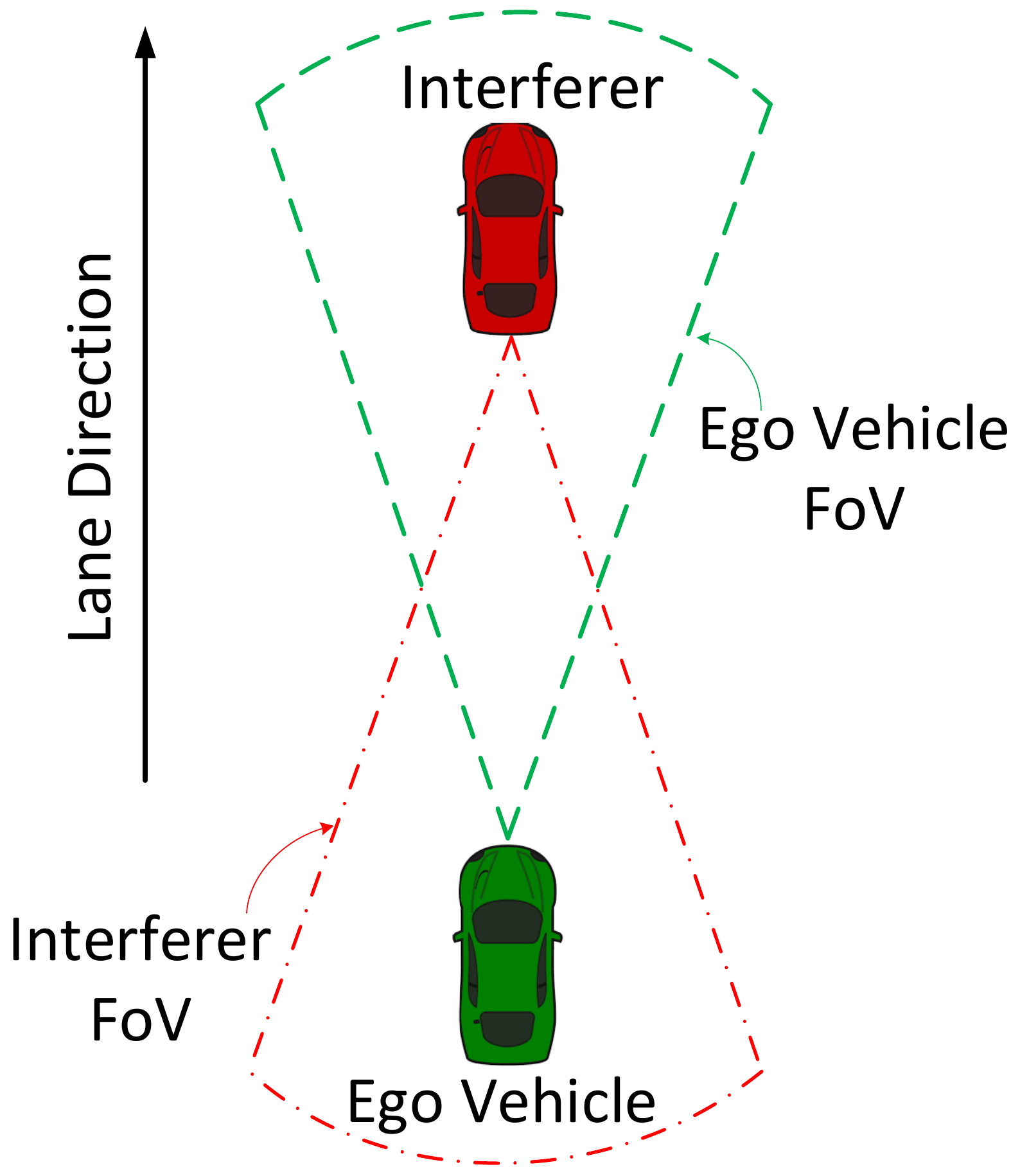}
        \caption{Adaptive Cruise Control}
        \label{fig:LRR_LRR}
    \end{subfigure}
    \begin{subfigure}{0.24\textwidth}
        \centering
        \includegraphics[width=0.99\textwidth]{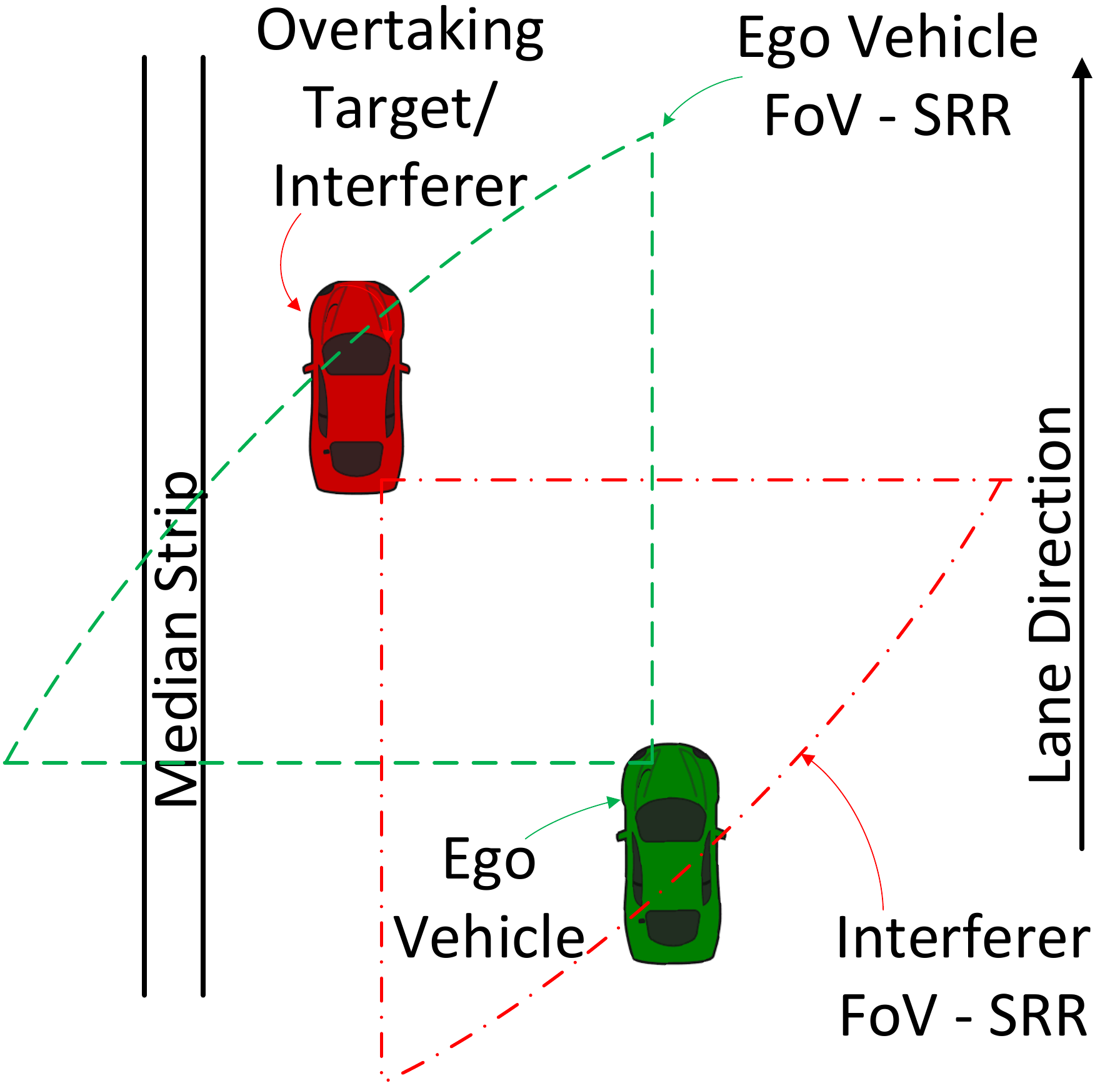}
        \caption{Overtaking Scenario}
        \label{fig:SRR_SRR}
    \end{subfigure}
    \caption{Automotive scenarios with mutual interference.}
    \label{fig:ScenariosInterference}
\end{figure*}
The idea is to generate a code sequence using \gls{PECS} algorithm
towards designing a transmission scheme for \gls{PMCW} and evaluate its performance by comparing it with the \gls{FMCW} waveform transmission when interference is introduced in the \gls{FoV} of the victim sensor.
\figurename{~\ref{fig:PhaseCmprsn_FMCW_PMCW}} shows the phase-time plot comparison of both the waveforms.
In this plot, the pulse length, $T_p$ (i.e. chirp time period - FMCW and sequence period - PMCW) and
\gls{PRF}, $f_p$ of both the type of waveforms is matched for simplicity of the concept which results in equal Doppler response of the target.
\gls{PECS} code sequence has input parameters: quadratic phase $Q = 2$, code length $N = \frac{T_p}{\tau_{chip}}$ and varying sub-sequence length $M$.
Thus, the \gls{MI} occurs within a single pulse time period $T_p$ amongst FMCW waveforms but it is less likely going to occur in PECS waveforms. 
Even though the initial phase in \gls{FMCW} waveform from two different sensors would start from any random initialization, they tend to interfere as they exhibit similar quadratic phase behavior for the whole chirp time.
This phenomenon is not seen in \gls{PECS} waveforms 
since any sensor can choose different sub-sequence lengths and hence result in different quadratic phase coefficients resulting
in minimal cross-correlation with other transmitting sequences in its \gls{FoV}.

In \figurename{~\ref{fig:ScenariosInterference}} few automotive scenarios are described where \gls{MI} may occur. 
As mentioned in the introduction, two types of interference are witnessed using \gls{FMCW} i.e. Similar-slope and sweeping slope interference.
The similar-slope interference occurs when the sensors operate with same specifications in order to sense the environment for the following applications:
\begin{itemize}
    \item whenever a sensor is mounted for \gls{BSD} application in ego-vehicle \figurename{~\ref{fig:BSD}} and another vehicle with \gls{SRR} mounted on front side enters the \gls{FoV}.
    \item when the ego vehicle is equipped with \gls{ACC} function, it follows the trajectory of a car driving ahead. The target vehicle may be equipped with a rear mounted radar and would lie in the \gls{FoV} of the front mounted radar on the ego vehicle \figurename{~\ref{fig:LRR_LRR}}.
    \item Front side mounted radar from the Ego-Vehicle may get interfered by another overtaking vehicle equipped with a rear side mounted radar \figurename{~\ref{fig:SRR_SRR}}.
\end{itemize}

In \figurename{~\ref{fig:LRR_SRR}}, Sweeping slope interference may occur when a side looking radar of the oncoming target vehicle is in the \gls{FoV} of the ego vehicle's front looking radar, provided both are operating in the same band (i.e. $77-81$GHz).
\subsubsection{Simulation results}
To demonstrate the efficacy of the proposed algorithm, we simulated an automotive driving scenario for the cases described in \figurename{~\ref{fig:ScenariosInterference}}. Here, radar sensing performance was evaluated independently in two different cases:
\begin{itemize}
    \item when the ego-vehicle is operating with \gls{FMCW} waveform and \gls{MI} (i.e. similar-slope and sweeping slope) is observed from the target. 
    \item the ego-vehicle is operating with \gls{PMCW} waveform using \gls{PECS} algorithm based code sequence and encounters \gls{MI} from another target in its \gls{FoV} with the same scheme.
\end{itemize}
\begin{table}[!htbp]
    \caption{Radar sensor parameters and Motion information for ego-vehicle and interfering vehicle}
    \begin{center}
            \begin{tabular}{|M{1em}|M{20em}|M{5em}|}
            \hline
            & \textbf{parameter} & \textbf{value} \\ \hline\hline
            \multirow{10}{1em}{\rotatebox{90}{FMCW \& PMCW params  }} & Operating Frequency & $79$GHz   \\ 
           & Antenna Gain  & $10$dB          \\ 
           & Range Resolution & $1$m         \\ 
           & Transmit Power (Victim / Interferer) & $12$dBm \\ 
           & Bandwidth (Victim)    & $150$ MHz       \\
           & Bandwidth (Sweeping-slope Interferer)    & $75$ MHz       \\
           & Bandwidth (Similar-slope Interferer)    & $148.5$ MHz       \\
           & Pulse length ($T_p$)    & $60$ us         \\ 
           & PRF ($f_p$)          & $16.66$ kHz     \\ 
           & Number of Pulses & $256$        \\ 
           & Chip Time     & $6.66$ ns       \\ 
           & Code Length   & $4500$          \\ \hline \hline
           \multirow{5}{1em}{\rotatebox{90}{Motion Info}}& Target Range  & $30$m               \\ 
           & Target Speed  & $20$kmph            \\ 
           & Target RCS    & $10$dBsm            \\ 
           & Interferer Range & $50$m            \\ 
           & Interferer Speed & $40$kmph         \\ \hline
            \end{tabular}\label{Table:RadarSensorParams}
    \end{center}
\end{table}

\begin{figure}[!htbp]
    \centering
    \begin{subfigure}{.5\textwidth}
        \centering
        \includegraphics[width=.9\linewidth]{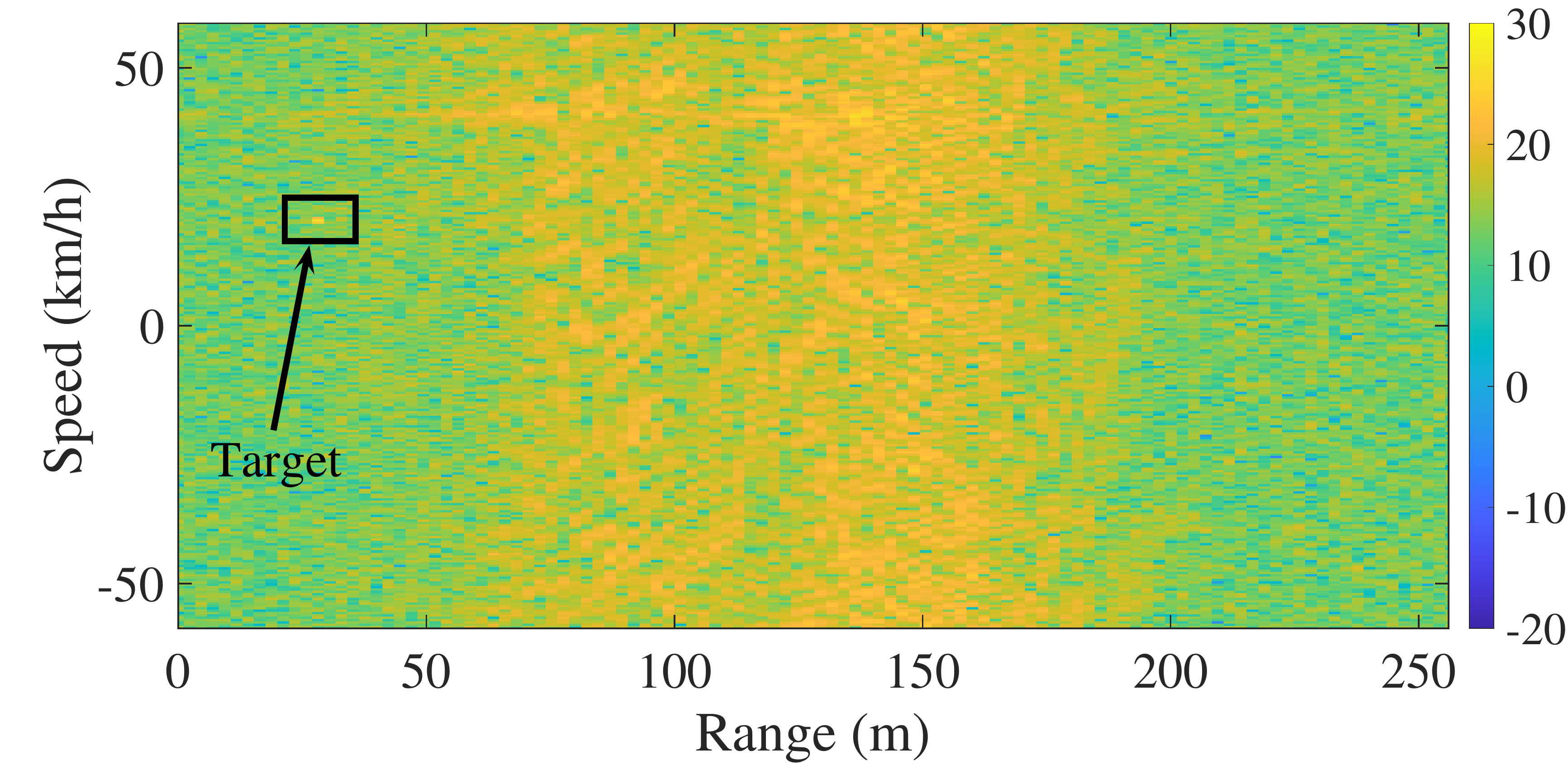}
        \caption{Similar slope interference in FMCW waveform}
        \label{fig:SimSlopeInterf}
    \end{subfigure}
    \begin{subfigure}{.5\textwidth}
        \centering
        \includegraphics[width=.9\linewidth]{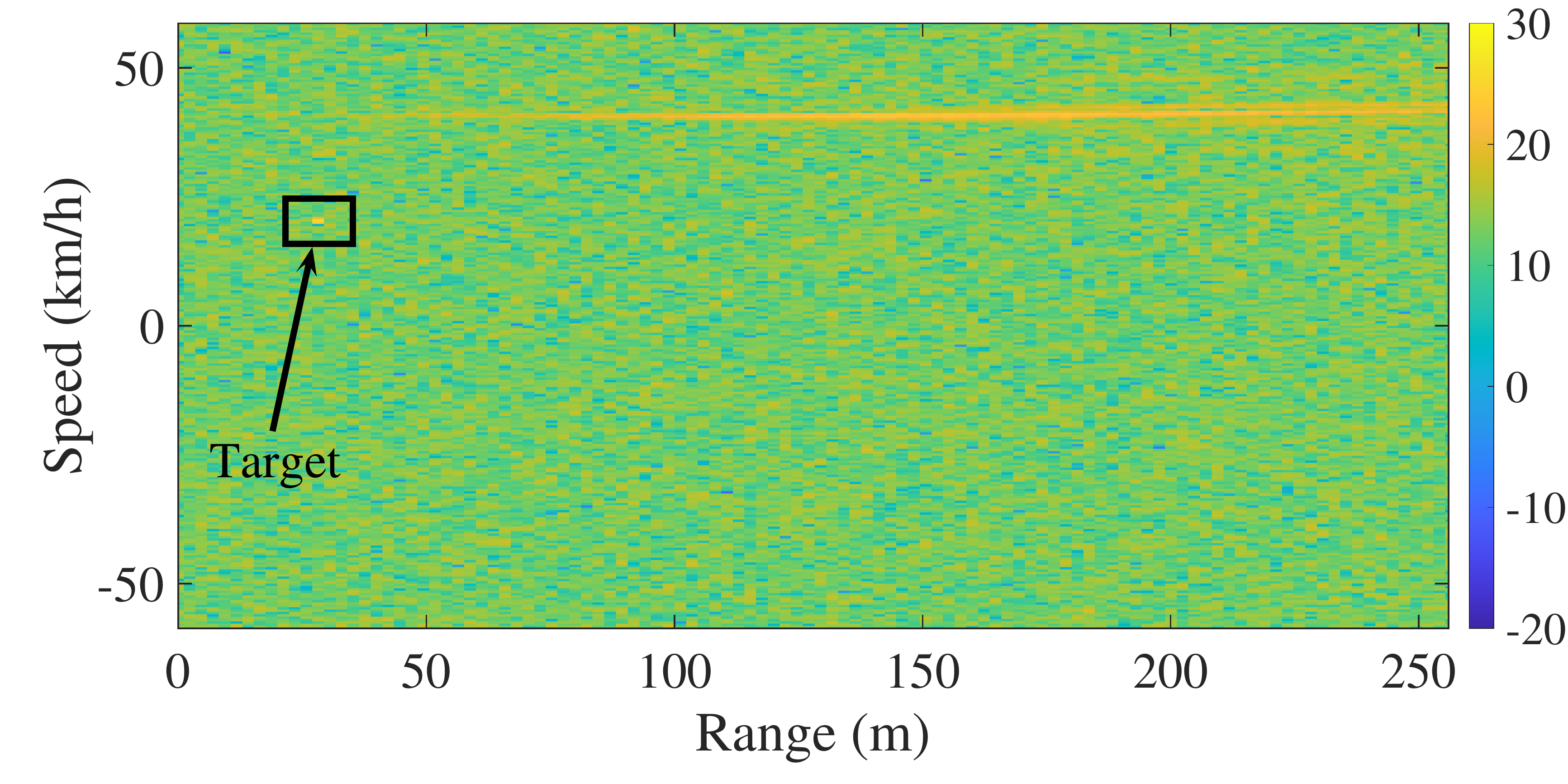}
        \caption{Sweeping slope interference in FMCW waveform}
        \label{fig:SweepSlopeInterf}
    \end{subfigure}
    \begin{subfigure}{.5\textwidth}
        \centering
        \includegraphics[width=.9\linewidth]{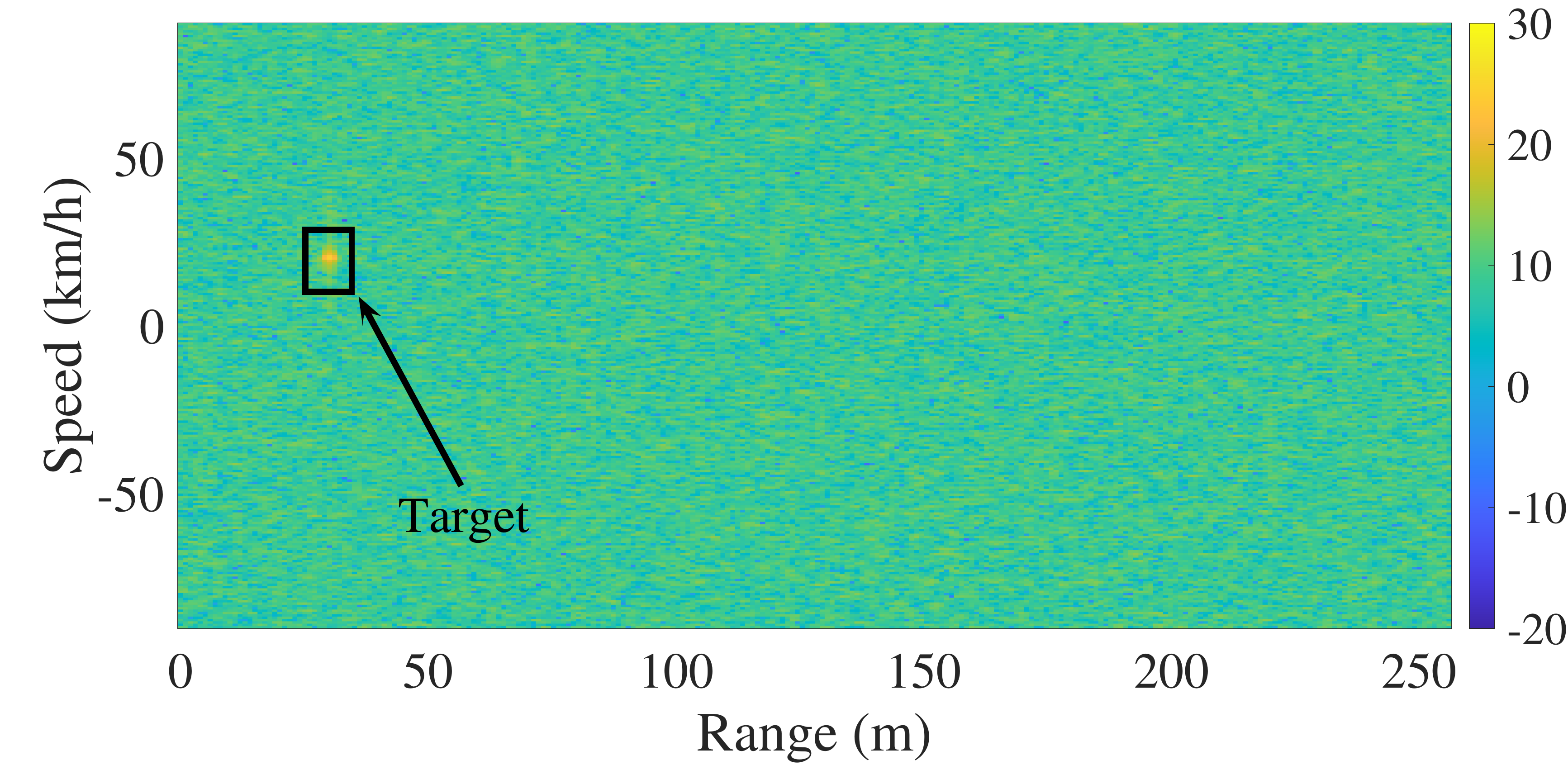}
        \caption{PMCW-PECS waveform based Interference}
        \label{fig:PECSInterf}
    \end{subfigure}
    \begin{subfigure}{.5\textwidth}
        \centering
        \includegraphics[width=.9\linewidth]{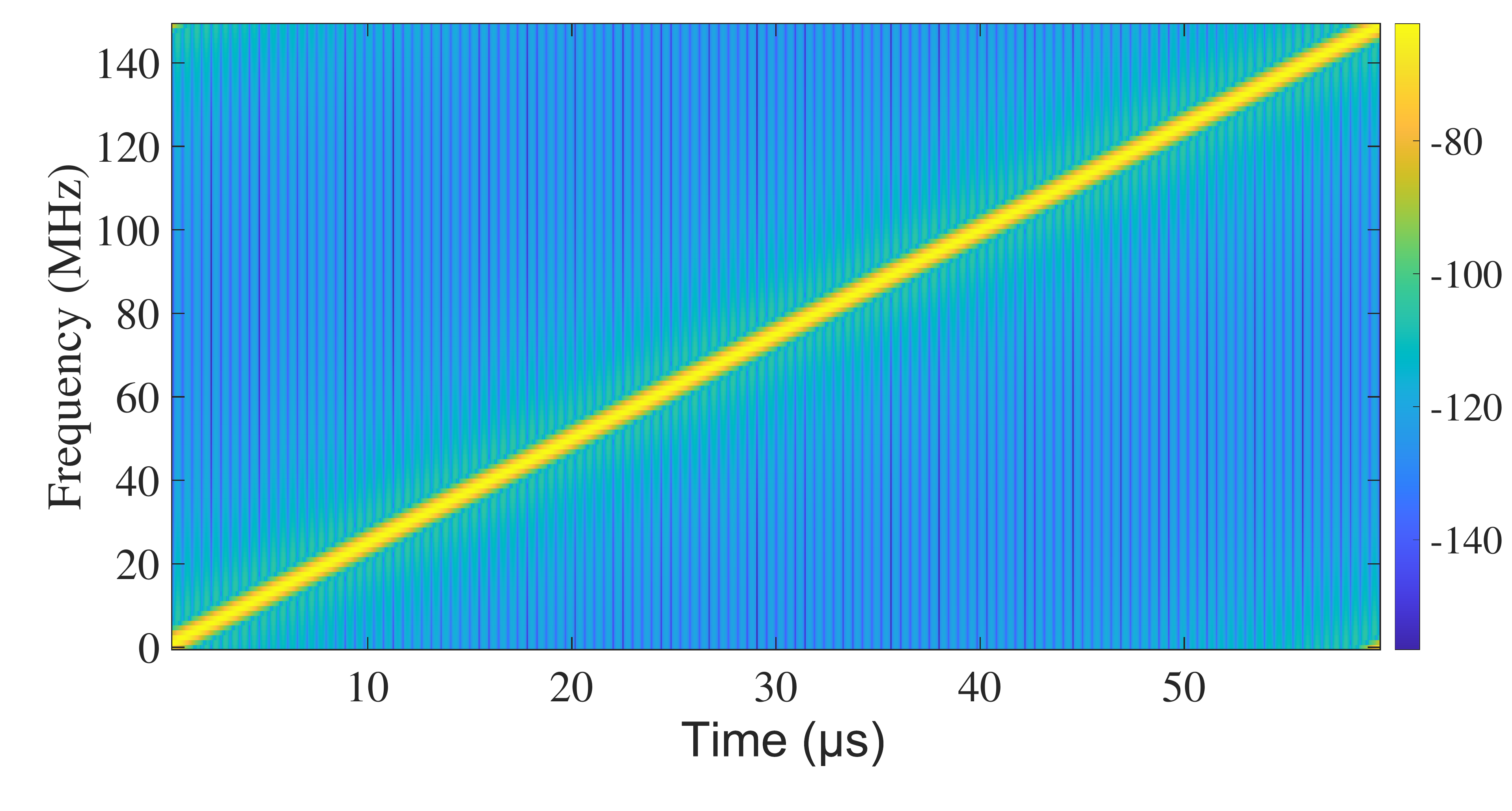}
        \caption{FMCW waveform spectrogram}
        \label{fig:FMCWSpctrgrm}
    \end{subfigure}
    \begin{subfigure}{.5\textwidth}
        \centering
        \includegraphics[width=.9\linewidth]{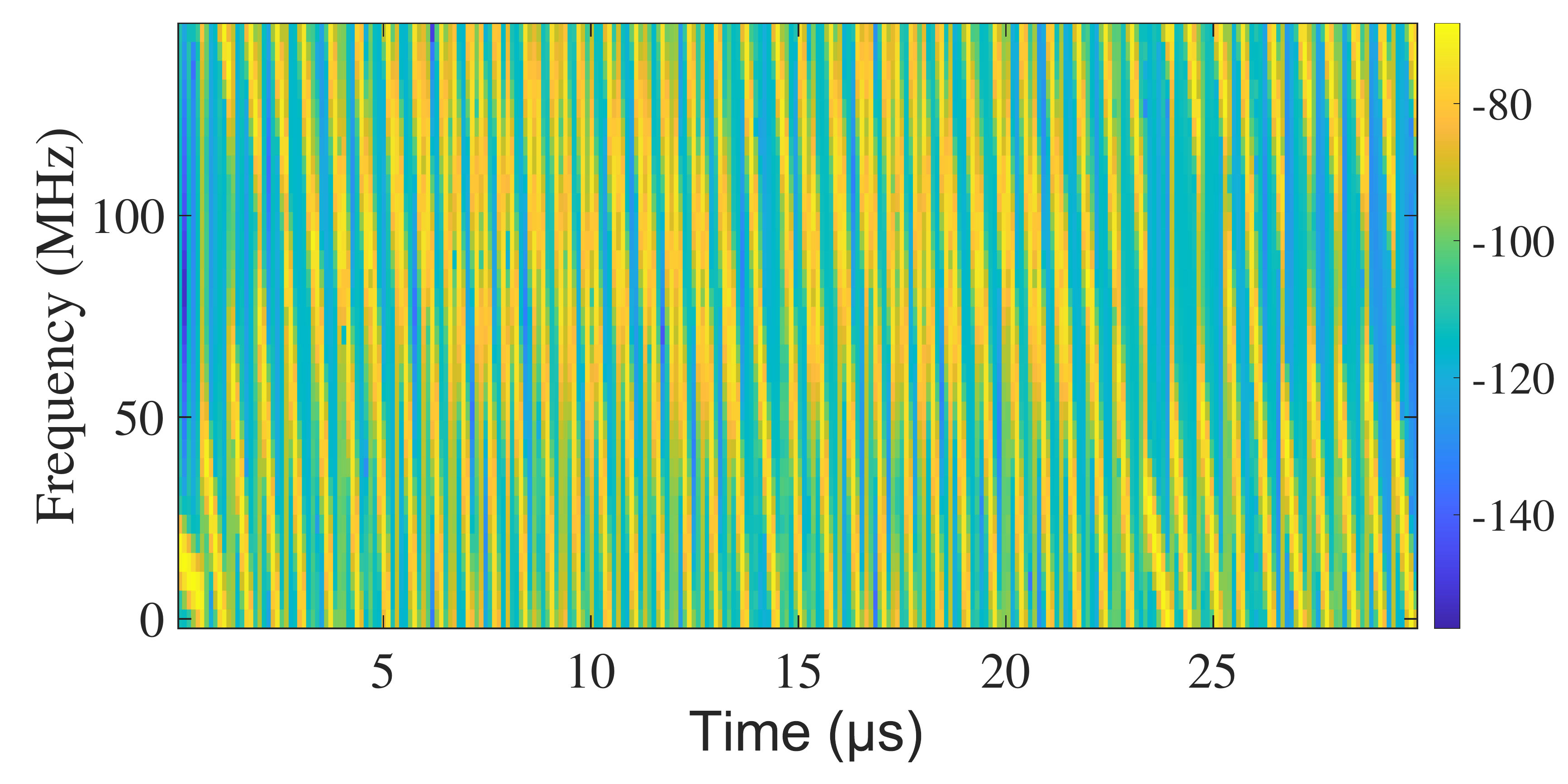}
        \caption{PMCW-PECS waveform spectrogram}
        \label{fig:PECSInterfSpctrgrm}
    \end{subfigure}
    \caption{scenario with FMCW and PMCW type waveforms being used by the Ego-Vehicle and target respectively (Power values on \textit{colorscale} are in dBm).}
    \label{fig:CmprsnFMCW_PMCW_Intrfrnce}
\end{figure}
The radar sensing parameters for both the \gls{FMCW} and \gls{PMCW} waveforms are described in the Table \ref{Table:RadarSensorParams}. In order to have a fair comparison, we have matched the \gls{TBP} and Doppler characteristics (pulse length, \gls{PRF}) of both the waveforms.
\figurename{~\ref{fig:CmprsnFMCW_PMCW_Intrfrnce}} shows the comparison of the range-Doppler map for similar-slope and sweeping slope interference in \gls{FMCW} and PMCW waveforms. 
In FMCW waveforms, the \gls{SINR} of the victim sensor in an interference free scenario is $13.06$dB.
When a single interferer is introduced, the similar-slope interference observed by the victim-sensor (driving scenarios mentioned previously in \figurename{~\ref{fig:BSD}, \ref{fig:LRR_LRR} \& \ref{fig:SRR_SRR}}) witnessed a \gls{SINR} of $10.8$dB (refer \figurename{~\ref{fig:SimSlopeInterf}}). Similarly with single interferer, in the case of sweeping slope interference \figurename{~\ref{fig:SweepSlopeInterf}} the \gls{SINR} was $12.9$dB. 
Although, these results have been derived after applying mitigation technique (i.e. varying \gls{PRF} at the victim sensor), the impact of interference on the detection capability of the victim sensor is evident and weak targets may not be detected in such a scenario. 
In the current case, the target, interferer position and Doppler have been deliberately kept different for purposes of plotting but if both have similar values, the detectability of target becomes even more arduous in the case of similar-slope interference.

On the other hand for \gls{PMCW} waveform derived from \gls{PECS}, both the victim and the interferer sensor transmit an optimized \gls{PECS} code sequence which is Doppler tolerant and unique. 
Noteworthy fact here is that both the victim and interferer are operating at the same bandwidth (i.e. same range resolution), center frequency and transmit power without causing \gls{MI} which proves the robustness of algorithm.
As is evident from \figurename{~\ref{fig:PECSInterf}}, there is no interference peak present in the range-Doppler spectrum and the \gls{SINR} in the case of single interferer is $20.7$dB better than FMCW in interference free scenario.
Further to this result, the reduction in the \gls{SINR} due to increase in the number of interferers was analyzed.
The results from the simulation show that as the number of interferers increase from one to five, the \gls{SINR} decreases to $16.8$dB.
As the presence of more than $10$ interferers in a certain driving scenario is very unlikely, therefore the interference results were limited to $10$ interferers and \gls{SINR} in this case was $14.65$dB.

The spectrum occupancy of \gls{FMCW} and \gls{PECS} based PMCW waveform is mentioned in \figurename{~\ref{fig:FMCWSpctrgrm}} and \figurename{~\ref{fig:PECSInterfSpctrgrm}}.
We observed that in the duration of single chirp (where chirp time period = sequence period), the frequency variation of \gls{FMCW} waveform is linearly increasing upto $150$MHz whereas in \gls{PECS} waveform multiple ramps of linearly increasing frequency can be seen in the time period. 
This again ascertains the fact that the frequency variation is linear (hence the quadratic phase variation). Randomness in the ramp slopes within a chirp duration and start phase is the reason for the uniqueness of each waveform and can be considered as an additional degree of freedom in the system design which leads to interference immune waveform. Thus, \gls{PECS} can be considered a new approach in this domain.

\subsubsection{Statistical Analysis}
\begin{figure}[!htbp]
    \centering
    \includegraphics[width=.95\linewidth]{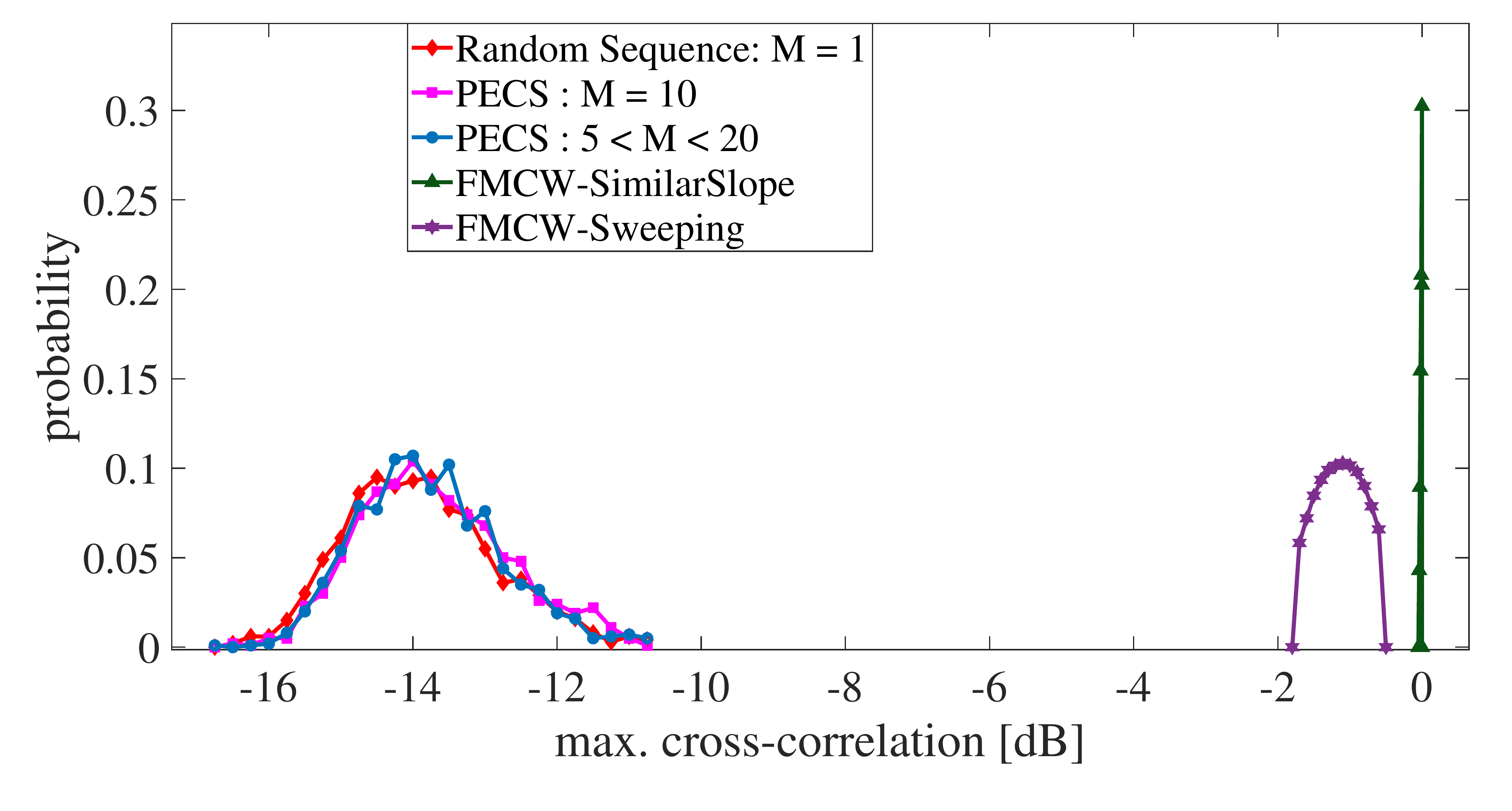}
    \caption{Statistical Interference Analysis - Probabilistic comparison of code sequences of length $N=100$ generated using (a) \gls{PECS} with fixed length sub-sequences $M=10$ (b) \gls{PECS} with random length sub-sequences $M_{min}=5$ and $M_{max}=20$ (c) completely random sequence.}
    \label{fig:StatisticalAnalysis}
\end{figure}

Further, in order to understand the interference immunity of different waveforms, a statistical validation of this concept was performed and its comparison was evaluated. 
In the first category, two independent sequences with same properties ($N = 100$, $M = 10$, $Q = 3$ and $p=10$) were generated using \gls{PECS} algorithm.
In the second and third category, two independent sequences were designed with such characteristics that they cause similar-slope and sweeping slope \gls{FMCW} interference, respectively. 
A statistical analysis of maximum cross-correlation was performed for $10^3$ trials and the results are given in \figurename{~\ref{fig:StatisticalAnalysis}} for each of the category.
The $x$-axis represents normalized cross-correlation of two signals and the $y$-axis represents the probability of occurrence in $10^3$ trials. 
As the \gls{PECS} framework allows additional degree of freedom in choosing the sub-sequence length randomly, the analysis for the PECS category was performed with $M \in [M_{min}, M_{max}]$ where $M_{min} = 5$ and $M_{max} = 20$. This category does not impact the performance of the maximum cross-correlation and is similar to the one which can be achieved if two completely random sequences are correlated.
The cross correlation results obtained from similar-slope and sweeping slope interference are centered at $-1.5$dB and $-0.2$dB respectively, whereas for PECS based sequences cross-correlation is centered at $-14.1$dB. As is evident from the plot, an improvement of approximately $12$dB is expected if PMCW waveform with PECS based code sequences are used for radar sensing as compared to the FMCW waveform.
This proves the credibility and robustness of the concept.
\section{Conclusion}\label{Sec:Conlcusion}
A stable design procedure has been proposed for obtaining polyphase sequences synthesized with a constraint of polynomial phase behavior optimized for minimal \gls{PSL}/\gls{ISL} for any sequence length. 
Results shown in the text indicate the robustness of the method for various scenarios which are prone to interference and offers user additional degrees of freedom to adapt the input parameters in order to design unique waveforms.
Its application to automotive scenarios with dense interference prove the feasibility of its use in practical radar systems and provides one of the solution to the current interference issues in multi-sensor applications.
The algorithm performance was tested in comparison to the other techniques present in the literature and convincing results were observed.
In addition the technique can be used to improve the performance of the state-of-the-art algorithms be extending them with inclusion of PECS which offers additional design parameters for waveform design.
The algorithm is implemented by means of FFT and least squares operations and therefore it is computationally efficient. 

\appendices
\section{Chirplike phase codes} \label{ap:Chirplike}
A summary of chirplike phase codes is reported in \tablename{~\ref{tab:ChirplikeCodes}}. 
\begin{table*}[h]
    \caption{Expressions of each code and their \gls{AF} \cite{Levanon_Radar}}
    \centering
    \begin{center}
        \begin{tabular}{|M{10em}|M{25em}|M{20em}|}\hline
            \textbf{Type of Code}   &    \textbf{Phase Expression}    & \textbf{\gls{AF}}     \\\hline
            \textbf{Frank Sequence} &   
            \begin{equation*}
            \begin{aligned}
                \phi_{n,k} & = 2\pi \frac{(n-1).(k-1)}{L}\\ \\
                \text{for}~~ &  1 \leq n \leq L, ~~1 \leq k \leq L,
            \end{aligned}
            \end{equation*}
            & \includegraphics[width=5cm, height=3cm]{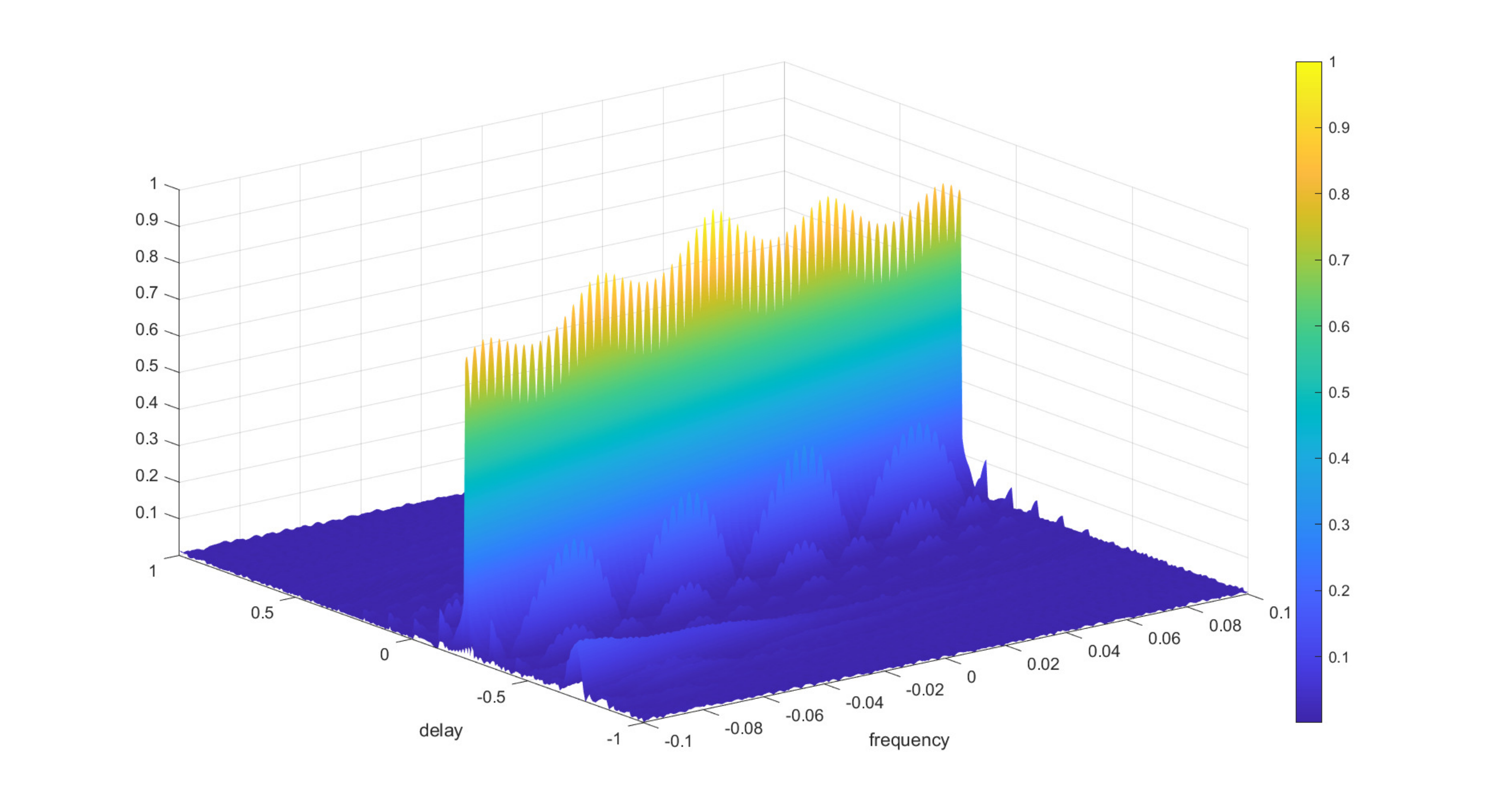}\\ \hline
            \textbf{$P_x$ Sequence} &  
                \begin{equation*}
                    \begin{aligned}
                        \phi_{n,k} &= 
                        \begin{cases}
                            \frac{2\pi}{L} \left[ \left(\frac{(L+1)}{2} - k\right) \left( \frac{(L+1)}{2} - n \right) \right], & \text{L even}\\ \\
                            \frac{2\pi}{L} \left[ \left(\frac{L}{2} - k\right) \left( \frac{(L+1)}{2} - n \right) \right], & \text{L odd}\\
                        \end{cases}
                        \\ \\~~~~~~~~~~~&\text{for}~~ 1 \leq n \leq L, ~~1 \leq k \leq L,
                    \end{aligned}
                \end{equation*}
            & \includegraphics[width=5cm, height = 3cm]{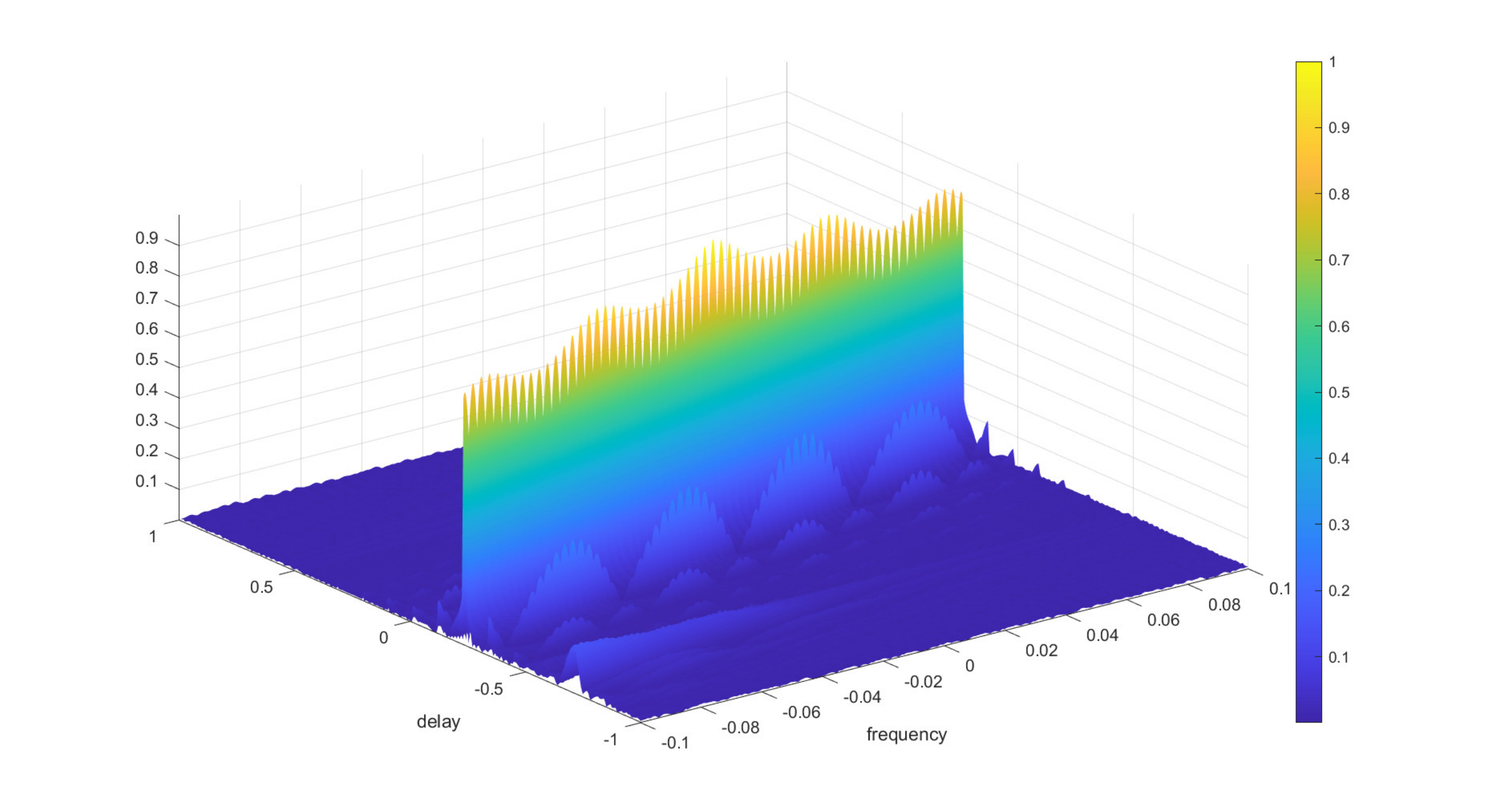} \\ \hline
            \textbf{$P_1$ Code}       
            &  
                \begin{equation*}
                    \begin{aligned}
                        \phi_{n,k} &= \frac{2\pi}{L} \left[ \left(\frac{(L+1)}{2} - n\right) \left( (n-1)L + (k-1) \right) \right]
                        \\ \\~~~~~~~&\text{for}~~ 1 \leq n \leq L, ~~1 \leq k \leq L,
                    \end{aligned}
                \end{equation*}
            & \includegraphics[width=5cm,height=3cm]{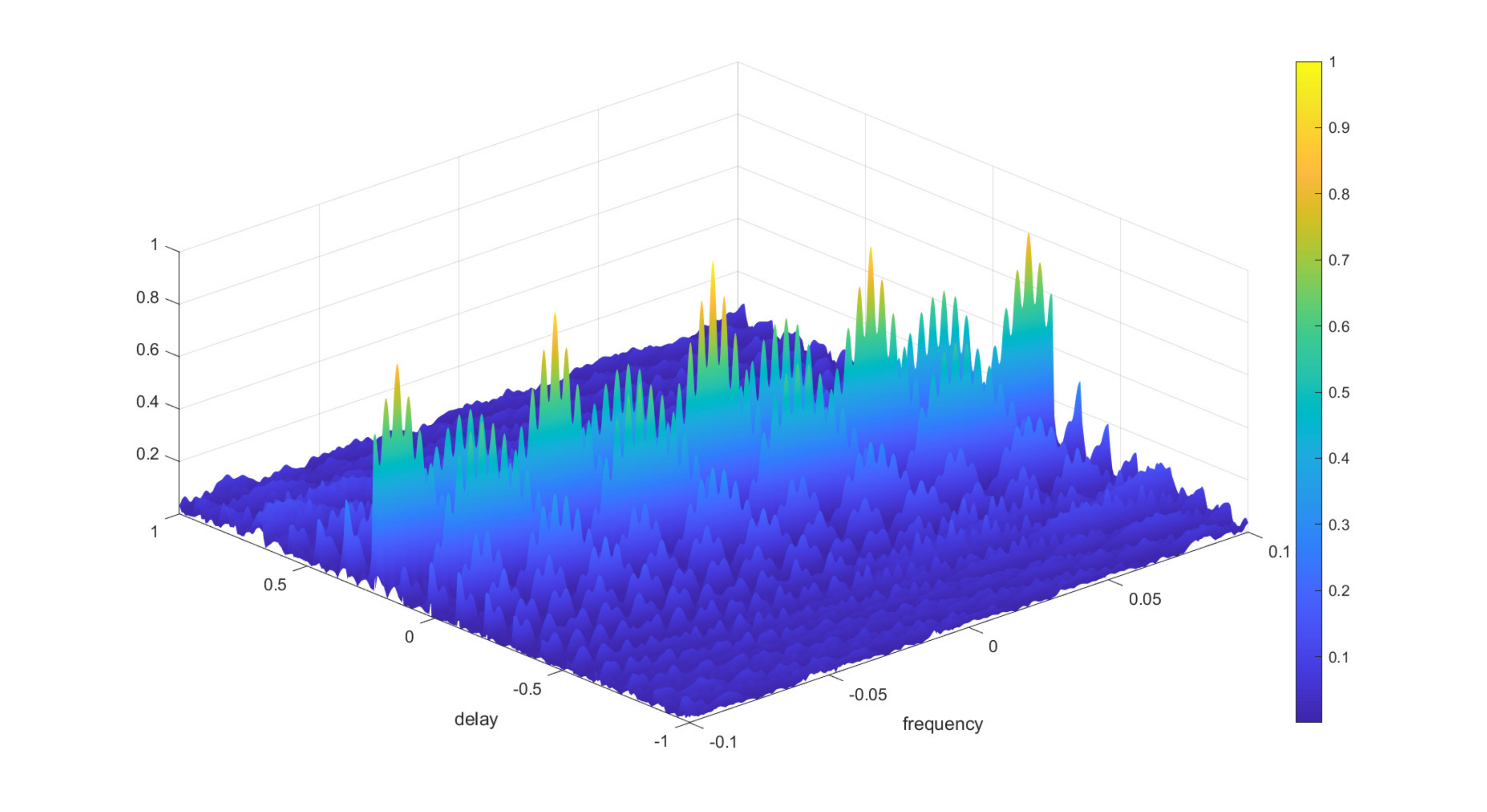} \\ \hline
            \textbf{$P_2$ Code}  
            &  
                \begin{equation*}
                    \begin{aligned}
                        \phi_m &= \frac{2\pi}{M} \left[ \frac{(m-1)^2}{2} \right] 
                        \\ \\\text{for}&~~ 1 \leq m \leq M,
                    \end{aligned}
                \end{equation*}
            & \includegraphics[width=5cm, height=3cm]{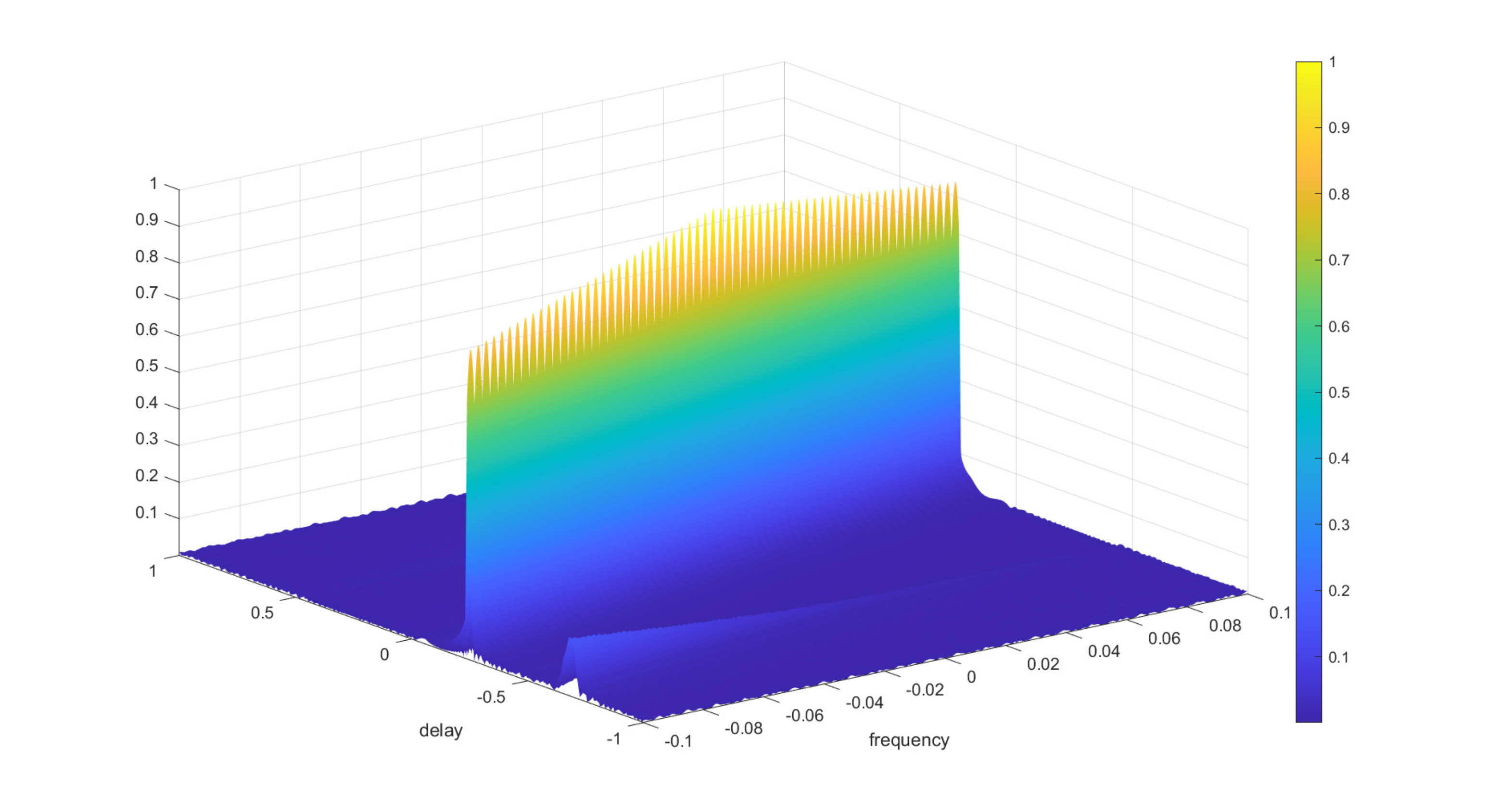} \\ \hline
            \textbf{$P_4$ Code}  
            &  
                \begin{equation*}
                    \begin{aligned}
                        \phi_m &= \frac{2\pi}{M} (m-1) \left[ \frac{m-1-M}{2} \right] 
                        \\ \\&\text{for}~~ 1 \leq m \leq M,
                    \end{aligned}
                \end{equation*}
            &  \includegraphics[width=5cm, height = 3cm]{images/P4_AF-eps-converted-to} \\ \hline
            \textbf{Zadoff Code} 
            &  
                \begin{equation*}
                    \begin{aligned}
                        \phi_m &= \frac{2\pi}{M} (m-1)\left[ r.\left( \frac{M-1-m}{2}\right) - q  \right] 
                        \\ \\\text{for}&~~ 1 \leq m \leq M, ~~0 \leq q \leq M \\
                        \text{where}&\text{ $M$ is any integer and $r$ is any}\\
                        &\text{ integer relatively prime to $M$},
                    \end{aligned}
                \end{equation*}
            & \includegraphics[width=5cm, height=3cm]{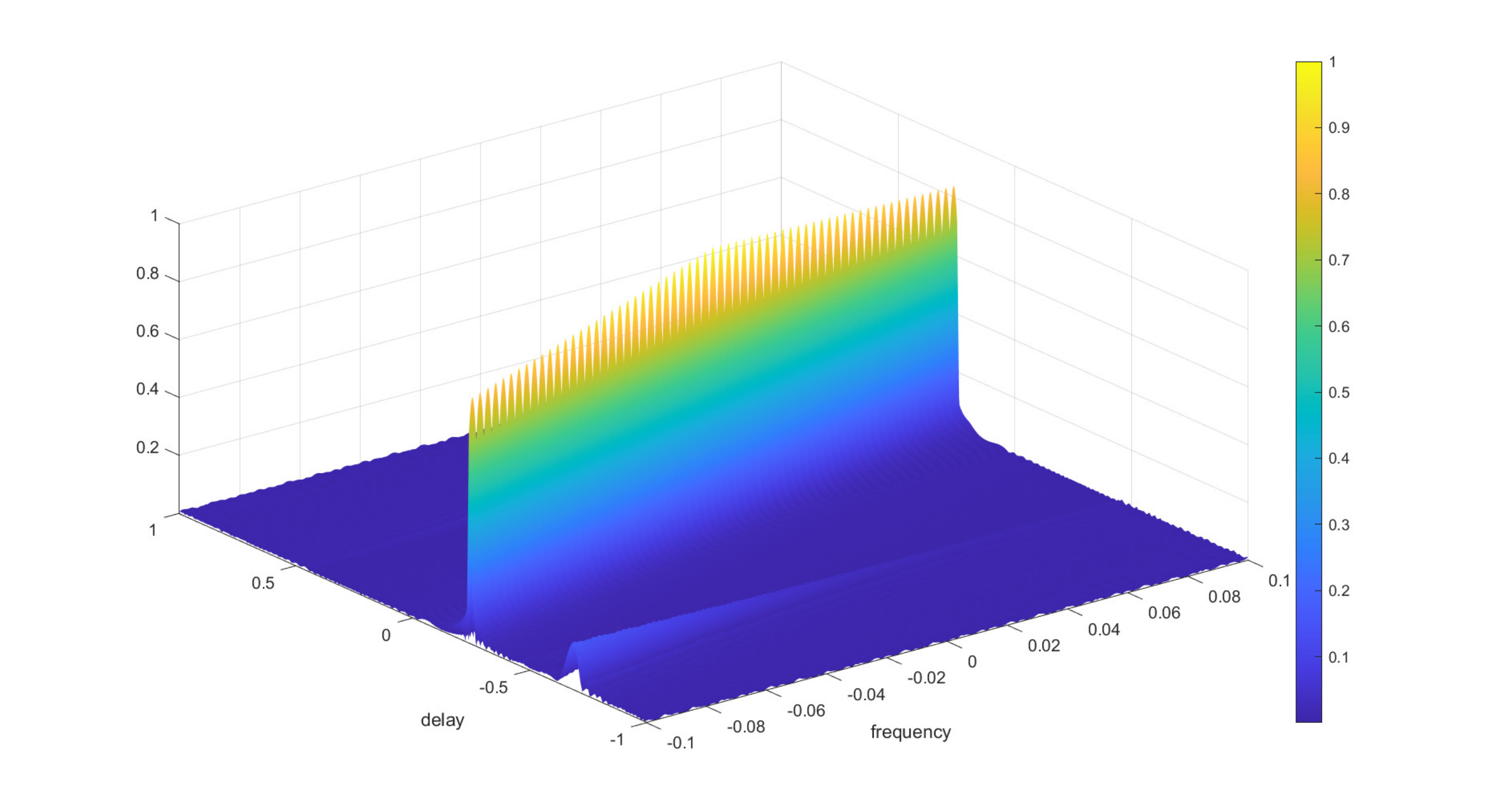}\\ \hline
            \textbf{Golomb Sequence} 
            &  
                \begin{equation*}
                    \begin{aligned}
                        \phi_m &= \frac{2\pi}{M} r'' \left[ \frac{(m-1)(m)}{2} \right] 
                        \\ \\~~~~\text{for}~~ &1 \leq m \leq M,\\
                        \text{where}&\text{ $M$ is any integer and}\\
                        \text{$r''$ is any}&\text{ integer relatively prime to $M$},
                    \end{aligned}
                \end{equation*}
            & \includegraphics[width=5cm, height=3cm]{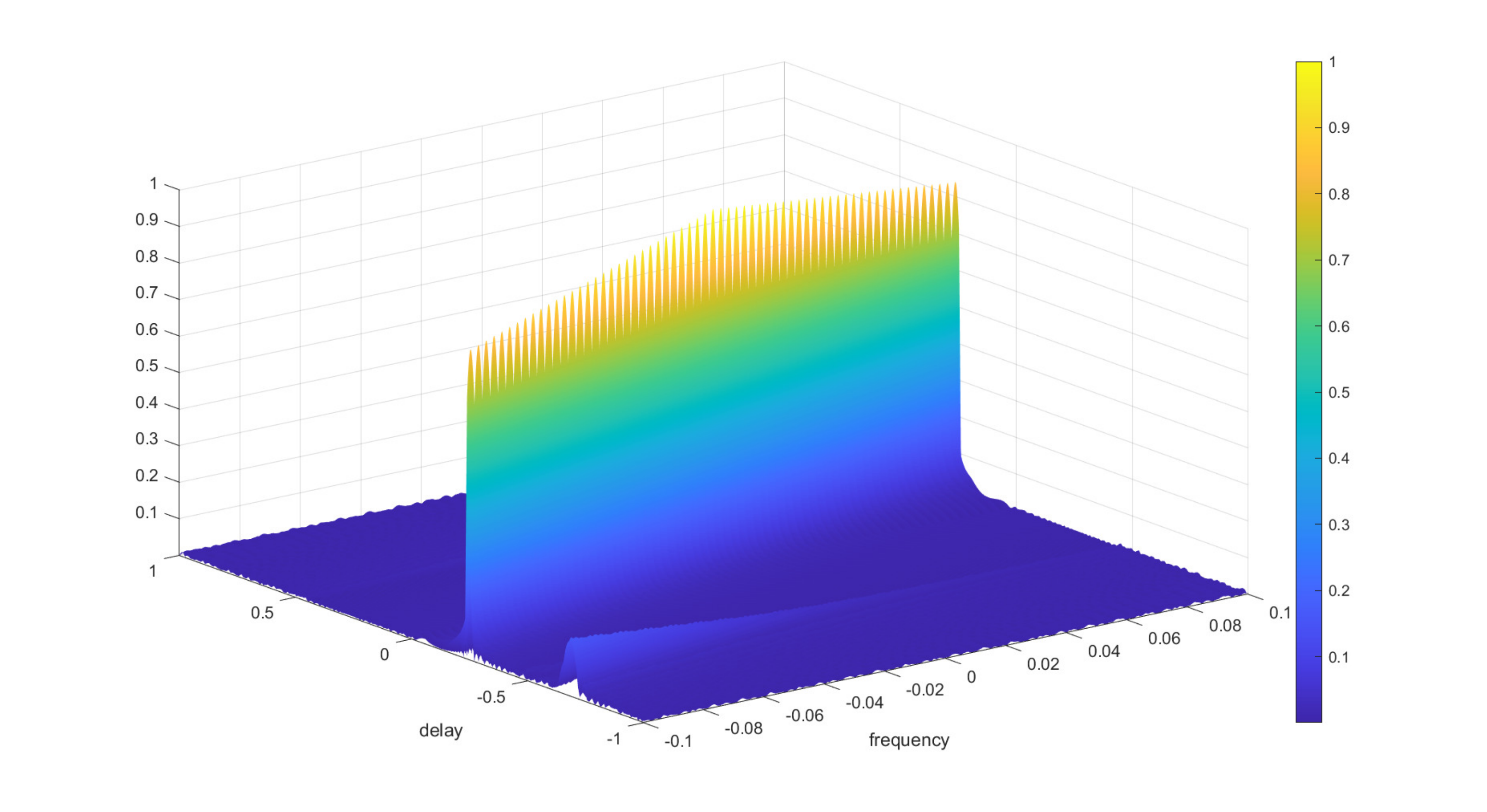}
            \\ \hline
        \end{tabular}
    \end{center}
    \label{tab:ChirplikeCodes}
\end{table*}
Note that Frank code is derived from the phase history of a linearly frequency stepped pulse. The main drawback of the Frank code is that it only applies for codes of perfect square length ($M = L^2$) \cite{Levanon_Radar}. 
P1, P2, and Px codes are all modified versions of the Frank code, with the DC frequency term in the middle of the pulse instead of at the beginning. 
Unlike Frank, P1, P2 and Px codes which are only applicable for perfect square lengths ($M = L^2$), the Zadoff code is applicable for any length. 

Chu codes are important variant of the Zadoff code, and Golomb, P3, and P4 codes are specific cyclically shifted and decimated versions of the Zadoff-Chu code. Indeed, as P1 and P2/Px codes were linked to the original Frank code, similarly, P3, P4 and Golomb polyphase codes are linked to the Zadoff-Chu code, and are given for any length. 

\section{Derivation of MM-PSL}\label{Ap:MPSL_Dervn}
We aim to obtain a minimizer of \eqref{eq:PSLOptC2}
iteratively  using \gls{MM} algorithm. We can majorize $|r_k|^p$ by a quadratic function locally \cite{7362231}. 
From the literature (see \cite{7362231,7967829} for more details), it is known that given $|r_k^{(i)}|$ at iteration $i$,
$|r_k|^p$ can be  majorized at $|r_k^{(i)}|$ over $[0,t]$ by 
\begin{equation}
    \Tilde{\alpha}_k|r_k|^2 + \Tilde{\beta}_k|r_k| + \Tilde{\alpha}_k\left|r_k^{(i)}\right|^2 - (p-1)\left| r_k^{(i)}\right|^p,
\end{equation}
where
\begin{equation} \label{eq:t_ak_beta}
    \begin{aligned}
    t & = \left(\sum_{k = 1}^{N-1}|\widetilde{r}_{k}^{(i)}|^{p}\right)^{\frac{1}{p}} \\
    \Tilde{\alpha}_k &= \frac{t^p - \left| r_k^{(i)}\right|^p - p\left| r_k^{(i)}\right|^{p-1} (t - \left| r_k^{(i)}\right|)}{(t - \left| r_k^{(l)}\right|)^2},\\
    \Tilde{\beta}_k &= p\left| r_k^{(i)}\right|^{p-1} - 2\Tilde{\alpha}_k\left| r_k^{(i)}\right|.
    \end{aligned}
\end{equation}
The surrogate function is then given by (ignoring the constant terms)
\begin{equation}\label{eq:PSLOptC3}
     \sum_{k=1}^{N-1}(\Tilde{\alpha}_k|r_k|^2 + \Tilde{\beta}_k|r_k|)
\end{equation}
The first term in the objective function is just the weighted \gls{ISL} metric with weights $w_k = \Tilde{\alpha}_k$, which can be majorized at $\mathbf{x}^{(i)}$ by (with constant terms ignored) \cite{7362231}
\begin{equation}\label{eq:Song1}
    \mathbf{x}^H \left( \boldsymbol{R} - \lambda_{max}(\boldsymbol{L})\mathbf{x}^{(i)} (\mathbf{x}^{(i)})^H \right)\mathbf{x},
\end{equation}
\begin{equation*}
    \begin{aligned}
    \boldsymbol{R} = \sum_{k = 1-N}^{N-1}& w_k r_{-k}^{(i)}\boldsymbol{U}_k, ~ \boldsymbol{L} = \sum_{k = 1-N}^{N-1} w_k \text{vec}(\boldsymbol{U}_k) \text{vec}(\boldsymbol{U}_k)^H,\\
    \lambda_{max}(\boldsymbol{L}) &= \max_k{\{w_k(N-k)|k = 1,\ldots,N-1\}},\\
    \end{aligned}
\end{equation*}
and $\boldsymbol{U}_k$, $k = 0,\ldots,N-1$ to be $N \times N$ Toeplitz matrix with the $k$-th diagonal elements being $1$ and $0$ elsewhere.
For the second term, since it can be shown that $\Tilde{\beta}_k \leq 0$, we have
\begin{equation}\label{eq:Song2}
    \sum_{k=1}^{N-1}\Tilde{\beta}_k |r_k| \leq \frac{1}{2}\mathbf{x}^H \left( \sum_{k=1-N}^{N-1} \Tilde{\beta}_k  \frac{r_k^{(i)}}{|r_k^{(i)}|} \boldsymbol{U}_{-k} \right) \mathbf{x}.
\end{equation}
By adding the two majorization functions, i.e. (\ref{eq:Song1}) and (\ref{eq:Song2}), and other simplifications as given in \cite{7362231}, we derive the majorizer of (\ref{eq:PSLOptC3}) as
\begin{equation}
\mathbf{x}^H \left( \widetilde{\boldsymbol{R}} - \lambda_{max}(\boldsymbol{L}) \mathbf{x}^{(i)} (\mathbf{x}^{(i)})^H \right) \mathbf{x}
\end{equation}
where $\hat{w}_{-k} = \hat{w}_k = \Tilde{\alpha}_k + \frac{\Tilde{\beta}_k}{2|r_k^{(i)}|} = \frac{p}{2} |r_k^{(i)}|^{p-2}, k = 1,\ldots, N-1$ and $\widetilde{\boldsymbol{R}} = \sum_{k =1-N}^{N-1} \hat{w}_k r_{-k}^{(i)}\boldsymbol{U}_k$.
Finally, after performing one more majorization step as mentioned in \cite{7362231}, we derive another majorizer as
\begin{equation}\label{eq:PSLfinalOptDervn}
|| \mathbf{x} - \mathbf{y}||_2
\end{equation}
where $\mathbf{y} = \left( \lambda_{\text{max}}(\boldsymbol{L})N + \lambda_u \right) \mathbf{x}^{(i)} - \widetilde{\boldsymbol{R}}\mathbf{x}^{(i)}$,
which forms the basis of our problem in \eqref{eq:PSLfinalOpt}.

\bibliographystyle{IEEEtran}
\bibliography{ref1}

\end{document}